\documentclass[a4paper,12pt]{article}

\usepackage{natbib}
\usepackage{graphicx}
\usepackage{txfonts}
\usepackage{hyperref}
\usepackage{authblk}
\usepackage{subfig}
\usepackage{makecell}
\usepackage[]{hyperref}
\hypersetup{colorlinks=true,citecolor=blue}
\usepackage[utf8]{inputenc}
\usepackage[left=2cm,right=2cm,top=2cm,bottom=2cm]{geometry}
\DeclareUnicodeCharacter{00A0}{ }

\def\Msun{M_\odot}
\def\Rsun{R_\odot}

\def\Rp{R_{\rm p}}

\def\Rs{R_{\star}}

\def\Rearth{R_\oplus}

\usepackage[normalem]{ulem}
\usepackage{color}
\definecolor{green}{RGB}{0,200,0}

\title{A review of possible planetary atmospheres in the TRAPPIST-1 system}
\author[1]{Martin Turbet}
\author[1]{Emeline Bolmont}
\author[1]{Vincent Bourrier}
\author[2]{Brice-Olivier Demory}
\author[3]{J\'er\'emy Leconte}
\author[4]{James Owen}
\author[5]{Eric T. Wolf}

\affil[1]{Observatoire Astronomique de l'Universit{\'e} de Gen{\`e}ve, 51 chemin de P{\'e}gase, 1290 Sauverny, Switzerland}
\affil[2]{University of Bern, Center for Space and Habitability, Gesellschaftsstrasse 6, CH-3012, Bern, Switzerland}
\affil[3]{Laboratoire d'astrophysique de Bordeaux, Univ. Bordeaux, CNRS, B18N, all\'ee Geoffroy Saint-Hilaire, 33615 Pessac, France}
\affil[4]{Astrophysics Group, Department of Physics, Imperial College London, Prince Consort Rd, London, SW7 2AZ, United Kingdom}
\affil[5]{Laboratory for Atmospheric and Space Physics, University of Colorado, Boulder, CO 80309 USA}

\date{}

\begin{document} 





\maketitle
\newpage
\section*{Abstract}

TRAPPIST-1 is a fantastic nearby ($\sim$~39.14~light years) planetary system made of at least seven transiting terrestrial-size, 
terrestrial-mass planets all receiving a moderate amount of irradiation. To date, this is the most observationally favourable system 
of potentially habitable planets known to exist.
   Since the announcement of the discovery of the TRAPPIST-1 planetary system in 2016, a growing number of techniques and approaches 
have been used and proposed to characterize its true nature. Here we have compiled a state-of-the-art overview of all the observational 
and theoretical constraints that have been obtained so far using these techniques and approaches. 
The goal is to get a better understanding of whether or not TRAPPIST-1 planets can have atmospheres, and if so, what they are made of. 
For this, we surveyed the literature on TRAPPIST-1 about topics as broad as irradiation environment, planet formation and migration,  
orbital stability, effects of tides and Transit Timing Variations,  transit observations, stellar contamination, density measurements, 
and numerical climate and escape models. Each of these topics adds a brick to our understanding of the likely - or on the contrary unlikely - 
atmospheres of the seven known planets of the system. We show that (i) Hubble Space Telescope transit observations, (ii) bulk density measurements 
comparison with H$_2$-rich planets mass-radius relationships, (iii) atmospheric escape modelling, and (iv) gas accretion modelling altogether offer 
solid evidence against the presence of hydrogen-dominated -- cloud-free and cloudy -- atmospheres around TRAPPIST-1 planets. This means that the 
planets are likely to have either (i) a high molecular weight atmosphere or (ii) no atmosphere at all. There are several key challenges ahead to 
characterize the bulk composition(s) of the atmospheres (if present) of TRAPPIST-1 planets. The main one so far is characterizing and correcting 
for the effects of stellar contamination. Fortunately, a new wave of observations with the James Webb Space Telescope and near-infrared high-resolution 
ground-based spectrographs on existing very large and forthcoming extremely large telescopes will bring significant advances in the coming decade.


\section{Introduction}
\label{section_introduction}

Nearly 25 years after the first detection of an exoplanet orbiting a solar-type star \citep{Mayor:1995}, several thousand extrasolar planets have been detected at a frenetic rate \citep{Schneider:2011,Akeson:2013}. While the science of exoplanets initially focused mainly on the detection of exoplanets, it is gradually moving towards their characterisation. A large number of space missions (e.g. Hubble Space Telescope, James Webb Space Telescope, ARIEL) and ground-based instruments (e.g. HARPS, VLT-ESPRESSO, ELT-HIRES) mounted on large telescopes are in the process of thoroughly characterizing the atmospheric composition, chemistry, clouds, and many other properties of large warm exoplanets which are the most amenable for in-depth characterization. This opens up the field of comparative exoplanetology. 
To a lesser extent, the detection rate of small possibly temperate exoplanets, much more difficult to observe, 
has also exploded in recent years. Nearly 40 exoplanets with a mass and/or radius similar to that of the Earth,
and with incident fluxes close to that received on Earth, have been detected so far. 
However, the vast majority of these planets are inaccessible to our telescopes for 
the characterization of their atmosphere and surface. 
The TRAPPIST-1 system -- at the heart of this review -- provides us with a natural laboratory to characterize for the first time, in a few years only, the atmospheres and surfaces of temperate rocky planets outside the solar system. The exploration of TRAPPIST-1 is likely to revolutionize, through comparative planetology, all the knowledge we have accumulated so far about the evolution of the atmospheres and habitability of terrestrial planets.

\medskip

TRAPPIST (the TRansiting Planets and PlanestIsimals Small Telescope; \citealt{Gillon:2011,Gillon:2013}), a small 60-cm ground-based telescope located at the ESO La Silla Observatory in Chile, 
monitored the brightness of the star 2MASS J23062928-0502285 (a.k.a. EPIC 246199087, or simply TRAPPIST-1) for 245~hours over 62 nights from 17 September to 28 December 2015. 
The analysis of the light curves measured during these observation series \citep{Gillon:2016} led to the detection of 
two transit-like signatures with amplitudes close to 1$\%$ named TRAPPIST-1b and c, and a tentative detection of a third planet for which the orbital period was not known. Starting 19 September 2016, nearly 20 days of quasi-continuous photometric monitoring with NASA's Spitzer Space Telescope\footnote{This photometric campaign was complemented by TRAPPIST (South and North), LT/IO:O, UKIRT/WFCAM, WHT/ACAM, SAAO-1m/SHOC, VLT/HAWK-1, HCT/HFOSC and HST/WFC3 observations, listed by decreasing number of hours of observation \citep{Gillon:2017}.} identified that the third signal measured by TRAPPIST was in fact a combination of multiple signals due to the presence of several additional planets in the system \citep{Gillon:2017}, named TRAPPIST-1d, e, f and g. These observations also led to the detection of an orphan transit, indicating the possible presence of a seventh planet in the system, named TRAPPIST-1h. The existence of this seventh planet was later confirmed \citep{Luger:2017} through a 79 consecutive days observation campaign (starting 15 December 2016) with the NASA's Kepler Space Telescope in its two-reaction wheel mission (a.k.a. K2; \citealt{Howell:2014}). Follow-up observations \citep{Delrez:2018,Ducrot:2018,Burdanov:2019,Ducrot:2020} later not only confirmed the existence of at least seven temperate, terrestrial-size planets around the star TRAPPIST-1, but also helped to better constrain their main properties (summarized in Table~\ref{table_planets}).

    \begin{table*}[ht!]
    \centering 
    \tiny
    \setlength{\tabcolsep}{4.5pt} 
    \renewcommand{\arraystretch}{1.5} 
    \begin{tabular}{ccccccccc}        
    \hline
    \textbf{Stellar Parameters} & & & \textbf{Value}& & & \textbf{Reference} & & \\
    \hline
    \textbf{Star} & & & \textbf{TRAPPIST-1} & & & &  \\
    Mass $M_{\star}$ ($M_{\odot}$)& & & $0.089 \pm 0.007 $ & & & 1 & & \\
    Radius $R_{\star}$ ($R_{\odot}$) & & & $0.121 \pm 0.003 $ & & & 1 & & \\
    Luminosity   $L_{\star}$ ($L_{\odot}$) & & & $0.00055 \pm 0.00002 $ & & & 1,4 & & \\
    Effective temperature $(K)$ & & & $2516 \pm 41 $ & & & 1 & & \\
    Metallicity, [Fe/H] $(dex)$ & & & $0.04 \pm 0.08 $ & & & 2 & & \\
    Age [Gigayears] & & & $7.6 \pm 2.2 $ & & & 3 & & \\
    Distance [light years] & & & $39.14 \pm 0.01 $ & & & 4 & & \\
    \hline
    \textbf{Planets}  & \textbf{b} & \textbf{c} & \textbf{d} & \textbf{e} & \textbf{f} & \textbf{g} & \textbf{h} & \textbf{References} \\    
    Periods (days) & 1.51088 & 2.42179 & 4.04976 & 6.09981 & 9.20587 & 12.35381 & 18.76727 & 5 \\
    & $\pm$ 0.00001 & $\pm$ 0.00001 & $\pm$ 0.00004 & $\pm$ 0.00006 & $\pm$ 0.0001 & $\pm$ 0.0001 & $\pm$ 0.00004 & 6 \\
    
    Transit impact \\ parameter b ($R_{*}$) & 
    $0.16  \pm 0.08 $  & 
    $0.15 \pm 0.09 $ & 
    $0.08 \pm 0.1$ & 
    $0.24 \pm 0.05$ & 
    $0.34 \pm 0.04$  & 
    $0.41 \pm 0.03$ & 
    $0.39 \pm 0.04$ & 5 \\
    
    Transit duration (min) & 
    36.2 $\pm$ 0.1 & 
    42.3 $\pm$ 0.1 &
    49.3 $\pm$ 0.4 & 
    55.9 $\pm$ 0.4 & 
    63.2 $\pm$ 0.4 & 
    68.5 $\pm$ 0.4 &  
    76.9 $\pm$ 1 & 5 \\
    
    Inclination i ($^{\circ}$) & 
    $89.6 \pm 0.2$ & 
    $89.7 \pm 0.2$ & 
    $89.9 \pm 0.1$ & 
    $89.73\pm 0.05$  & 
    $89.72 \pm 0.03$ & 
    $89.72 \pm 0.02$ & 
    $89.80 \pm 0.02$ & 5 \\
    
    Semi major axis \\ a ($10^{-3}AU$) & 
    $11.54775$ & 
    $15.8151$ & 
    $22.2804$ & 
    $29.2829$ & 
    $38.5336$ & 
    $46.8769$ &
    $61.9349$ & 6 \\
    & $\pm 6 \times$10$^{-5}$ & 
    $\pm 2 \times$10$^{-4}$ & 
    $\pm 4 \times$10$^{-4}$ & 
    $\pm 4 \times$10$^{-4}$ & 
    $\pm 5 \times$10$^{-4}$ & 
    $\pm 3 \times$10$^{-4}$ &
    $\pm 8 \times$10$^{-4}$ & \\

    Eccentricity \\ e ($10^{-3}$) & 
    $6.22 \pm 3$ & 
    $6.54 \pm 2$ & 
    $8.37 \pm 0.9$ & 
    $5.10 \pm 0.6$ & 
    $10.1 \pm 0.7$ & 
    $2.08 \pm 0.5$ &
    $5.67 \pm 1$ & 6 \\
    
    Irradiation $S_{p}$ ($S_{\odot}$) & 
    $4.11  \pm 0.14$ &  
    $2.19 \pm 0.08$  & 
    $1.10 \pm 0.04$ & 
    $0.638 \pm 0.02 $ & 
    $0.369 \pm 0.013$ & 
    $0.250 \pm 0.009$ & 
    $0.143 \pm 0.005$ & 1,6 \\
    
    Equilibrium \\ temperature $T_{eq}$ (K)$^{a}$  & 
    $396.5 \pm 3.7$ & 
    $338.8 \pm 3.1$ &  
    $285.4 \pm 2.6$ & 
    $249.0 \pm 2.3$ & 
    $217.0 \pm 2.0$ & 
    $196.8 \pm 1.8$ & 
    $171.2 \pm 1.5$ & \\
    
    Dayside Equilibrium \\ temperature $T_{eq}$ (K)$^{b}$ & 
    $506.0 \pm 4.7$ & 
    $433.0 \pm 4.1$ &  
    $364.8 \pm 3.4$ & 
    $318.1 \pm 3.0$ & 
    $277.4 \pm 2.6$ & 
    $251.5 \pm 2.3$ & 
    $218.8 \pm 2.0$ & \\
    
    Radius $R_{p}$ ($R_{\oplus}$) & 
    $1.12 \pm 0.02$ & 
    $1.11 \pm 0.02$ &  
    $0.80 \pm 0.02$ & 
    $0.93 \pm 0.02$ & 
    $1.05 \pm 0.02$  & 
    $1.14 \pm 0.04$  & 
    $0.78 \pm 0.04$ & 7\\
    
    Mass $M_{p}$ ($M_{\oplus}$) & 
    $1.02 \pm 0.14$ & 
    $1.16 \pm 0.13$ &  
    $0.30 \pm 0.04$ & 
    $0.77 \pm 0.08$ & 
    $0.93 \pm 0.08$  & 
    $1.15 \pm 0.1$  & 
    $0.33 \pm 0.05$ & 6\\
    
    Density $\rho_{p}$ ($\rho_{\oplus}$) & 
    $0.73 \pm 0.09$ & 
    $0.85 \pm 0.08$ &  
    $0.59 \pm 0.07$ & 
    $0.96 \pm 0.07$ & 
    $0.80 \pm 0.04$  & 
    $0.78 \pm 0.04$  & 
    $0.70 \pm 0.12$ & 6,7\\
    \hline                                   
    \end{tabular}
    \caption{Updated stellar and planetary parameters of the TRAPPIST-1 system along with their 1~$\sigma$ uncertainty. 
The equilibrium temperatures were derived assuming a null albedo. They can be rescaled to a non-zero albedo $A$ by being multiplied by $(1-A)^{1/4}$. 
The period uncertainties were derived using consecutive transit timing variations (TTVs) of \citet{Grimm:2018}. $^{a}$ for a null albedo; $^{b}$ for a 
null albedo, a synchronous rotation, and no atmosphere at all. \newline \textbf{References}: (1) \citet{vangrootel:2018}, (2) \citet{Gillon:2016}, 
 (3) \citet{Burgasser:2017}, (4) \citet{Kane:2018} using Gaia DR2 parallaxes, (5) \citet{Delrez:2018}, 
 (6) \citet{Grimm:2018}, (7) \citet{Ducrot:2020} using Spitzer IRAC channel 2 (4.5~$\mu$m) transit depths.}
\label{table_planets} 
    \end{table*}

The TRAPPIST-1 system is exceptional because it is -- through a series of techniques that will be discussed in this review paper -- the most observationally favourable system of potentially habitable planets (i.e. planets that could have liquid water on their surface and can therefore have the preconditions for life as we know it on Earth) known to exist. This mostly results from a subtle combination of (i) its proximity (39.14 light years from us), (ii) the fact that planets are transiting (frequently) in front of their star, and (iii) the extremely small radius of the ultra-cool dwarf host star TRAPPIST-1.

Since the discovery of the TRAPPIST-1 system was announced in 2016 \citep{Gillon:2016}, a flourishing number of multidisciplinary scientific works have been carried out (about 170 peer-reviewed publications per year in 2018 and 2019; source: NASA/ADS) to obtain information on the true nature of the TRAPPIST-1 system. The main purpose of this review is to set the stage of what we think we have learned so far about this system and what this implies for the presence and nature (if any) of TRAPPIST-1 planetary atmospheres. The second purpose of this review is to discuss the future opportunities available -- with increasingly large telescopes and increasingly performant instruments -- to characterize the nature of the TRAPPIST-1 planets, particularly through their atmosphere, and to identify the potential challenges that lie ahead.

Firstly, we review in Section~\ref{section_stellar_environment} previous works on the stellar environment (irradiation, stellar activity) in order to identify the context in which the planets of the TRAPPIST-1 system and their atmospheres have evolved. Secondly, we present in Section~\ref{section_orbital_architecture} previous works carried out on the orbital architecture of the system, which contains key information on (1) how the planets were formed (including how much volatile they accreted in the first place), (2) their mode of rotation and (3) their masses. These are three key pieces of information for interpreting the nature of the TRAPPIST-1 planets and their possible atmospheres. Thirdly, we gather and then discuss in Section~\ref{section_transit} all existing multi-wavelengths transit observations (with HST, Spitzer, K2 and ground-based telescopes) of TRAPPIST-1 planets. These observations can not only help us to eliminate a number of hypotheses about the compositions of TRAPPIST-1 planetary atmospheres , but also to identify key challenges for their spectroscopic characterization with the future generation of large telescopes. Fourthly, we review in Section~\ref{section_numerical_modelling} the theoretical and numerical advances that have been made -- using sophisticated numerical atmospheric and escape models -- in recent years on the atmospheres of planets orbiting ultra-cool stars, and what this implies for the range of possible compositions of planetary atmospheres in the TRAPPIST-1 system. Fifthly, we provide in Section~\ref{section_future_prospects} an overview of the near and far-future prospects to detect and characterize (if present) these atmospheres. While there are many challenges ahead, the prospects for future characterization are extremely promising. Finally, the most important conclusions of this review are summarized in Section~\ref{section_conclusions}.

\section{Constraints from the stellar environment}
\label{section_stellar_environment}

A fundamental characteristic of the TRAPPIST-1 planets is that they orbit a very small, very cold and very low mass star. The evolution of the luminosity and the activity of such stars have severe consequences on the evolution of planetary atmospheres, which we review in this section.

\subsection{Temporal evolution of TRAPPIST-1 luminosity and runaway greenhouse}

Ultra-cool stars such as TRAPPIST-1 can stay for hundreds of millions of years in the Pre Main Sequence (PMS) phase, 
a phase during which their luminosity can decrease possibly by several 
orders of magnitude \citep{Chabrier:1997,Baraffe:1998,Baraffe:2015}. During this PMS phase, 
planets are exposed to strong irradiation, which make them very sensitive to atmospheric processes such 
as hydrodynamical escape \citep{Vidal-Madjar:2003,Lammer:2003} or runaway greenhouse \citep{Ramirez:2014c}, indicating that all the common -- so-called volatile -- molecular species (e.g. H$_2$O, SO$_2$, NH$_3$, CO$_2$) and most of their byproducts must be in gaseous form in the atmosphere.

\begin{figure}
    \centering
\includegraphics[width=\linewidth]{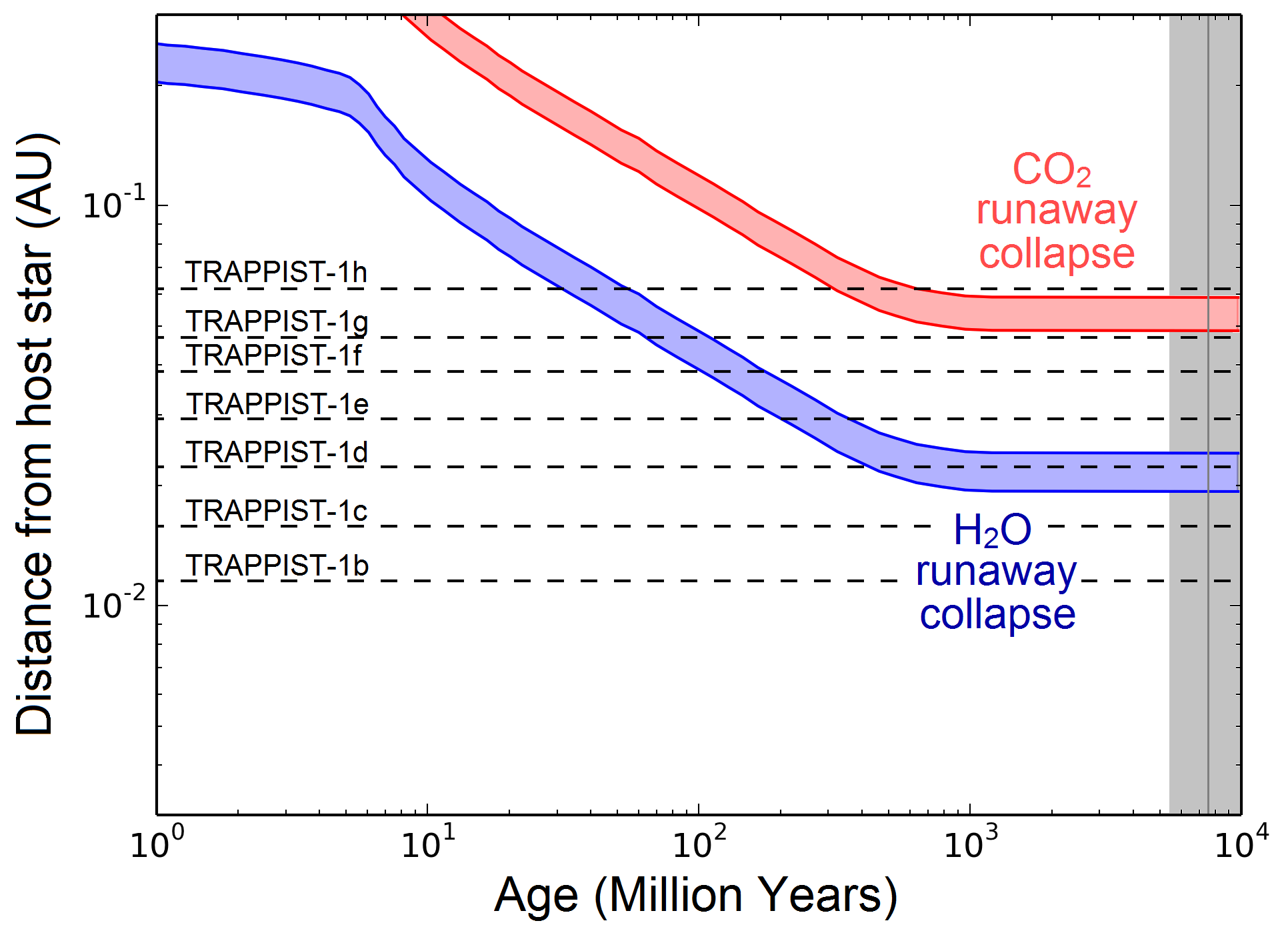}
\caption{Architecture of the TRAPPIST-1 system and evolution of the runaway greenhouse/atmospheric collapse limit for water (a.k.a. the traditional inner edge of the Habitable Zone) and carbon dioxide. The spread of the runaway greenhouse/atmospheric collapse for water was calculated assuming a synchronous planet (i.e. at 1.4$\times$ the bolometric flux received on Earth; see \citealt{Yang:2014b}.) 
and a non-synchronous planet (i.e. at 0.9$\times$ the bolometric flux received on Earth, using the results of the 1-D calculations of \citealt{Kopparapu:2013,Kopparapu:2013erratum}). 
Note that the 1.4$\times$F$_\oplus$ limit is a conservative estimate according to the results of \citealt{Kopparapu:2016} showing this threshold could vary depending on the metallicity of the host star, because the stellar mass-luminosity relationship depends on the metallicity, and thus does the rotation rate of the planet. Moreover, this runaway greenhouse estimate has been calculated assuming a cold start which is likely not a good approximation for planets orbiting such ultra-cool dwarfs, and that are thought to have started hot.
The spread of the runaway greenhouse/atmospheric collapse for CO$_2$ was calculated based on the results of \citet{Turbet:2018aa} that the irradiation limit at which CO$_2$ cannot accumulate in the atmosphere of TRAPPIST-1 planets is located between the orbit of TRAPPIST-1g and h. This result is discussed in more details in the Section~\ref{subsection_CO2_atm} of the manuscript.
These lines were drawn assuming a luminosity derived from evolutionary models for a 0.089~$\Msun$ M-dwarf \citep{vangrootel:2018}. For reference, we added the estimated age of 7.6~$\pm$2.2~Gigayears based on \citet{Burgasser:2017}. The figure was adapted from \citet{Bourrier:2017b}.}
\label{trappist-1_luminosity_evolution}
\end{figure}

As an illustration, and following \citet{Bolmont:2017} and \cite{Bourrier:2017b}, Figure~\ref{trappist-1_luminosity_evolution} shows how the limit at which all water (in blue) should be vaporized in a planetary atmosphere, as a function of time. In other words, for planets located closer to TRAPPIST-1 than the blue curve, water is expected to be unstable in condensed (solid or liquid) form and should form a steam atmosphere. This limit is also known as the runaway greenhouse \citep{Ingersoll:1969,Kasting:1988,Pier:10book,Goldblatt:2012}. 

First, Figure~\ref{trappist-1_luminosity_evolution} illustrates the fact that TRAPPIST-1b, c, and also maybe TRAPPIST-1d\footnote{Whether or not TRAPPIST-1d is below or beyond the runaway greenhouse irradiation limit for water depends on subtle cloud and atmospheric circulation feedbacks \citep{Yang:2013,Yang:2014,Kopparapu:2016,Wolf:2017,Turbet:2018aa}.}, have spent their entire life in a state where water can only be present in the form of steam. Second, the four outermost planets of the system are compatible with the presence of surface water in liquid or icy form today, but all four planets must have spent a significant fraction of their lives in a state where all water was trapped in vapour form in the atmosphere.

For illustration, and based on the results of \citet{Turbet:2018aa}, we added in Figure~\ref{trappist-1_luminosity_evolution} the limit (in red) at which all available CO$_2$ should be vaporized in a planetary atmosphere (or collapse on the surface, respectively), as a function of time. The conditions required for atmospheric collapse were studied in depth with numerical and analytical models \citep{Joshi:1997,Wordsworth:2015apj,Koll:2016,Auclair-Desrotour:2020} and applied specifically to the TRAPPIST-1 system in \citet{Turbet:2018aa}. The only planets of the system sensitive to strong CO$_2$ atmospheric collapse today are TRAPPIST-1g and h. CO$_2$ collapse can theoretically occur on the other planets but it requires some special conditions \citep{Turbet:2018aa} discussed in Section~\ref{subsection_CO2_atm} and Fig.~\ref{n2_co2_collapse}, the main condition being that the planets must be depleted of non-condensable gases. Other common gases such as CH$_4$, O$_2$, CO or N$_2$ are too volatile to be sensitive to atmospheric collapse on the TRAPPIST-1 planets \citep{Turbet:2018aa}. If present, these volatile species should be in the atmosphere (i.e. not trapped on the surface).

\subsection{Stellar activity and atmospheric loss}
\label{sub_section_stellar_activity}

Knowing the XUV (i.e. from X to UV) irradiation of TRAPPIST-1 is crucial because it affects the stability and erosion of planetary atmospheres \citep{Lammer:2003,Bolmont:2017}, controls photochemical reactions in the upper atmosphere \citep{Rugheimer:2015,Arney:2017,Chen:2019}, and can further influence the development and survival of life on a planet surface \citep{Rugheimer:2015b,Omalley:2017, Ranjan2017}. As a M8-type star, TRAPPIST-1 is thought to be a very active star, with a strong X/Extreme UV (EUV) flux \citep{Wheatley:2017,Bourrier:2017b} and frequent, intense flaring events \citep{Vida:2017}. However, little is known about the FUV emission of these ultra-cool dwarfs. In fact TRAPPIST-1 is the coldest exoplanet host star for which FUV emission has been measured, via measurement of its Lyman-$\alpha$ line with HST/STIS \citep{Bourrier:2017}. The comparison between this measurement and that of TRAPPIST-1 X-ray emission \citep{Wheatley:2017} further show that the stellar chromosphere is only moderately active compared to its transition region and corona.

\begin{figure*}
    \centering
\includegraphics[width=\linewidth]{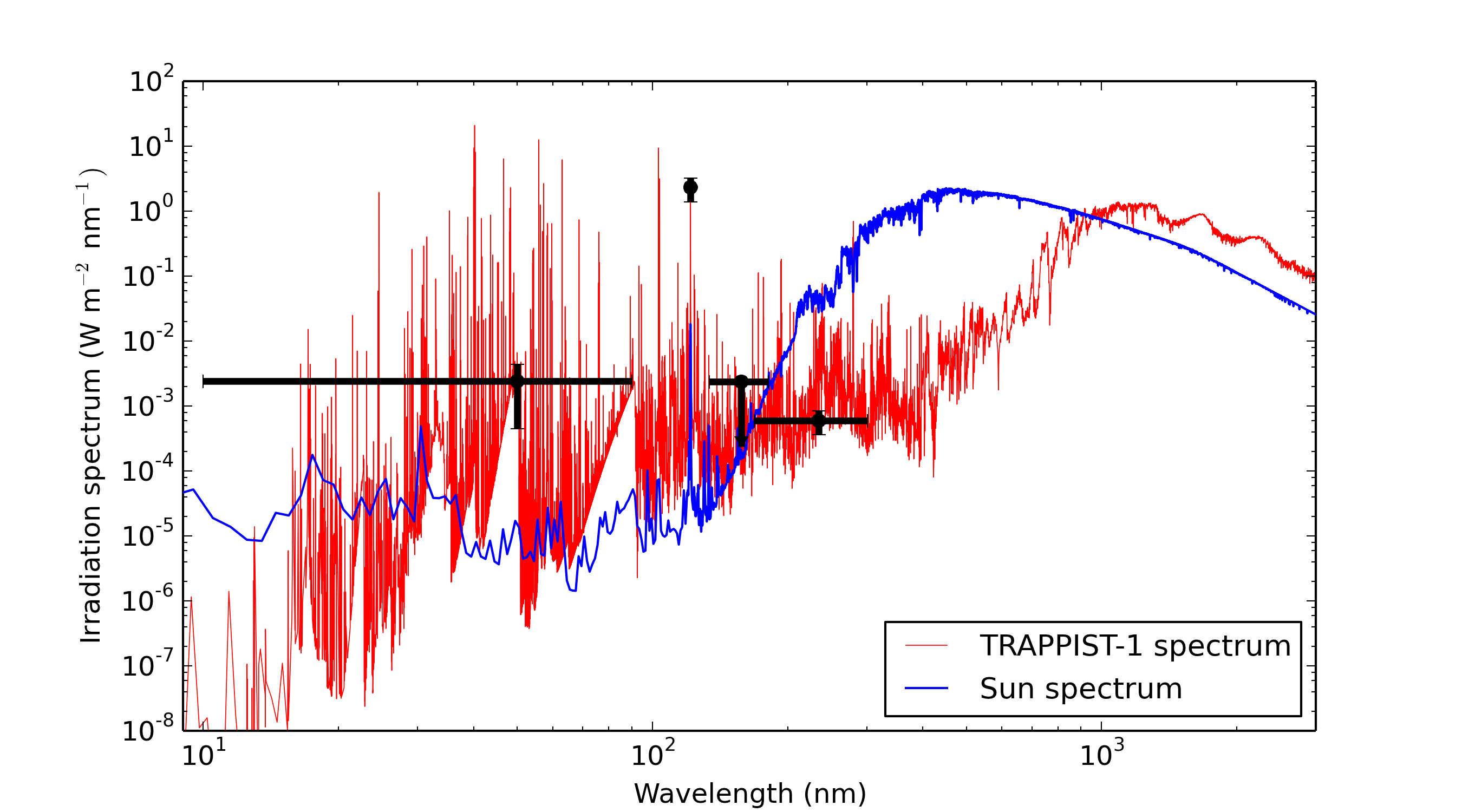}
\caption{This figure shows irradiation spectra emitted by the star TRAPPIST-1 (red line) and the Sun (blue line). Both spectra were normalized to a total bolometric flux of 1366~W~m$^{-2}$, i.e. the mean irradiation received at the top of the atmosphere of present-day Earth. The solar spectrum (blue line) is the solar reference spectrum (SOLAR-ISS) taken from \citet{Meftah:2018}. The TRAPPIST-1 spectrum (red line) is calculated in \citet{Peacock:2019} (scenario 1A). Black data points are described in the main text (Section~\ref{sub_section_stellar_activity}).}
\label{trappist-1_spectrum}
\end{figure*}

Based on (i) HST/STIS Lyman-$\alpha$ observations of TRAPPIST-1 \citep{Bourrier:2017,Bourrier:2017b}, (ii) XMM-Newton X-ray observations of 
TRAPPIST-1 \citep{Wheatley:2017}, (iii) constraints from GALEX far-UV and mid-UV photometry survey (partly based on the 
work of \citealt{Schneider:2018}) on a sample of very nearby, similar late (M8) stars, 
and (iv) PHOENIX Models\footnote{available on \url{https://phoenix.ens-lyon.fr/simulator/index.faces}.}, 
\citet{Peacock:2019} constructed full emission spectra of TRAPPIST-1 from X to far-infrared wavelengths. 
Figure~\ref{trappist-1_spectrum} shows a calculated emission spectrum of TRAPPIST-1 \citep{Peacock:2019}, 
normalized to a total bolometric flux of 1366~W~m$^{-2}$, i.e. the mean irradiation received at the top of the 
atmosphere of present-day Earth. This corresponds to the stellar flux that an hypothetical planet located at $\sim$~0.023~AU 
of the star TRAPPIST-1 (between the orbits of planet d and e) would receive. Black data points correspond (by increasing wavelength) to:
\begin{enumerate}
    \item F$_{\textrm{XEUV}}$ (10-90~nm) estimates based on (i) scaling of XMM newton measurements of X-ray emission of TRAPPIST-1 \citep{Wheatley:2017} into EUV emission using the F$_{\textrm{EUV}}$/F$_{\textrm{X}}$ scaling relationship of \citet{Chadney:2015}; (ii) scaling of Ly$\alpha$ line measurements of TRAPPIST-1 using HST/STIS observations \citep{Bourrier:2017,Bourrier:2017b} into EUV emission using the F$_{\textrm{EUV}}$/F$_{\textrm{Ly}\alpha}$ scaling relationship of \citet{Linsky:2014}, summed with the observed X-ray flux of \citet{Wheatley:2017}.
    \item Ly$\alpha$ line (121.44-121.7~nm) measurements based on HST/STIS observations of TRAPPIST-1 \citep{Bourrier:2017,Bourrier:2017b}. Note that the synthetic spectrum of TRAPPIST-1 (red line) was plotted at very high resolution between 121.44 and 121.7~nm to make the comparison with the HST/STIS data easier.
    \item GALEX FUV (134.0-181.1~nm) upper estimate measurements of the emission of three very nearby M8-M8.5 very low mass stars (2MASS 12590470-4336243, 2MASS 10481463-3956062, 2MASS 18353790+3259545), i.e. that are of a spectral type similar to TRAPPIST-1. The three stars have an estimated age of $\sim$~5~Gigayears \citep{Schneider:2018,Peacock:2019}, which is within a few gigayears of the estimated age of TRAPPIST-1 \citep{Burgasser:2017}.
    \item GALEX NUV (168.7-300.8~nm) measurements of the emission of the same three very nearby M8-M8.5 very low mass stars.
\end{enumerate}
Note that the infrared part of the calculated TRAPPIST-1 spectrum of \citet{Peacock:2019} (red line) is consistent with the IRTF/SpeX near-infrared measurements of \citet{Gillon:2016}.

Figure~\ref{trappist-1_spectrum} shows that while the thermal emission of TRAPPIST-1 (T$_{\textrm{eff}}$~$\sim$~2516K) is shifted to higher wavelength than the Sun (T$_{\textrm{eff}}$~=~5778K), the non thermal emission of TRAPPIST-1 is significantly higher than that of the Sun (relative to the total bolometric flux) for wavelengths lower than 150~nm. In other words, at constant bolometric flux, the X-EUV and more energetic photon flux is much larger for TRAPPIST-1 than Sun-like stars. Moreover, this strong high-energy photon flux was likely much higher in the past \citep{Bourrier:2017b,Fleming:2020}. In fact, because the XUV irradiation is expected to decrease faster with time than the bolometric irradiation, the relative fraction of XUV irradiation was likely even much larger than today. This is an important aspect because it indicates that the atmospheric escape processes were likely much more efficient in the TRAPPIST-1 system than in the solar system.  This energetic photon flux is indeed likely to drive strong atmospheric escape, possibly hydrodynamically-driven \citep{Roettenbacher:2017,Bourrier:2017b}. Repeated measurements of TRAPPIST-1 Lyman-$\alpha$ line with HST/STIS show that the stellar UV emission varies over timescales of a few months \citep{Bourrier:2017,Bourrier:2017b}, suggesting similar variability in the strength of atmospheric escape and highlighting the need for long-term monitoring of the system.  

Furthermore, flaring events likely add an additional, significant component to the high-energy emission of TRAPPIST-1. \citet{Vida:2017} measured using K2 data that TRAPPIST-1 frequently produces flaring events of integrated intensities ranging from 1.26$\times$10$^{23}$ to 1.24$\times$10$^{26}$~J. As a reference, the energy brought by the most intense flaring event reported by \citet{Vida:2017} corresponds to the integrated bolometric emission of TRAPPIST-1 during $\sim$~10~minutes. For comparison, the most intense known solar flares have an integrated intensity $\sim$~10$^{25}$~J \citep{Shibata:2011}, which corresponds to the integrated bolometric emission of the Sun during $\sim$~0.03~second. While flares can have a substantial effect on atmospheric erosion, and possibly even on photochemistry \citep{Segura:2010}, they should have a minimal effect on the direct warming of the surface and atmosphere of TRAPPIST-1 planets.

In addition, atmospheric stripping by the strong stellar winds of TRAPPIST-1 is thought to be efficient \citep{Garraffo:2017,Dong:2017,Dong:2018,Dong:2019,Fraschetti:2019} for planets orbiting such a low mass star.

Consequently, it is likely that the planets of the TRAPPIST-1 system all have lost a significant fraction of their initial atmosphere, and may possibly have completely lost it. The fact that the planets of the TRAPPIST-1 system today have an atmosphere or not results from a competition between (i) the efficiency and duration of the atmospheric escape processes and (ii) the amount of volatiles (i.e., which can form an atmosphere; this can for example be water, carbon dioxide, methane, nitrogen, etc.) initially present on the planet and later brought in by degassing and by cometary or asteroid impacts.

\section{Constraints from the orbital architecture of the TRAPPIST-1 planetary system}
\label{section_orbital_architecture}

\begin{figure*}[htbp] 
\centering
\subfloat{ \includegraphics[scale=.185,trim = 0cm .cm 0.2cm 0.cm, clip]{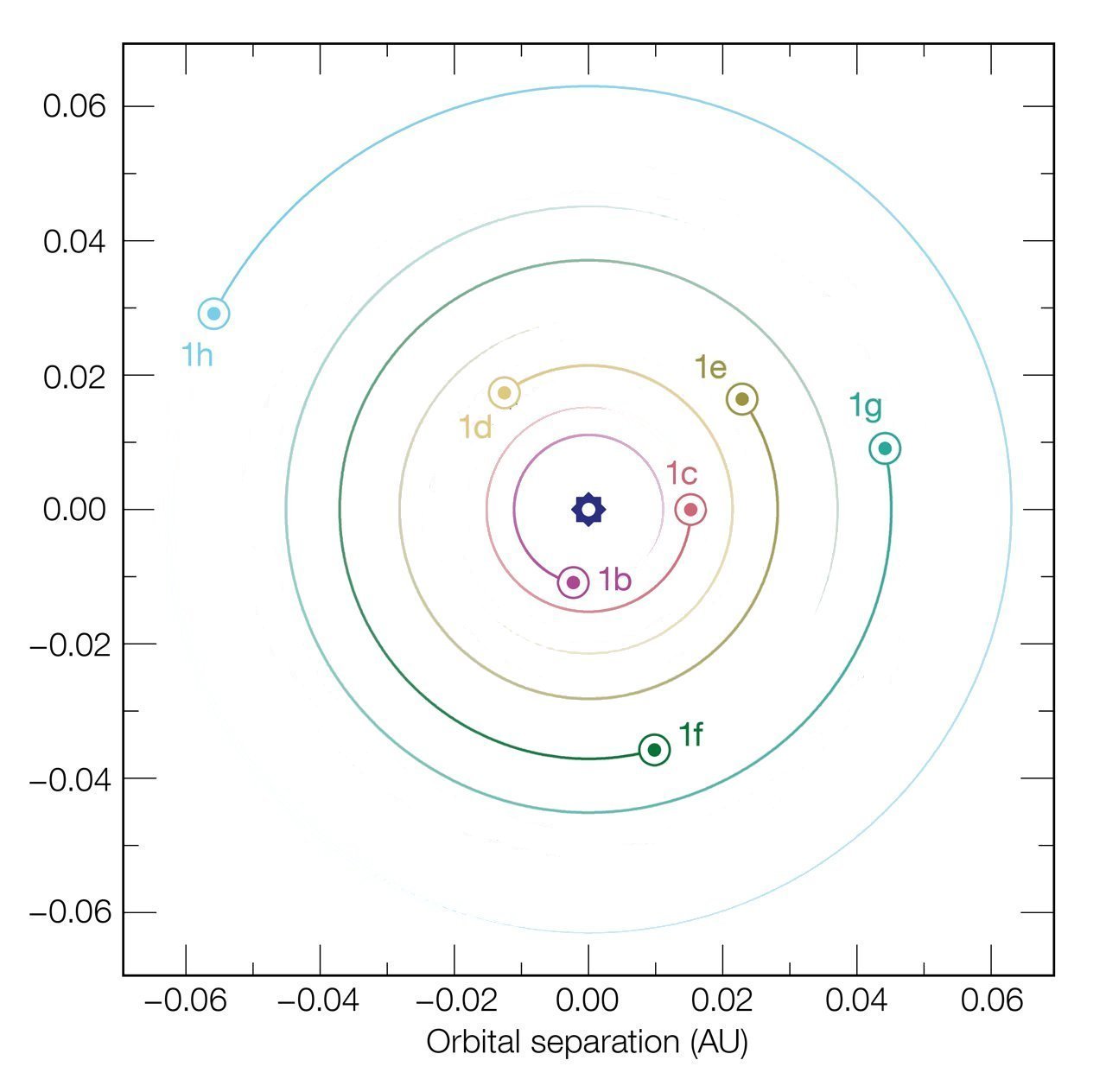} }
\subfloat{ \includegraphics[scale=.395,trim = 1.75cm .cm 0.cm 0.cm, clip]{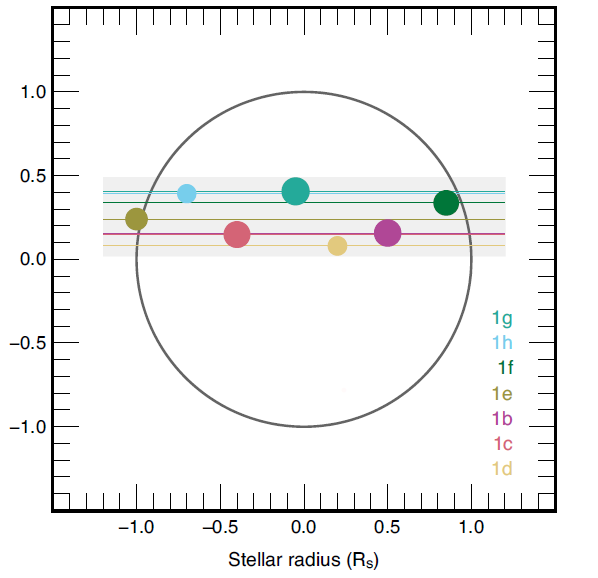} }
\caption{Representation of the TRAPPIST-1 system viewed from above (left panel, figure adapted from \citealt{Luger:2017}) or seen edge-on with the seven planets transiting in front of their star (right panel; figure taken from \citealt{Delrez:2018}).}
\label{orbital_architecture_trappist-1}
\end{figure*}

The orbital architecture of the TRAPPIST-1 planetary system is very peculiar (see Fig.~\ref{orbital_architecture_trappist-1}). First, it is extremely compact. All seven planets are confined within $\sim$~0.06~AU from their host star \citep{Luger:2017}. Secondly, all planets have a highly circularized orbit, with eccentricities lower than 0.01 for all planets \citep{Gillon:2017,Grimm:2018}. Thirdly, the system is very coplanar \citep{Luger:2017occultation,Delrez:2018}. Last but not least, all the planets form a resonant chain, and therefore have strong mutual gravitational interactions. Each pair of adjacent planets (bc, cd, de, ef, fg, gh) have period ratios near small integer ratios, and each triplets of adjacent planets (bcd, cde, def, efg, fgh) follow three-body Laplace resonances \citep{Luger:2017}. We discuss below how we can take advantage of this peculiar orbital architecture (i) to infer the formation and migration history of the TRAPPIST-1 system, (ii) to infer masses of the planets, first through orbital stability analysis, then through Transit Timing Variations (TTVs) analysis, and (iii) to predict the possible rotational states of TRAPPIST-1 planets, all three of which have implications on the possible atmospheres of TRAPPIST-1 planets today.

\subsection{Formation and migration scenarios for the TRAPPIST-1 system}

In the initial discovery papers \citep{Gillon:2017,Luger:2017} as well as in following discussions on the system’s formation and evolution \citep{Ormel:2017,Tamayo:2017,Papaloizou:2018,Coleman:2019,Schoonenberg:2019,Brasser:2019}, it has been proposed that the TRAPPIST-1 planets had to undergo migration to end up being locked into resonances.
Specifically, several authors \citep{Ormel:2017, Coleman:2019} have argued that the most likely formation scenario for the TRAPPIST-1 system is (i) planets formed far away from their host star, likely exterior to the water ice line and (ii) planets migrated inwards (in a timescale of $\sim$~10$^{6}$~years) in resonant convoys to reach their present location, very close to their host star. The inner edge of the disk would provide a migration barrier \citep{Masset:2006} such that the planets pile up into chains of mean motion resonances \citep{Terquem:2007,Ogihara:2009,Cossou:2014,Izidoro:2017}.

Meanwhile, \citet{Macdonald:2018} argued that long-distance migration is not the only plausible explanation for the formation of the TRAPPIST-1 resonant chain. In fact, they showed using orbital numerical simulations that if TRAPPIST-1 planets have formed (quasi) in situ, then either short-distance migration or eccentricity damping could have naturally lead the system toward a resonant chain similar to TRAPPIST-1 system.

Whether TRAPPIST-1 planets formed in situ or beyond the ice line has severe consequences regarding the amount of volatile species (e.g. water) that the planets were able to accrete in the first place. If planets formed in situ, then planets are likely dry today due to strong atmospheric erosion; if planets formed beyond the ice line, then planets are likely volatile-rich (and water-rich) because even atmospheric erosion should be insufficient to remove $>$~1-10$\%$ of the total planetary mass in volatile for these planets \citep{Tian:2015,Bolmont:2017,Bourrier:2017b}.

In fact, even if the first scenario -- in which planets formed far from their host star and then migrated inward -- is correct, it is possible that the volatile content remains very different between the inner and outer planets of the system. First, it is possible that the inner and outer planets have migrated in several distinct groups \citep{Papaloizou:2018} -- that merged afterward -- and have thus been formed at different locations of the protoplanetary disk, with different bulk compositions. Then, it is possible that the seven planets were each formed mainly from planetesimal accretion or pebble accretion \citep{Coleman:2019,Schoonenberg:2019} which would lead to a scatter in TRAPPIST-1 planets volatile bulk composition.

Last but not least, even if the planets formed in situ and all volatiles on the surface and atmosphere were stripped through atmospheric erosion, secondary outgassing or late-stage volatile delivery could still have been brought through cometary or asteroid impacts \citep{Kral:2019,Dencs:2019}. Impact volatile delivery holds mostly for outer TRAPPIST-1 planets, for which impactor velocities are expected to be low enough that volatile delivery dominates over impact erosion mechanisms \citep{Kral:2019}. 

We note there are some ongoing efforts to characterize the outer part of the TRAPPIST-1 system \citep{Marino:2020} to better constrain the whole architecture of the system and possibly look for exocometary or exoasteroid belts.

\subsection{Stability of the TRAPPIST-1 system}

The fact that the TRAPPIST-1 system is observed today with its near-integer period ratios after $\sim$~8~billion years \citep{Burgasser:2017} suggests that the orbital architecture of the system is long-lived.

Despite this apparent long-term stability, initial N-body simulations aimed at reproducing the TRAPPIST-1 system \citep{Gillon:2017} were unstable on a very short timescale ($\sim$~0.5~million years) even when including the eccentricity damping effect of tides (which only delayed the instability by a few million years). 

In contrast, \citet{Tamayo:2017} prepared N-body simulations of various planetary systems similar to TRAPPIST-1 and let them evolve through disk migration to form resonant chains of planets. They found that, even without accounting for tidal dissipation, most physically plausible resonant chains of planets were stable on timescales of at least 50~Myr, i.e. two orders of magnitude larger than in \citet{Gillon:2017}. This result shows that when a TRAPPIST-1-like resonant chain of terrestrial-mass planets is formed, it is generally very stable over time \citep{Tamayo:2017}. However, the exact stable orbital configuration of a resonant planet chain depends strongly on the parameters (orbital periods, masses, radii, eccentricities, etc.) of the planets. This indicates that the stability of a given observed resonant chain of planets such as TRAPPIST-1 is highly dependent on the initial planet parameters assumed. Therefore, it is likely that the N-body simulations of \citet{Gillon:2017} were unstable because the selected planet properties of the planets were far enough from reality.

\citet{Quarles:2017} took advantage of this result to estimate the masses of TRAPPIST-1 planets. For this, they performed thousands of N-body simulations of TRAPPIST-1 with planet properties perturbed from the observed values and then identified those that were stable for millions of years. With this stability analysis, \citet{Quarles:2017} identified self-consistent orbital solutions (i.e. that are stable on the long-term) from which they derived a posterior distribution of masses for each of the seven TRAPPIST-1 planets. Theses masses are provided in Table~\ref{mass_planets}. \citet{Makarov:2018} confirmed -- using the planet properties of \citet{Quarles:2017} -- that the stability of the system was greatly improved.

\begin{table*}
\centering
\normalsize
\setlength{\tabcolsep}{4.5pt}
\caption{Estimates of TRAPPIST-1 planet masses derived using (i) stability analysis, and (ii) TTV analysis. TTV masses and stability masses are all compatible within 1~$\sigma$. However, the TTV estimates are much more precise than the stability ones. Stability masses were derived from \citet{Quarles:2017}$^a$, while TTV masses were derived from \citet{Grimm:2018}$^b$. Mass estimates are provided in Earth mass units, and with 1~$\sigma$ uncertainties.}
\begin{tabular}{lcr}
\hline
Planet & Stability$^a$ masses & TTV$^b$ masses\\
\hline
T1b  & 0.88$^{+0.62}_{-0.53}$ & 1.017$^{+0.154}_{-0.143}$\\ 
T1c  & 1.35$^{+0.61}_{-0.59}$ & 1.156$^{+0.142}_{-0.131}$\\
T1d  & 0.42$^{+0.25}_{-0.21}$ & 0.297$^{+0.039}_{-0.035}$\\
T1e  & 0.55$^{+0.51}_{-0.35}$ & 0.772$^{+0.079}_{-0.075}$\\
T1f  & 0.68$^{+0.17}_{-0.18}$ & 0.934$^{+0.080}_{-0.078}$\\
T1g  & 1.39$^{+0.76}_{-0.69}$ & 1.148$^{+0.098}_{-0.095}$\\
T1h  & 0.47$^{+0.26}_{-0.26}$ & 0.331$^{+0.056}_{-0.049}$\\
\hline
\end{tabular} 
\label{mass_planets} 
\end{table*}

\subsection{Transit Timing Variations}

In tightly packed planetary systems such as TRAPPIST-1, the continuous exchange of angular momentum between gravitationally interacting planets causes them to accelerate and decelerate along their orbits. This makes in turn their transit times occur early or late compared with a Keplerian orbit, possibly in a detectable way. Detecting these changes in transit times is known as the Transit Timing Variations (TTV) technique \citep{Holman:2005,Agol:2005}.

\begin{figure}
    \centering
\includegraphics[width=\linewidth]{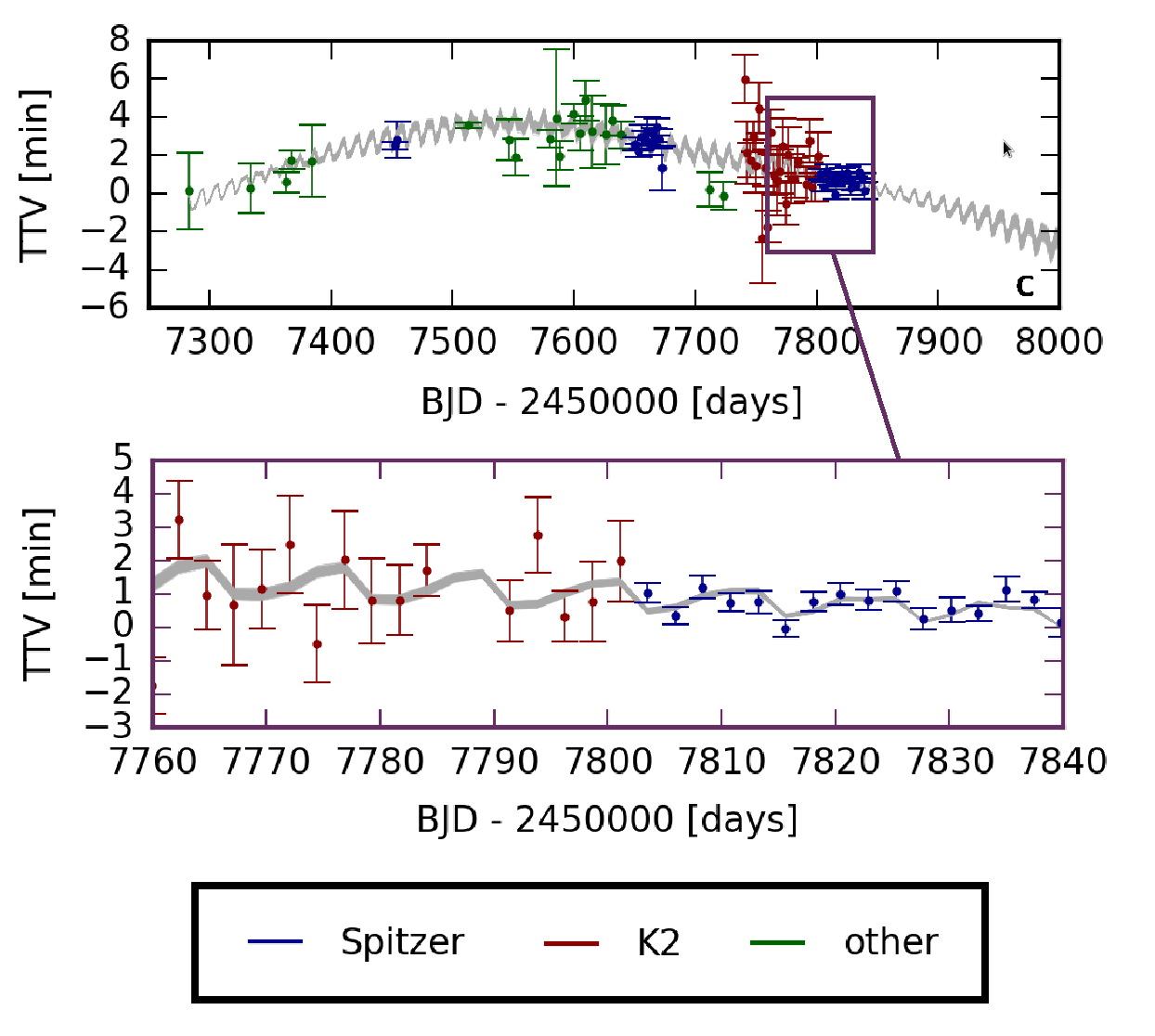}
\caption{Measured transit times of TRAPPIST-1c (with corresponding 1~$\sigma$ uncertainties) are indicated by coloured symbols, according to the origin of the data (Spitzer, K2 or other telescopes). The grey line indicates the spread of TTV fits obtained for one thousand distinct MCMC calculations samples \citep{Grimm:2018}. The low-frequency TTV component is visible in the top panel, and the high-frequency (chopping) TTV component is visible in the bottom panel. A detailed list of all transits is given in the appendix of \citet{Grimm:2018}. This figure was adapted from Fig.~2 of \citet{Grimm:2018}, which also shows the TTVs of the 6 other planets in the system.}
\label{ttv_T1c}
\end{figure}

All TRAPPIST-1 planets exhibit transit timing variations \citep{Gillon:2017,Grimm:2018} owing to gravitational pulls by their closest neighbours. The TTV signal for each planet is dominated primarily by interactions with adjacent planets, and these signals have the potential to be particularly large because each planet is near first-order mean motion resonance with its neighbours. In the TRAPPIST-1 system, the TTV amplitudes range in magnitude from 2~min to 1~hour \citep{Gillon:2017,Grimm:2018} depending on the planet. 

To get accurate TTV data, it is necessary to derive precise timings for the transits of the planets. This requires transit observations with a large aperture telescope (to get as much photons as possible) and a stable photometry, to get the best possible time-resolved SNR of the transit light curve and thus the transit timing. 
TTVs have two components:
\begin{enumerate}
    \item a low-frequency component, visible in the upper panel of Fig.~\ref{ttv_T1c}. To capture this TTV component, regular transit observations are needed over several years.
    \item a high-frequency component -- aka chopping \citep{Holman:2010,Deck:2015} -- visible in the lower panel of Fig.~\ref{ttv_T1c}. The periodicity of chopping pattern encodes the timespan between successive conjunctions of pairs of successive planets whose amplitudes yield the masses of adjacent perturbing planets. To capture this TTV component, continuous transit observations are needed over several tens of days.
\end{enumerate}
For the TRAPPIST-1 system, the most accurate TTV datasets were obtained with Spitzer (Spitzer Proposals ID 13067, 13175, 14223) space mission (see Fig.~\ref{ttv_T1c}). However, transit observations with ground-based telescopes were also useful to constrain the long-term (or low-frequency) component of the TTV.

TTV datasets were then analyzed through inversion, a process through which observed transit times are fit using a model of gravitationally interacting planets in order to determine the system parameters \citep{Carter:2012}. This specifically includes the determination of planetary masses, relative to the mass of the star. TRAPPIST-1 TTVs were first modeled in \citet{Gillon:2017} and then completed in \citet{Grimm:2018}, for which a much larger amount of TTV data (a total of 284 individual transit timings) has been included. \citet{Grimm:2018} used a sophisticated algorithm to compare and fit the outcome of a large number of N-body simulations of the TRAPPIST-1 system -- initialized with a wide range of planet parameters -- with the TTV data. Using this methodology, they were able to derive masses estimates for the TRAPPIST-1 planets. Theses masses estimates are provided in Table~\ref{mass_planets}. TTVs masses and Stability Analyses masses are all compatible within 1~$\sigma$~error bars. However, we recommend the use of TTVs masses which have a much better measurement accuracy. These masses can be used (see Section~\ref{section_numerical_modelling}) to evaluate the planet bulk densities and gather information on their possible atmospheres. We note that significant efforts are currently being made to use all observational data -- including all the latest Spitzer observation campaigns -- to estimate the masses of the TRAPPIST-1 planets through TTVs as accurately as possible (Agol et al., in preparation). 

\subsection{Effect of tides on TRAPPIST-1 planets}
\label{subsection_tides}

All observed TRAPPIST-1 planets are inside an orbital distance
of $\sim$~0.06~AU. For such close-in planets, tidal interactions are expected to be strong and to influence the orbital and rotational dynamics of the system. 

In fact, the TRAPPIST-1 system principally evolves due to the gravitational tides raised by the star in the planets (the planetary tide). Simple order of magnitude calculations show that the tides exerted by the planets in the star are today a priori negligible \citep{Turbet:2018aa}. Moreover, the contribution of tidal torques between planets has been shown to be overall very small \citep{Wright:2018,Hay:2019}, compared to the planetary tides. Planetary tides act to (i) slow down the rotation rate of the planets, (ii) reduce their obliquity, and (iii) circularize their orbits. It can trap the planets into spin-orbit resonances, possibly down to the synchronous rotation.

\citet{Turbet:2018aa} estimated the evolution timescales for the rotation to range from 10$^{-4}$~Myr for TRAPPIST-1b to 7~Myr for TRAPPIST-1h. For the obliquity, the evolution timescales range from 10$^{-3}$~Myr for planet-b to 80~Myr for planet-h. These timescales, computed assuming a tidal dissipation for the planets to be a tenth of the dissipation of the Earth -- i.e. close to an ocean-less earth --, are highly uncertain. However, considering the estimated age of the system of $\sim$~8~Gigayear \citep{Burgasser:2017}, it is reasonable to assume that all planets have reached a state of near tidal equilibrium, with small obliquities and a slow rotation. The exact rotation will depend on the presence and strength of other processes able to balance the braking effect of tides.

Indeed, it is now known that some other processes can sometimes act to avoid the synchronous state. A first possibility is that thermal tides in the atmosphere can create a strong enough torque to balance the stellar tidal
torque on the mantle, as is expected to be the case on Venus \citep{Chapman:1970,Ingersoll:1978,Correia:2001,Leconte:2015,Auclair:2017}. For this process to be efficient, the planet must be close enough from the star so that tides in general are able to affect the planetary spin, but far enough so that bodily tides are not strong enough to overpower atmospheric tides. In the TRAPPIST-1 system, this zone rests well beyond the position of the seven known planets (see Fig. 3 of \citealt{Leconte:2015}). Atmospheric tides should thus be unable to significantly affect the spins of TRAPPIST-1 planets.

Another possibility for planets on an eccentric enough orbit is capture into a higher order spin-orbit resonance \citep{Goldreich:1966}, i.e. higher than the synchronous rotation. Using the formalism of \citet{Ribas:2016}, \citet{Turbet:2018aa} calculated that the eccentricity of a given planet in the TRAPPIST-1 system must be above $\sim$~10$^{-2}$ to be possibly captured in a higher order spin-orbit resonance. 
Probability of capture becomes greater than 10$\%$ only for an
eccentricity greater than 0.03 (compatible with the calculations of \citet{Makarov:2012}, although made for a different system). 
Simulations of the system dynamics accounting for tides and planet-planet interactions \citep{Turbet:2018aa} seem to show that such eccentricities are on the very high end of the possible scenarios. This was also confirmed in the TTV analysis of \citet{Grimm:2018} showing that all TRAPPIST-1 planets most likely have eccentricities equal or below 10$^{-2}$. Recently, \citet{Auclair:2019} developed a tidal model for ocean planets and showed that, although resonances of oceanic modes are likely to decrease the critical  eccentricity for which eccentric planets can be trapped in 3:2 spin-orbit resonance, this should not directly affect the end spin-state of the TRAPPIST-1 planets because their eccentricities are likely too low.

In summary, none of the two aforementioned processes should be strong enough to counteract bodily tides so that all TRAPPIST-1 planets are mostly likely today in a synchronous-rotation state.

However, it has been recently shown by \citet{Vinson:2019} using a pendulum spin model introduced in \citet{Vinson:2017} that due to planet mutual interactions, some of the TRAPPIST-1 planets may be pushed in a specific spin state of the synchronous rotation with (i) significant libration of the spin state and/or (ii) complete circulation of the spin state. In the numerical simulations of \citet{Vinson:2019}, the timescale for the spin libration and/or circulation is on the order of several Earth years, i.e. on the order of hundreds of orbits of TRAPPIST-1 planets. They also noted that these libration and/or circulation spin-states are quasi-stable and that TRAPPIST-1 planets can shift from one state to another on the order of 10$^3$-10$^5$~years. The exact timescales depend on the planet considered, and tidal dissipation factors assumed.

Whether the planets are in a classic synchronous state, in a higher order spin-orbit resonance, or in a synchronous state with strong libration, possibly even circulation, would have profound implications for the possible climates and atmospheres of TRAPPIST-1 planets (see an example in Fig.~1 of \citealt{Turbet:2016}).

Last but not least, it is important to mention that tidal heating is likely the dominant source of internal heating at least for the innermost TRAPPIST-1 planets \citep{Turbet:2018aa,Barr:2018,Makarov:2018,Dobos:2019}. For instance, \citet{Turbet:2018aa} calculated that the mean surface tidal heating flux on TRAPPIST-1b is within 0.7-40~W~m$^{-2}$, depending on the tidal dissipation factor assumed and eccentricity calculated. Similar orders of magnitude for the tidal heating flux were obtained by \citealt{Papaloizou:2018} ($<$~5~W~m$^{-2}$), \citealt{Barr:2018} (between 0.8 and 4~W~m$^{-2}$), and \citealt{Dobos:2019} (between 0.1 and 2~W~m$^{-2}$). Differences are due to different tidal dissipation factors assumed and different eccentricities calculated and/or assumed. Interestingly, the viscosity of the mantle depends on its physical state (e.g. temperature), which itself depends on the rate of heating. \citet{Makarov:2018} thus argued that in planets with strong potential heating (such as TRAPPIST-1b) the tidal dissipation might be determined by a feedback loop (between the physical state of the planetary interior and the heating rate) rather independently from the exact eccentricity.

All these tidal heat flux estimates are likely much larger for TRAPPIST-1b than expected radiogenic heating. On Earth, the typical radiogenic heating is $\sim$~0.08~W~m$^{-2}$ \citep{Davies:2010,Davies:2013} but it is likely to be lower on TRAPPIST-1 planets given the age of the system \citep{Burgasser:2017}, about twice that of the solar system. Uncertainties on the initial inventory of thermally important radioactive elements as well as on the stellar age are high \citep{Burgasser:2017}, which may affect this conclusion. Last, we note that the mechanism of electromagnetic induction heating proposed by \citet{Kislyakova:2017} should have a negligible contribution to the surface heat flux. The  maximum  induction  heating  estimated for  TRAPPIST-1  planets  by  \citet{Kislyakova:2017} yields a surface flux of 2$\times$10$^{-2}$~W~m$^{-2}$.

In summary, tidal heating is expected to be the dominant interior heating process for TRAPPIST-1 inner planets, but likely not for outer ones. The tidal heating flux may be high enough for innermost planets that it could melt the mantle and possibly trigger intense volcanic activity and thus outgassing of volcanic gases. However, and for all planets, tidal heating is at least two orders of magnitude lower than the stellar flux they received, indicating that the direct surface and atmospheric warming from tides is negligible.

\section{Constraints from transit observations}
\label{section_transit}

\begin{figure*}
    \centering
\includegraphics[width=\linewidth]{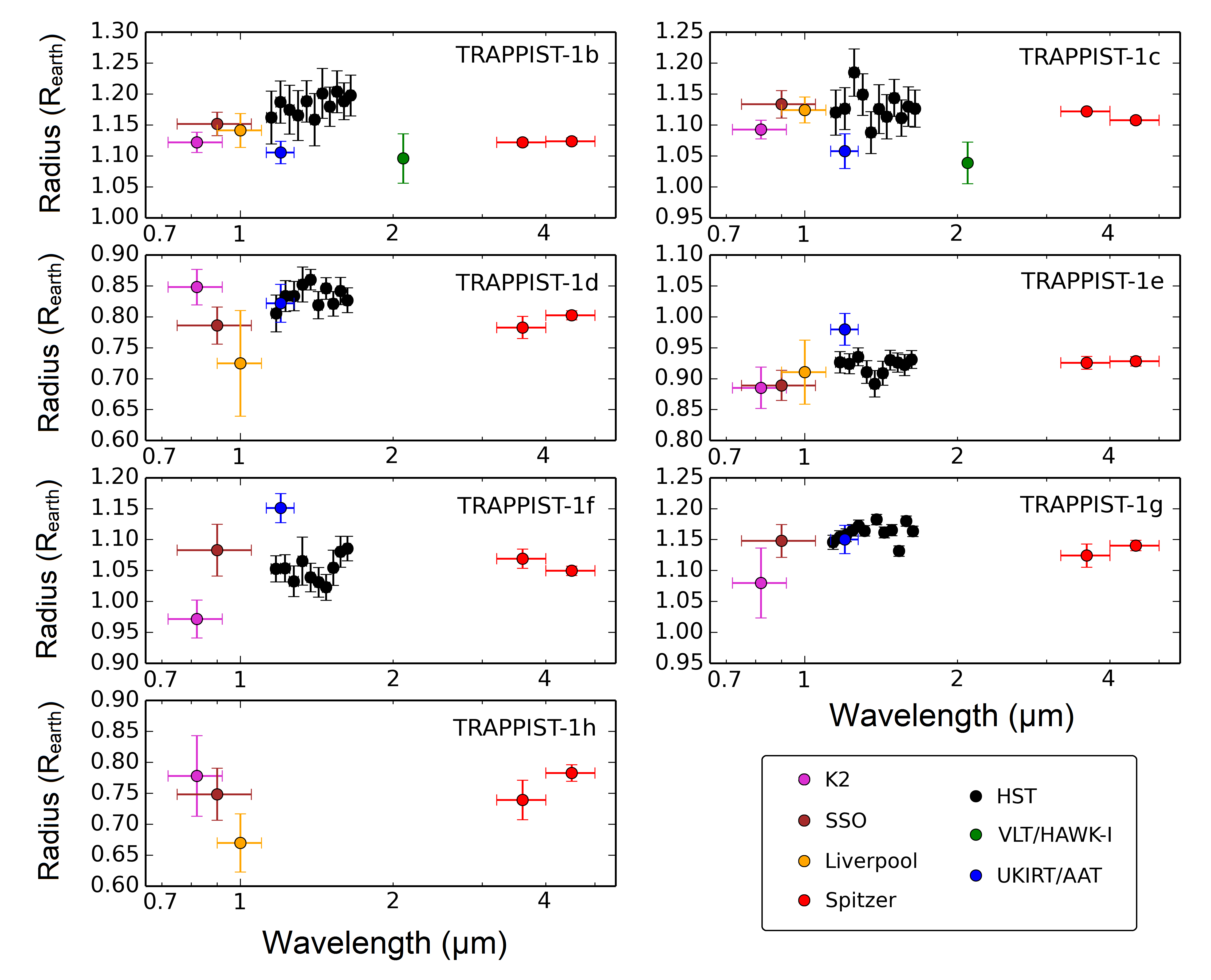}
\caption{This figure shows transit spectra (in Earth radius units) of the seven TRAPPIST-1 planets. These spectra were constructed using the transit depth measurements obtained with Spitzer \citep{Gillon:2017,Delrez:2018,Ducrot:2018,Ducrot:2020}, HST \citep{Dewit:2016,Dewit:2018,Wakeford:2019}, K2 \citep{Luger:2017, Ducrot:2018}, SPECULOOS-South Observatory aka SSO and Liverpool telescope \citep{Ducrot:2018}, VLT/HAWK-I, UKIRT and AAT \citep{Burdanov:2019}. These transit depths were then converted into transit radii using the TRAPPIST-1 stellar radius estimate of \citet{vangrootel:2018}, i.e. $\Rs$~=~0.121$\Rsun$~$\pm$0.003. The absolute error bars on the planetary radii due to the uncertainty on the radius of the star (about 2.5~$\%$ according to \citealt{vangrootel:2018}) have not been applied here, because it is the relative uncertainties that are of interest here. Note that some of the transit observations (e.g. ground-based observations, HST/WFC3 observations) may not have a reliable absolute monochromatic baseline level \citep{Ducrot:2018,Ducrot:2020}.}
\label{transit_spectra}
\end{figure*}

Since the initial discovery of the TRAPPIST-1 system, many ground and space-based large-aperture telescopes have been used to measure transits of all the seven TRAPPIST-1 planets in a large range of wavelengths. Figure~\ref{transit_spectra} summarizes all the transit observations that have been published as of December 2019  \citep{Dewit:2016,Gillon:2017,Bourrier:2017,Bourrier:2017b,Dewit:2018,Delrez:2018,Ducrot:2018,Luger:2017,Burdanov:2019,Wakeford:2019,Ducrot:2020} as a function of the wavelength of observation, producing the most complete transmission spectra of TRAPPIST-1 planets to date.
At least three effects can explain radius variations with wavelength: (i) atomic/molecular absorptions by TRAPPIST-1 planetary atmospheres, (ii) instrumental biases and (iii) contamination by the stellar activity (e.g. presence of spots). Below, we review the information gathered and lessons learned from these multi-wavelengths transit observations.

\subsection{No cloud-free hydrogen-dominated atmospheres on most TRAPPIST-1 planets}
\label{section_H2_transit}

Hubble Space Telescope (HST) observations of the transits of TRAPPIST-1 planets \citep{Dewit:2016,Dewit:2018} have brought the strongest constraint so far on the possible atmospheres of TRAPPIST-1 planets. Transits were observed on HST using the WFC3/IR instrument (1.1-1.7~$\mu$m) first on TRAPPIST-1bc \citep{Dewit:2016} and then on TRAPPIST-1defg \citep{Dewit:2018,Wakeford:2019}. Improvements in the data reduction pipeline of the HST transit observations were proposed later in \citet{Zhang:2018}, which reported a net increase in the efficiency of HST observations by $\sim$~25$\%$.

\citet{Dewit:2016} and \citet{Dewit:2018} produced synthetic spectra of H$_2$-dominated cloud-free atmospheres and compared them with real HST data (see black data points in Fig.~\ref{transit_spectra}). They showed that the lack of prominent features in the HST spectra rules out cloud-free (and haze-free) hydrogen-dominated atmospheres for TRAPPIST-1b, c, d, e, f at 12, 10, 8, 6, and 4~$\sigma$, respectively. 
For example, \citet{Dewit:2016} showed that the expected amplitude of the 1.4~$\mu$m water feature in a hydrogen-dominated, low molecular weight atmosphere is $\sim$~2000~ppm (in transit depth) for TRAPPIST-1b and c, corresponding to planetary radius variation $\sim$~0.15-0.20$\Rearth$ which are not seen in HST transit observations. \citet{Dewit:2018} and \citet{Moran:2018} calculated that the amplitude of the same feature is less than 1000~ppm (or less than 0.07$\Rearth$) for TRAPPIST-1g, mostly because the atmosphere is colder, thus reducing the atmospheric scale height H$~=~\frac{R T}{M g}$, where R is the perfect gas constant, T the atmospheric temperature, M the mean molar mass of the atmosphere, and g the gravity. As a result, a clear hydrogen-rich atmosphere cannot be firmly ruled out for TRAPPIST-1g with HST observations only \citep{Dewit:2018,Moran:2018}.

\citet{Moran:2018} then performed atmospheric calculations to explore if more sophisticated models of hydrogen-rich atmospheres (including higher metallicity, clouds, photochemical hazes) could also be ruled out by HST observations. They determined that H$_2$-rich atmospheres (with solar metallicity\footnote{Metallicity refers to the overall heavy-element abundance. An atmosphere with solar metallicity therefore has the same composition of heavy elements (in particular carbon and oxygen, which can be converted into water vapour and methane) as the Sun.}) with high altitude clouds (at pressures of 12~mbar or lower) are consistent with the HST observations for TRAPPIST-1d and e. Moreover, they found that HST observations cannot rule out (at 3~$\sigma$) hydrogen-dominated atmosphere (with a cloud layer at 0.1 bar) with 300, 100 and 60$\times$ solar metallicity for TRAPPIST-1d, e and f, respectively. This stems from the fact that high metallicity hydrogen-dominated atmosphere have a much larger mean molecular weight, and thus a lower atmospheric scale height and therefore reduced atmospheric feature amplitudes.

In conclusion, most of TRAPPIST-1 planets are unlikely to have an extended hydrogen-dominated atmosphere. However, this possibility cannot be completely ruled out by the HST/WFC3 observations, because either (i) a very high altitude cloud cover or (ii) very high metallicity H$_2$-dominated atmospheres could in principle both fit HST/WFC3 observations. Furthermore, Lyman-$\alpha$ observations obtained with HST/STIS showed marginal flux decrease at the time of TRAPPIST-1 b and c transits, which could hint at the presence of extended hydrogen exospheres around these planets \citet{Bourrier:2017,Bourrier:2017b}. More HST/STIS observations of TRAPPIST-1 planetary transits (HST Proposal ID 15304, PI: Julien de Wit) are expected to put more precise constraints in a near-future.

\subsection{Possible indications of stellar contamination}

\begin{figure*}
    \centering
\includegraphics[width=\linewidth]{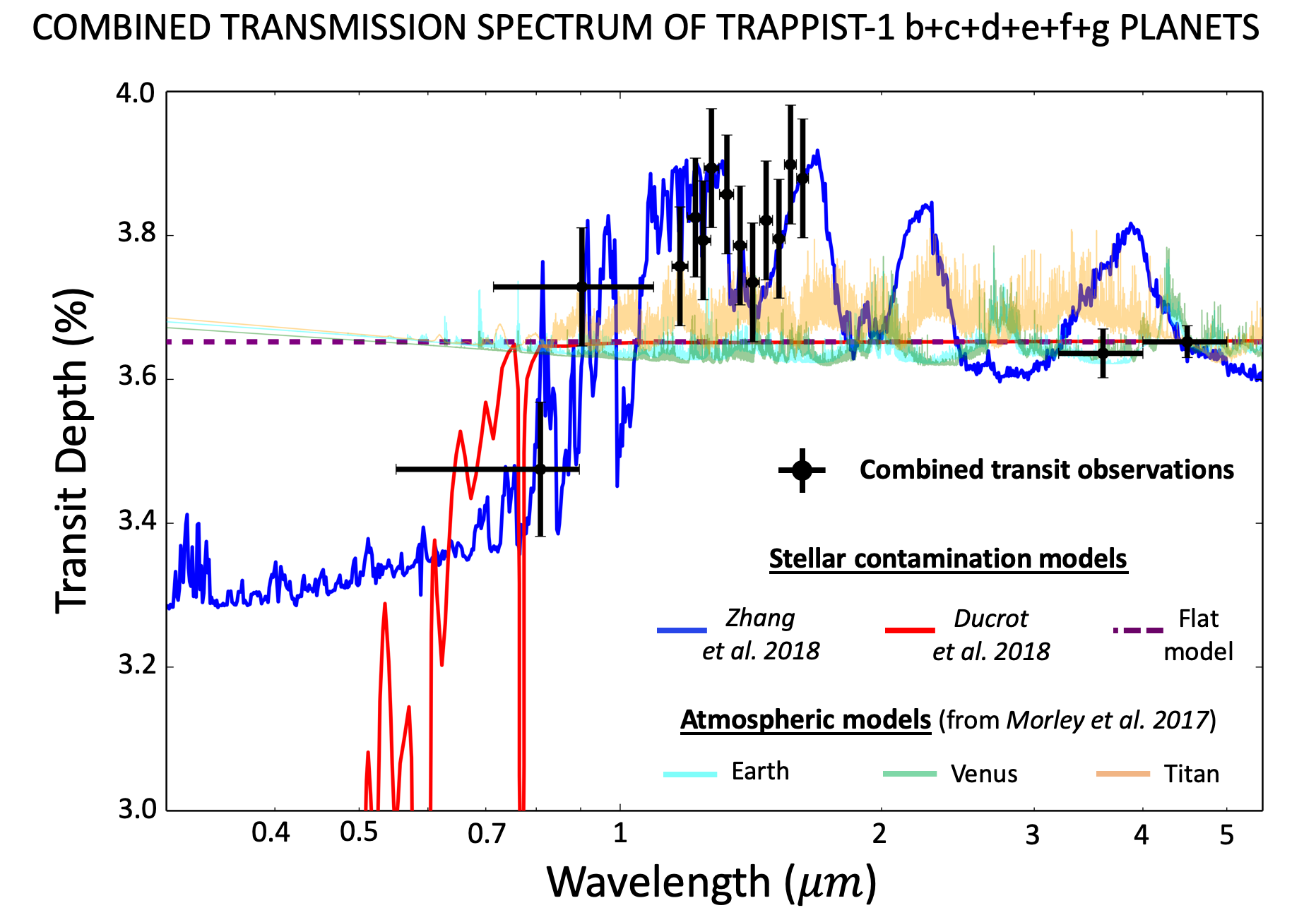}
\caption{Stellar contamination models fit to the K2+SPECULOOS-South+HST/WFC3+Spitzer/IRAC channels 1 and 2 \citep{Dewit:2016,Gillon:2017,Luger:2017,Delrez:2018,Dewit:2018,Ducrot:2018,Ducrot:2020} combined TRAPPIST-1 transmission spectra for planets b+c+d+e+f+g (black points and error bars, in transit depth $\%$ units). The blue stellar contamination spectrum \citep{Zhang:2018} corresponds to a three components model with (i) a photosphere (T~=~2400~K), (ii) hot faculae (T~=~3000~K) covering 50$\%$ of the projected stellar disk and cold spots (T~=~2000~K) covering 40$\%$ of it. The red stellar contamination spectrum \citep{Morris:2018a,Ducrot:2018} corresponds to a two-component model with (i) a photosphere (T~=~2500~K) and (ii) a few very bright spots (T~=~5300~K). The dashed purple stellar contamination spectrum corresponds to a flat model (i.e. no stellar contamination), which also corresponds to the best fit scenario in \citet{Wakeford:2019}. For each contamination spectrum, a small offset was added to ensure that each spectrum is compatible with the Spitzer/IRAC 4.5~$\mu$m transit measurement. Note that in the three contamination models (blue, red, and dashed purple lines), the signal (when fitted) is assumed to be fully stellar, i.e. no contribution from wavelength-dependent absorption by planetary atmospheres. The pale (i) cyan, (ii) green and (iii) orange lines correspond to combined synthetic transmission spectra from planetary atmospheres all made of (i) Earth-like, (ii) Venus-like and (iii) Titan-like compositions. These combined spectra were computed by summing the synthetic spectra of TRAPPIST-1b+c+d+e+f+g from \citet{Morley:2017}. They assume no stellar contamination.}
\label{contamination_spectrum_v2}
\end{figure*}

No significant absorption features have been detected so far in any individual transmission spectra (see Fig.~\ref{transit_spectra}). Yet, it is possible to increase the SNR of transit observations by combining the transmission spectra of all planets (see black dots in Fig.~\ref{contamination_spectrum_v2}). Part of the variations seen in the combined transit spectrum may be due to the presence of atmospheric absorptions, but also likely to the presence of heterogeneities in the photosphere of the star TRAPPIST-1. This includes the presence of both occulted and unocculted (cold) spots and/or (hot) faculae.

When a planet transits in front of its host star, the transit depth of the planet is measured by taking the difference of the disk-integrated stellar spectrum between in and out of transit. It is usually assumed in this calculation that the disk-integrated spectrum is identical to the light incident on the planet and its atmosphere. In reality, however, the planet is occulting only a small region within the transit chord, and only at a given time. Due to the presence of spots, faculae, and even latitudinal temperature gradients, the spectrum of this small region may differ significantly from the disk-averaged spectrum of the star. As a result, the presence of heterogeneities in the stellar photosphere can bias transit observations.

\citet{Rackham:2018} recently called into question the fidelity of the transit observations -- more specifically of the HST measurements \citep{Dewit:2016,Dewit:2018} -- of TRAPPIST-1 planets, because of possible contamination of the transmission spectra by the presence of spots and faculae in the photosphere of the star TRAPPIST-1. This contamination is also known as the 'transit light source effect'. For this, \citealt{Rackham:2018} (see their equations~1-3) developed a simple stellar contamination spectrum model based on three components (photosphere, unocculted spot, unocculted faculae) each with a given temperature and spatial coverage. Each component can be fitted with a combination of three synthetic stellar spectra (e.g. PHOENIX spectra) at three different temperatures. The original model of \citet{Rackham:2018} assumes that no heterogeneities (e.g. spots or faculae) are present within the transit chord; or, if they are, that they can be identified in the light curve and properly taken into account. However, because the precision of observations may not allow stellar surface heterogeneities within the transit chord to be reliably detected, \citet{Zhang:2018} (final published version; see their equation~7) proposed an extension to the model of \citet{Rackham:2018} by taking into account the presence of spots and faculae in the transit chord. They also allowed the covering fraction of spots and faculae in the transit chord to differ from the whole-disk values.
Based on the formalism of \citet{Rackham:2018}, two stellar contamination scenarios have been proposed so far:
\begin{enumerate}
    \item The first model \citep{Rackham:2018,Zhang:2018} assumes a typical spot size of thousands of km, or $\sim$~0.04$\%$ of the projected stellar disk of TRAPPIST-1 (i.e. very similar in size to large spot groups on the Sun's photosphere). This assumption is based on the observation that long-baseline monitoring of TRAPPIST-1 using the Spitzer Space Telescope \citep{Gillon:2017,Delrez:2018,Ducrot:2020} and HST/WFC3 \citep{Dewit:2016} shows no definitive evidence of spot crossing events; and that the spot size must be significantly lower than $\sim$~0.5$\%$ of the projected stellar disk otherwise it should have impacted the stellar variability of TRAPPIST-1 in HST/WFC3 and Spitzer/IRAC observations \citep{Rackham:2018}. Using this assumption, the stellar contamination model is based on three components (i.e. the photosphere is decomposed into three regions of three different sizes and temperatures) designed to fit existing transit data from K2, SPECULOOS-South, HST/WFC3 and Spitzer/IRAC channel 2 (4-5~$\mu$m), as illustrated by the blue curve in Fig~\ref{contamination_spectrum_v2}. Their complete transmission spectrum (blue line in Fig~\ref{contamination_spectrum_v2}) is fully consistent with stellar contamination, except with the more recent Spitzer/IRAC channel 1 (3.15-3.9~$\mu$m) presented in \citet{Ducrot:2020}. The three components of this model \citep{Zhang:2018} are (i) a photosphere (T~=~2400~K), (ii) hot faculae (T~=~3000~K) covering 50$\%$ of the projected stellar disk and cold spots (T~=~2000~K) covering 40$\%$ of it. Note that in the final model of \citealt{Zhang:2018} (published version, and not the first submitted arXiv version\footnote{\citealt{Zhang:2018} (first arXiv version) initially used the original formalism of \citealt{Rackham:2018} (see equation~3) to build their stellar contamination spectrum, i.e. assuming no heterogeneities -- spots or faculae -- are present within the transit chord. \citet{Ducrot:2018} compared the model of \citealt{Zhang:2018} (first arXiv version) and discarded it (by more than 10~$\sigma$) mostly using K2 observations in the visible wavelengths. Meanwhile, \citealt{Zhang:2018} (published version) updated their model to include the presence of spots and faculae in the transit chord, considering the precision of observations may not allow stellar surface heterogeneities within the transit chord to be reliably detected. In their revised model (blue line in Fig~\ref{contamination_spectrum_v2}; see also equation~7 of \citealt{Zhang:2018}), they assume the transit chord is composed of the same three spectral components as the integrated disk, but they allow their covering fractions to differ from the whole-disk values.}), the spot coverage is much lower in the transit chord that in the rest of the planet suggesting, (i) a latitudinal variation of spot coverage on the star, and (ii) that it is important to take this possibility into account in the fit.
    \item The second stellar contamination model \citep{Morris:2018a,Ducrot:2018} is based on the observation that the activity of the star TRAPPIST-1 as measured with K2 and Spitzer \citep{Morris:2018a} was best described by a two-component model with (i) a photosphere (T~=~2500~K) and (ii) a few very bright spots (T~=~5300~K) with a fractional lower limit coverage of $\sim$~0.005$\%$. In fact, \citet{Morris:2018a} argued that cold spots (if present) should produce variability in the Spitzer light curves that is yet absent in existing data, motivating therefore a two-component model without cold spots. Additionally, \citet{Ducrot:2018} reported that the photosphere of TRAPPIST-1 is most likely described by a base photosphere with a small fraction of hot faculae (T~$>$~4000~K). This stems from the fact that their stellar contamination model (red line in Fig~\ref{contamination_spectrum_v2}) is fully consistent with existing transit observations, including the recent Spitzer/IRAC channel 1 (3.15-3.9~$\mu$m) data presented in \citet{Ducrot:2020}, but not the HST (1.1-1.7~$\mu$m) data \citep{Dewit:2016,Dewit:2018} which they discarded from their analysis (see discussion hereafter). \citet{Ducrot:2018} noted that the stellar photosphere may in principle also be fitted with high latitude cold spots.
\end{enumerate}

Both models appear to be roughly consistent with stellar contamination in the combined transmission spectrum. However, depending on which model is correct, our ability to characterize in detail the possible atmospheres of TRAPPIST-1 planets with forthcoming large aperture telescopes such as JWST may be strongly affected. This is particularly critical in infrared wavelengths for which the two models have very different predictions. While the model of \citet{Morris:2018a} and \citet{Ducrot:2018} predicts a nearly flat contamination spectrum for wavelengths $>$~0.7~$\mu$m (red line in Fig~\ref{contamination_spectrum_v2}), the model of \citet{Rackham:2018} and \citet{Zhang:2018} predicts strong spectral variations (blue line in Fig~\ref{contamination_spectrum_v2}). The latter stellar contamination spectrum would alter the transit depths of the TRAPPIST-1 planets for planetary atmospheric species by roughly 1-15x the strength of planetary features, for wavelengths $>$~0.7~$\mu$m, thus significantly complicating JWST follow-up transit observations of this system. 
Moreover, this contamination would also introduce a bias as high as $\sim$~2.5$\%$ (more likely an overestimate) on planetary radii estimated with Spitzer IRAC channel 2 \citep{Rackham:2018}, and could thus bias the density estimates.

Below, we review the pros and cons of the two models:
\begin{enumerate}
    \item \textbf{Fit of the combined transmission spectrum:} Both models provide a reasonable fit of existing transit data (see red and blue lines in Fig~\ref{contamination_spectrum_v2}). However, if HST data is included in the fit, the model of \citet{Zhang:2018} performs very well and the model of \citet{Ducrot:2018} can be discarded. Without HST data, the model of \citet{Ducrot:2018} performs better than the model of \citet{Zhang:2018}.
    \item \textbf{HST data and the inverted 1.4~$\mu$m water vapour feature:} \citet{Ducrot:2018} reported that the HST observations \citep{Dewit:2016,Dewit:2018} may not have a reliable absolute monochromatic baseline level, due to orbit-dependent systematic effects. This is a good argument to remove the HST monochromatic data from the global fit of the stellar contamination model. However, the chromatic variation of transit depth within the 1.1-1.7~$\mu$m range can in principle be used separately as an additional constraint. While \citet{Dewit:2016,Dewit:2018} do not see any strong evidence for variations within the HST/WFC3 band, \citet{Zhang:2018} show -- by improving on the HST/WFC3 data analysis pipeline and summing the contributions of all planets from b to g -- the presence of a strong inverted water vapour feature (black data points between 1.1 and 1.7~$\mu$m, in Fig.~\ref{contamination_spectrum_v2}) in the six-planets combined transmission spectrum. \citealt{Zhang:2018} (see their Fig~11) also found that the same water vapour inverted feature is present in all possible five-planet combined transmission spectra, indicating it is not solely due to the spectrum of an individual planet. The presence of this water vapour inverted feature -- if confirmed -- is a strong evidence for the reliability of the stellar contamination model of \citet{Zhang:2018}.
    \item \textbf{Planetary radius bias:} The model of \citet{Rackham:2018} and \citet{Zhang:2018} predicts a $\sim$~2.5$\%$ radius bias in Spitzer/IRAC infrared wavelengths, while the model of \citet{Morris:2018a} and \citet{Ducrot:2018} does not predict any. \citet{Morris:2018b} re-evaluated the TRAPPIST-1 planets transit depths by estimating -- in the Spitzer transit light curves -- (i) the durations of the ingress/egress, (ii) the duration of the transit and (iii) the impact parameter. The method is detailed in \citealt{Morris:2018c} (see equations~5-8). This method, also known as the transit light-curve 'self-contamination' technique, is very weakly affected by the presence of stellar heterogeneities and can thus be used in principle to derive transit depth estimates that are not biased by the presence of heterogeneities in the stellar photosphere \citep{Morris:2018c}. By using this method, \citet{Morris:2018b} found consistent transit depth measurements between the traditional method and the light curve self-contamination method of \citet{Morris:2018c} and concluded on the absence of statistically significant evidence of stellar contamination in the Spitzer infrared wavelengths, thus supporting the model of \citet{Morris:2018a} and \citet{Ducrot:2018}. However, \citet{Zhang:2018} reported that the light curve self-contamination method has such large uncertainties (see Fig.~3 of \citealt{Morris:2018b}) that they do not have the sensitivity to probe the level of stellar contamination predicted by the model of \citet{Rackham:2018} and \citet{Zhang:2018}.
    \item \textbf{K2/I band/Spitzer stellar variability and size of stellar heterogeneities:} While the \citet{Rackham:2018} and \citet{Zhang:2018} model assumed that the heterogeneities (two components) in TRAPPIST-1 photosphere are similar in size to those in the Sun's photosphere, \citet{Morris:2018a} and \citet{Ducrot:2018} model assumed instead an empirically driven hypothetical spot distribution for TRAPPIST-1, consisting of (one component) a few small, bright (hot) spots. The former model is based on the assumption that the $\sim$~1$\%$ TRAPPIST-1 typical I-band variability \citep{Rackham:2018} indicates the absence of large/giant heterogeneities in TRAPPIST-1 photosphere. The latter model is based instead on the assumption that the variability in the K2 and Spitzer light curves is driven by rotational modulation due to starspots \citep{Morris:2018a}. While K2 (400-900~nm) light curves show a variability of $\sim$~1.25$\%$ with a period of 3.3~days \citep{Luger:2017} comparable to the typical I-band variability \citep{Rackham:2018}, Spitzer (4-5~$\mu$m IRAC channel) light curves show very little variability $\leq$~0.1~$\%$ \citep{Delrez:2018,Morris:2018a,Ducrot:2020}. This observation led \citet{Morris:2018a} to conclude that large ($\sim$~10$^4$km) dark spots should be absent in the stellar photosphere, otherwise they would produce much stronger variability in Spitzer light curves (see Fig.~2 in \citealt{Morris:2018a} and discussion in \citealt{Ducrot:2018}). A two-component photosphere model \citep{Morris:2018a} with a few very bright spots can however drive the modulation with a 3.3 day period in the K2 bandpass without generating a corresponding signal in the Spitzer 4.5~$\mu$m band, in agreement with the observations. 
    Note that these observations do not discard the model of \citet{Rackham:2018} and \citet{Zhang:2018} which assumed much smaller dark spots (a factor of 10 lower than in \citet{Morris:2018a}). This also means that the model of \citet{Morris:2018a} and \citet{Ducrot:2018} could be improved by adding a third component of small, dark spots. Meanwhile, \citet{Wakeford:2019} noted that the spot temperature posterior distribution of \citealt{Zhang:2018} (see their Fig.~13 and 20) is truncated to high temperature, potentially preventing it from reaching the parameter space of very hot bright spots found in \citet{Morris:2018a} and \citet{Ducrot:2018}.
    \item \textbf{No clear spot crossing events during transits:} Analyses of observations in the visible and near-IR carried out by K2, SPECULOOS and the Liverpool telescope do not show transit depth variability arising from stellar spot crossing events during transits \citep{Delrez:2018,Ducrot:2018,Ducrot:2020}. \citet{Ducrot:2020} identified a few outliers in the Spitzer transit light curves but they interpreted most of them as instrumental noise. This means that spots must be either cool and smaller than the size of the planets (which would reduce the probability and amplitude of a spot occultation) or reside outside of the transit chords \citep{Morris:2018a}. These observations are particularly constraining on the bright spots, that should be absent of the transit chords of TRAPPIST-1. This is surprising, given the transit chords represents $\sim$~56$\%$ of an hemisphere \citep{Delrez:2018}. 
\end{enumerate}

To conclude, it is still unclear which of the two models is the best predictor of the stellar contamination. In the model of \citet{Rackham:2018} and \citet{Zhang:2018}, the stellar contamination is so strong in the infrared (by comparison to the expected amplitude of the atmospheric molecular absorption features; see Fig.~\ref{contamination_spectrum_v2}) that it is undoubtedly the most important limitation to our ability to characterize the atmosphere of the TRAPPIST-1 planets with the James Webb Space Telescope. In the model of \citet{Morris:2018a} and \citet{Ducrot:2018}, the stellar contamination is so flat in the infrared wavelengths that observations made with the two channels of Spitzer/IRAC could already be used to discard some high mean molecular weight atmospheres. For instance, if the stellar contaminaton model of \citet{Morris:2018a} and \citet{Ducrot:2018} is correct, then we conclude based on Fig~\ref{contamination_spectrum_v2} (orange line, for Titan-like atmospheres) that all TRAPPIST-1 planets cannot be CH$_4$-dominated \citep{Ducrot:2020}. This stems from the fact that the 3.3~$\mu$m CH$_4$ band should otherwise produce a significant variation of transit depth between the two Spitzer/IRAC warm channels that is not observed \citep{Ducrot:2018,Ducrot:2020}.

With a few exceptions (fit of HST data, Spitzer IRAC channel 1 data), both models appear to be so far broadly compatible with observational data. Besides, and as supported by the recent work of \citet{Wakeford:2019}, it seems that the two models could be reconciled by (i) adding a small, dark spot component and (ii) allowing for a very hot bright spot component. Better constraints on the nature of stellar heterogeneities present in the photosphere of TRAPPIST-1 is key to characterizing TRAPPIST-1 planetary atmospheres.

\section{Constraints from numerical modelling}
\label{section_numerical_modelling}

In this section, we review the various theoretical and numerical efforts that have been recently undertaken to better constrain the nature of the possible atmospheres of TRAPPIST-1 planets. Below, we list all common gases in known planetary atmospheres (H$_2$, H$_2$O, CO$_2$, CH$_4$, N$_2$, O$_2$) and discuss whether or not these gases may have predominantly accumulated in the atmospheres of TRAPPIST-1 planets, based on the results of numerical 1D and 3D climate models, photochemical models, and atmospheric escape models.

\subsection{H$_2$/He envelopes}

\begin{figure*}
    \centering
\includegraphics[width=\linewidth]{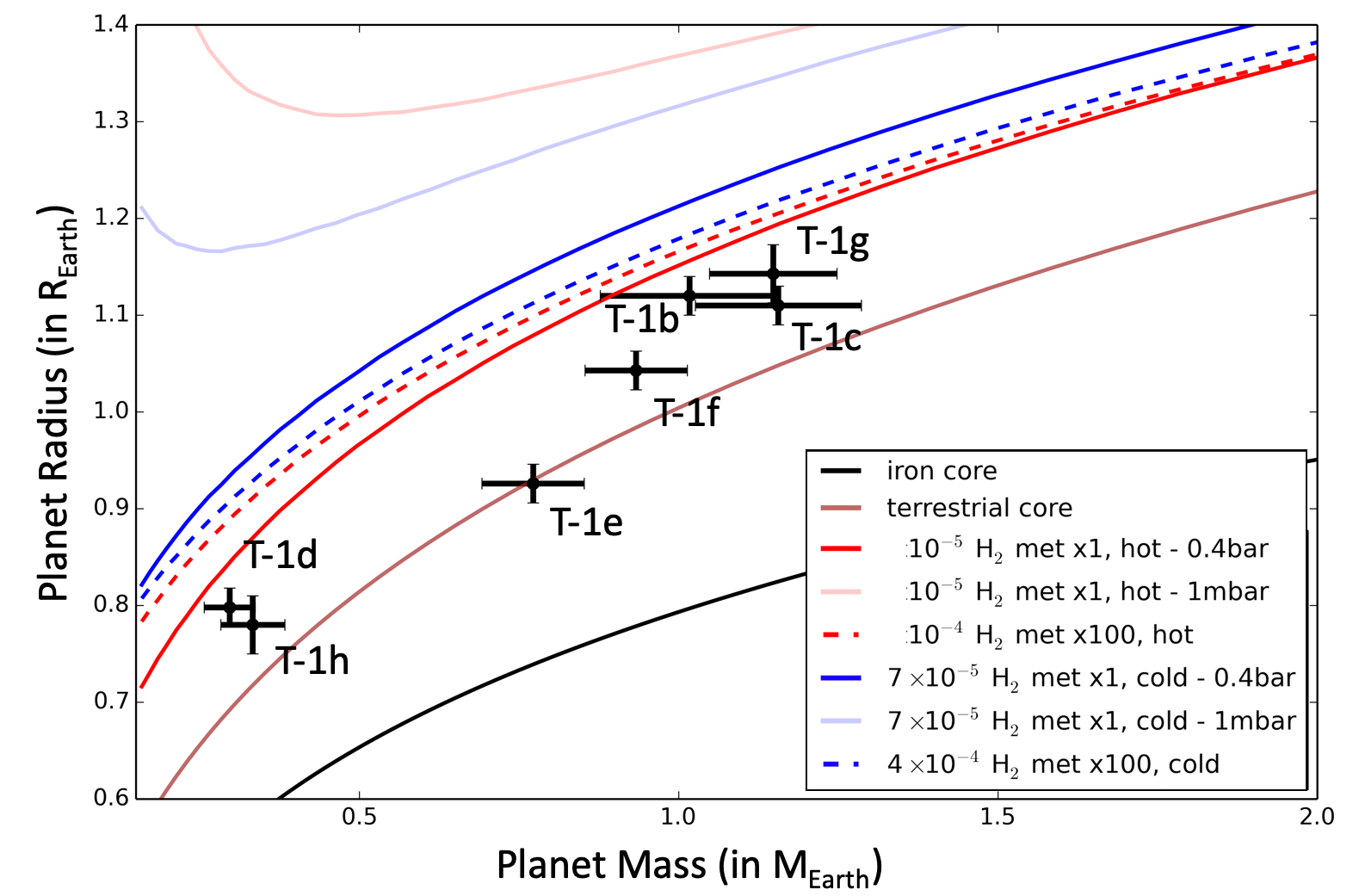}
\caption{Mass-radius relationships for various interior compositions and hydrogen envelope masses. The mass-radius relationships for planets endowed with hydrogen envelopes were constructed (i) assuming a core of terrestrial composition \citep{Zeng:2016}, (ii) endowed with a hydrogen envelope of 1x solar metallicity (solid lines) or 100x solar metallicity (dotted lines) for H$_2$O and CH$_4$. Red lines (and blue lines, respectively) indicate a scenario where the atmospheric temperature profile has been calculated in the irradition condition of TRAPPIST-1b (of TRAPPIST-1h, respectively), hence the name 'hot' (and 'cold', respectively). All transit radii were computed assuming a transit pressure of 0.4~bar, which is a conservative assumption based on the results of \citet{Grimm:2018} (Table~4). We also plotted the expected transit radii assuming a transit pressure at 1~mbar (pale red and blue lines). For comparison, we added the masses and radii of the seven TRAPPIST-1 planets measured from \citet{Grimm:2018} and \citet{Ducrot:2020}, with their associated 1~$\sigma$ error bars. For reference, we also added a terrestrial composition \citep{Zeng:2016} that ressembles that of the Earth, and a pure iron core composition \citep{Seager:2007}.}
\label{m-r_relationship_h2}
\end{figure*}

It has been argued \citep{Dewit:2016,Dewit:2018,Wakeford:2019} based on HST transit observations that most of TRAPPIST-1 planets are unlikely to be endowed with a hydrogen-dominated atmosphere, unless high altitude clouds/hazes are present or high metallicity atmospheres are considered \citep{Moran:2018}. This is discussed in more details in Section~\ref{section_H2_transit}.

Here we argue that hydrogen-dominated atmospheres are unlikely for TRAPPIST-1 planets, based also on their mass and radius estimates. Using the numerical atmospheric calculations of H$_2$-dominated atmospheres presented in \citet{Grimm:2018}, we constructed (see Fig.~\ref{m-r_relationship_h2}) mass-radius relationships for hydrogen-dominated atmospheres. For this, we assumed (i) a terrestrial core composition \citep{Zeng:2016}\footnote{Note that here and in the following subsection, we adopt a terrestrial core composition to construct mass-radius relationships. Although the core composition is not known (and is thus a highly degenerate quantity), it has been shown by \citet{Unterborn:2018} -- using the Hypatia catalog \citep{Hinkel:2014} -- that F-G-K stars of similar metallicity to TRAPPIST-1 have a Fe/Mg~=~1.72~$\pm$~0.46 in mass ratios, which gives a Fe/Si ratio of 1.49~$\pm$~0.4 (while fixing the Mg/Si mass ratio to 0.87). These values are similar to the solar mass ratio estimates of Fe/Si~=~1.69 and Mg/Si~=~0.89 \citep{Lodders:2009}, which motivates the use of a terrestrial core composition in the construction of the mass-radius relationships.}, 
(ii) hydrogen envelopes of various mass fraction (from 10$^{-6}$ to 10$^{-2}$), (iii) atmospheric temperature-pressure profiles based on the atmospheric numerical climate calculations of \citet{Grimm:2018}. For simplicity, we calculated mass-radius relationships based on two distinct atmospheric scenarios: first, a 'hot' scenario using the temperature-pressure profile calculated for a planet receiving the irradiation of TRAPPIST-1b, assuming 1x solar metallicity; and a 'cold' scenario, using the temperature-pressure profile calculated for a planet receiving the irradiation of TRAPPIST-1h, assuming 1x solar metallicity. At first order, numerical atmospheric models show \citep{Grimm:2018} that these temperature profiles can be described by an isothermal stratosphere (down to $\sim$0.1~bar) complemented with a dry convective troposphere (from $\sim$0.1~bar down to the surface).
The transit radius of the planet was then computed by integrating the hydrostatic equation (assuming a gravity varying with altitude) using the \citet{Saumon:1995} equation of state and assuming a transit pressure of 0.4~bar, the latter being a conservative estimate for all planets according to the results of \citealt{Grimm:2018} (Table~4).

In the most conservative scenario (high metallicity, cold atmosphere, transit pressure at 0.4~bar meaning the atmosphere is assumed to be cloud-free), Fig.~\ref{m-r_relationship_h2} shows that the maximum hydrogen-to-core mass fraction allowed (at 1~$\sigma$) by measured radii and masses of TRAPPIST-1 planets \citep{Grimm:2018,Ducrot:2020} is roughly 4$\times$10$^{-4}$. If the atmospheres were to be cloudy, then this maximum hydrogen content would be further reduced (e.g. down to $\sim$10$^{-4}$ for a cloud top at 1~mbar).

 According to \citet{Wheatley:2017}, TRAPPIST-1 planets receive between 3$\times$10$^3$ (for planet~b) and 10$^2$~erg~s$^{-1}$~cm$^{-2}$ (for planet~h) XUV flux from the star TRAPPIST-1 today. This results in mass loss rates between 5~$\times$~10$^8$ and 3~$\times$~10$^7$~g~s$^{-1}$ for TRAPPIST-1b and h, respectively, calculated in the energy-limited escape formalism \citep{Bolmont:2017,Bourrier:2017b}, using revised mass and radius estimates of \citet{Grimm:2018} and \citet{Ducrot:2020}. Furthermore, using the XUV flux history reconstruction of \citet{Bolmont:2017} and \citet{Bourrier:2017b}, we evaluate a cumulative (over 8~billion years) total hydrogen mass loss of 10$^{23}$~kg (i.e. 2~$\times$~10$^{-2}$ total mass fraction) and 10$^{22}$~kg (i.e. 5~$\times$~10$^{-3}$ total mass fraction) for TRAPPIST-1b and h, respectively. This is 2000 and 100$\times$, respectively, the amount of hydrogen required to fit the mass and radius of TRAPPIST-1b and h, assuming planets with a terrestrial core endowed with a high metallicity H$_2$-dominated envelope. As a result, the hydrogen-to-core mass fraction envelope required to fit the mass-radius relationships would be removed in $\sim$~100~million years or less only, which is significantly lower than the estimated age of the TRAPPIST-1 system \citep{Burgasser:2017}.

This escape rate should be even larger considering that a H$_2$-dominated atmosphere around a terrestrial-mass planet should be strongly extended (due to the reduction of gravity with altitude), and that the escape rate is expected to scale with $\Rp^3$ \citep{Erkaev:2007}. This atmospheric expansion is illustrated in the H$_2$-rich mass-radius relationships (see Fig.~\ref{m-r_relationship_h2}) that shows the difference of optical radius whether the atmosphere is assumed to be opaque at 0.4~bar (solid blue and red lines) i.e. cloud-free, or it is assumed to be opaque at 1~mbar (pale solid blue and red lines) i.e. with high altitude clouds. Assuming the XUV radius (i.e. the planetary radius at which the atmosphere becomes optically thick to XUV photons) occurs at a pressure of $\sim$~1~nanobar \citep{Murray-Clay:2009,Lopez:2012}, we evaluate (using equation~16 of \citealt{Grimm:2018}) that the XUV radius of TRAPPIST-1h is $\sim$~1.5$\times$ that of its optical radius (for 100$\times$ solar metallicity), increasing the escape rate by a factor of $\sim$~3. For TRAPPIST-1b (and assuming 1$\times$ solar metallicity), the XUV radius can increase by $\sim$~1.9 thus increasing the escape rate by a factor of $\sim$~7.

In summary, based on H$_2$-dominated mass-radius relationships for TRAPPIST-1 planets (Fig.~\ref{m-r_relationship_h2}), as well as hydrodynamical escape rate estimates \citep{Owen:2016,Wheatley:2017,Bolmont:2017,Bourrier:2017b}, H$_2$-dominated envelopes are unlikely to be stable around any of the TRAPPIST-1 planets.
Sustaining a hydrogen-rich atmosphere today is not theoretically impossible though, but it requires to consider a fine-tuned scenario where the planets were to be observed now exactly at the critical moment just before the complete loss of their initial hydrogen envelope (see \citealt{Owen:2016}, Fig.~13, for an example of hydrogen envelope evolution). However, it would not be possible to do this fine tuning for multiple planets in the same system. Besides, the absence of a large spread among TRAPPIST-1 planet densities is another argument against the presence of H$_2$-dominated atmospheres. This stems from the fact that any significant change of H$_2$ content -- e.g. arising from variations in the hydrogen-rich gas accretion rates during the planet formation phase \citep{Hori:2020} or from variations in the H$_2$ escape rates \citep{Bolmont2017,Bourrier:2017b} -- among TRAPPIST-1 planets should produce a very different density, which is not observed (see Fig.~\ref{m-r_relationship_h2}).

This argument, plus the HST/WFC3-IR transit measurements, are strong arguments against the presence of H$_2$-dominated envelopes around TRAPPIST-1 planets. This is also supported by recent calculations \citep{Hori:2020} showing the total mass loss of hydrogen-rich envelopes around TRAPPIST-1 planets is anyway likely higher than the amount of hydrogen-rich gas they can accrete during their formation.

\subsection{Water envelope}

\begin{figure*}
    \centering
\includegraphics[width=\linewidth]{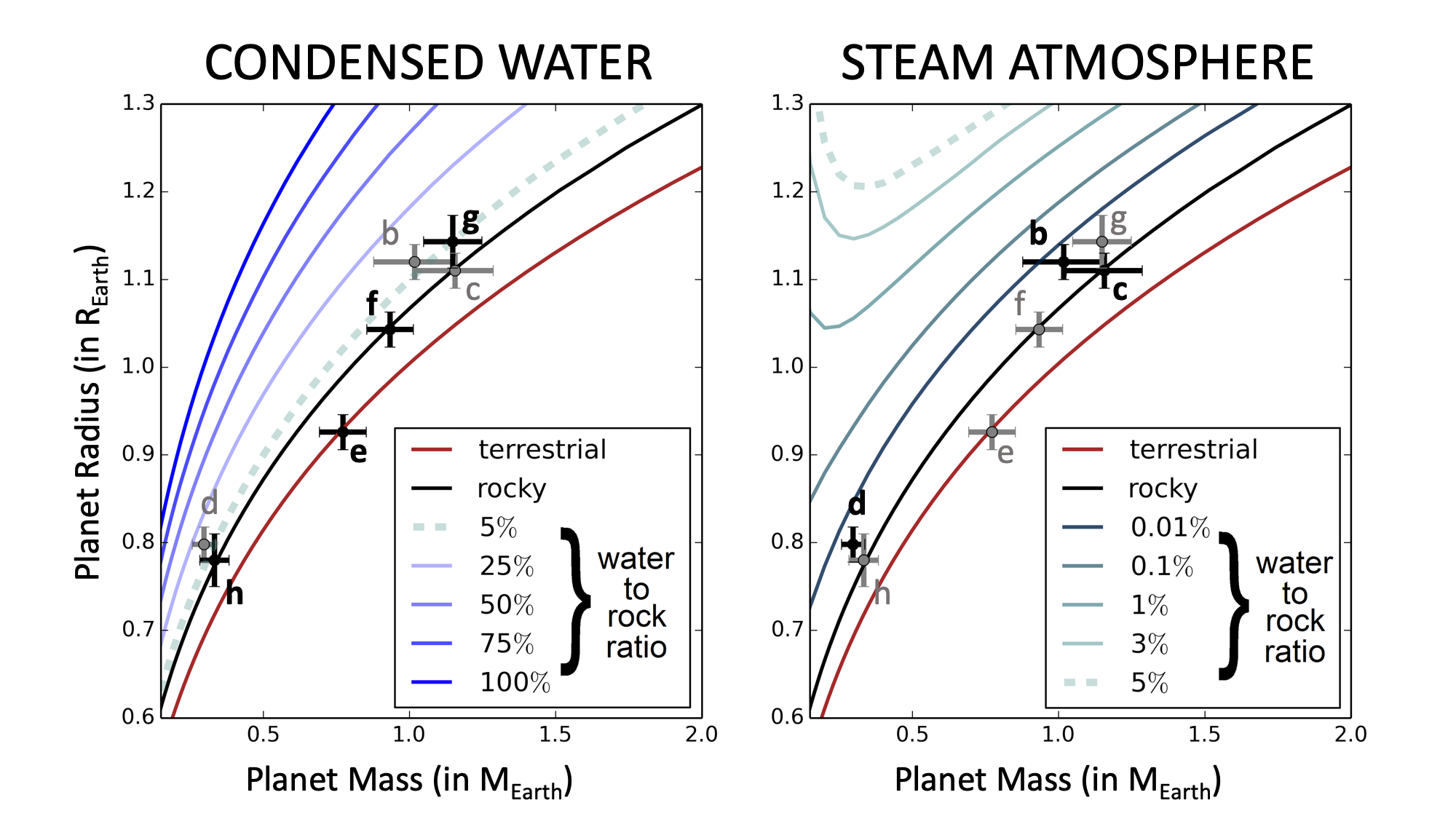}
\caption{Mass-radius relationships for various interior compositions and water content, assuming water is in the condensed form (left panel) and water forms an atmosphere (right panel). The rocky composition 
mass-radius relationship assumes a pure MgSiO$_3$ interior and 
was taken from \citet{Zeng:2016}. The water-rich mass-radius 
relationships for water in condensed form (left panel) were derived using the data from \citet{Zeng:2016}. 
The water-rich mass-radius relationships for water in gaseous form (right panel) were calculated in \citet{Turbet:2020aa}. 
All mass-radius relationships with water were built assuming a pure MgSiO$_3$ interior. 
For comparison, we added the measured positions of the seven 
TRAPPIST-1 planets measured from \citet{Grimm:2018,Ducrot:2020}, with their associated 1~$\sigma$ error bars. 
Based on the irradiation they receive compared to the theoretical runaway greenhouse limit (see Fig.~\ref{m-r_relationship_h2}), 
TRAPPIST-1e, f, g and h should be compared with mass-radius relationships on the left, while TRAPPIST-1b, c and d should be compared with those on the right. 
To emphasize this, we indicated on each panel in black the planets (and their associated 
1~$\sigma$ error bars) for which mass-radius relationships (with water) are appropriate. 
In contrast, we indicated on each panel in grey the planets (and their associated 
1~$\sigma$ error bars) for which mass-radius relationships (with water) are not appropriate. 
For reference, we also added a terrestrial composition that ressembles that of the Earth. Note that mass-radius relationships for steam planets (right panel) can be easily built following the procedure described in Appendix~D of \citet{Turbet:2020aa}. The figure was adapted from \citet{Turbet:2020aa}.}
\label{m-r_relationship_h2o}
\end{figure*}

Based on the masses and radii measurements of \citet{Grimm:2018} and \citet{Ducrot:2020}, at least three planets of the system (TRAPPIST-1b, d and g) are not compatible with a bare rock composition at 1~$\sigma$. This stems from the fact that these three planets have a measured bulk density that is lower than a pure Mg/Si planet (i.e. the lightest rocky planet that can be constructed), indicated by the black line in Fig.~\ref{m-r_relationship_h2o}. 
If confirmed, this result is a major achievement because it shows that these three planets must be enriched with volatiles and most likely with water (the most abundant of them) to lower the planetary density. 

\citet{Dorn:2018} showed using a Bayesian inference analysis that possible water mass fractions may range from 0 to 25$\%$. More recently, \citet{Turbet:2020aa} showed that the water content of the three innermost planets TRAPPIST-1b, c and d may in fact have been overestimated. This stems from the fact that these planets receive more irradiation than the runaway greenhouse limit (see Fig.~\ref{trappist-1_luminosity_evolution}) for which water has been shown to be unstable in condensed form and should rather form a thick, extended H$_2$O-dominated atmosphere. For the planet to exhibit the same radius, this H$_2$O-dominated atmosphere needs to be much less massive than a denser condensed water envelope \citep{Turbet:2019aa,Turbet:2020aa}. Taking into account this atmospheric effect, \citet{Turbet:2020aa} showed that TRAPPIST-1b, c and d should not have more (assuming a terrestrial core composition) than 2, 0.3 and 0.08$\%$ of water, respectively.

\citet{Bourrier:2017b} investigated -- using the formalism of \citet{Bolmont:2017} -- the history of hydrodynamic water loss from the planets in the energy-limited regime. They found that planet g and those closer in could have lost more than 20~Earth oceans through hydrodynamic escape, if the system is as old as $\sim$~8~Gigayear \citep{Burgasser:2017}. However, TRAPPIST-1e to h might have lost less than three Earth oceans if hydrodynamic escape stopped once they entered the habitable zone, i.e. when diffusion of water to the upper atmosphere (and then photodissociation of water) becomes the limiting process of water loss. Late-stage outgassing could also have brought significant amounts of water for the outer planets after they entered the Habitable Zone. Altogether, this indicates that,while the three inner planets are likely dry today, the outer planets of the TRAPPIST-1 system may have retained some water in their atmosphere or on their surface today.

\subsection{O$_2$ atmospheres}

O$_2$ is the natural molecule to discuss after H$_2$O because it has been shown that O$_2$ could easily accumulate from the photodissociation of H$_2$O molecules and the subsequent loss of lighter H atoms.

\citet{Luger:2015} calculated that as much as 10$^3$-10$^4$~bar of O$_2$ could have possibly accumulated in the atmosphere of planets orbiting very low mass stars such as TRAPPIST-1. However, the exact amount of O$_2$ that may accumulate in the atmosphere of TRAPPIST-1 planets depends on:
\begin{enumerate}
    \item \textbf{how much water is present initially.} \citet{Tian:2015b} showed that the more water is present initially on the planet, the more O$_2$ should build up on the planet. This means in particular that if the TRAPPIST-1 planets formed dry, then the accumulation of O$_2$ should be very limited. This assumes no atmosphere-interior interactions.
    \item \textbf{how efficiently hydrogen atoms can drag oxygen atoms with them.} \citet{Bolmont:2017} found that, in most of the configurations of their calculations, oxygen atoms are dragged away by the escaping hydrogen atoms, thus reducing the O$_2$ buildup to hundreds of bar, at maximum (see also \citealt{Johnstone:2019escape}). This is about two orders of magnitude lower than calculated in \citet{Luger:2015}. Furthermore, dragged along oxygen atoms could cool the outflows, resulting in reduced escape rates.
    \item \textbf{how efficient atmosphere-interior interactions are.} \citet{Wordsworth:2018} showed that during the runaway greenhouse phase where all water of the planet is trapped in the atmosphere, the surface temperature due to the greenhouse effect of water is so high \citep{Kopparapu:2013,Goldblatt:2013,Turbet:2019aa} that the surface and mantle are expected to melt. In this case, most oxygen produced from H$_2$O photolysis can be absorbed by the magma ocean, thus limiting the O$_2$ atmospheric buildup \citep{Hamano:2013,Schaefer:2016,Wordsworth:2018}. In the scenarios of \citet{Wordsworth:2018}, O$_2$~buildup is maximum if the water content is low enough that the planet reaches the point where (i) it is beyond the runaway greenhouse meaning all water is vaporized, but (ii) it has low enough water (i.e. roughly 1~Earth water ocean content) that the surface is solidified and cannot absorb O$_2$ anymore. While in this worst case scenario, no more than $\sim$~10$^2$~bar of O$_2$ should build up, in most cases \citet{Wordsworth:2018} show the O$_2$ buildup should be limited to 1~bar (50~bar, respectively) for TRAPPIST-1h (TRAPPIST-1b, respectively). Finally, \citet{Way:2020} recently proposed that the surface of a planet may not even need to be a magma ocean for large quantities of O$_2$ to be absorbed. O$_2$ may be indeed absorbed by magma released through large scale resurfacing processes as seen on Venus in the past several hundred million years.
\end{enumerate}

While there are multiple theoretical uncertainties in the above dependencies that have yet to be fully worked out, the conclusion of these works is that O$_2$-dominated atmospheres are one of the most serious candidates for the composition of TRAPPIST-1 planetary atmospheres \citep{Lincowski:2019}.

\subsection{CH$_4$/NH$_3$ atmospheres}

In the same way that H$_2$O is expected to be efficiently photodissociated in a water-dominated atmosphere, the photodissociation of CH$_4$ and NH$_3$ (which are not sensitive to the condensation cold trap at the radiation levels received on the TRAPPIST-1 planets ; see \citealt{Turbet:2018aa}) by UV radiation may play a significant role in shaping the TRAPPIST-1 planet atmospheres. High energy cross-section of CH$_4$ peaks around 80~nm and is significantly higher in the 20-150~nm wavelength range than for wavelengths just beyond 150~nm by at least 6 orders of magnitude \citep{Keller-Rudek:2013,Arney:2017}. In the 20-150~nm wavelength range, the emission of the star TRAPPIST-1 is much stronger than the Sun (see Fig.~\ref{trappist-1_spectrum}), relative to their total bolometric emission. This indicates that the photodissociation rate of CH$_4$ (with a photodissociation energy threshold $\sim$~277~nm) is expected to be very strong around a star like TRAPPIST-1. 

It is, for example, estimated that it should take roughly
10~My for Titan to remove all the methane (0.07 bar) from
the atmosphere \citep{Yung:1984} and that as much as $\sim$~30 bar could have been photochemically removed in the last 4~billion years. If the photodissociation rate scales linearly with the incoming X/EUV flux, as much as 10$^2$-10$^4$~times (averaged over the surface) more methane could have been photochemically destroyed in the atmosphere of TRAPPIST-1 planets \citep{Turbet:2018aa}. Over the expected age of the system of 7.6$\pm$2.2~Gy \citep{Burgasser:2017}, the planets could have lost between $\sim$~120~bar --Titan’s limit, including the gravity correction-- and 10$^6$~bar of CH$_4$ --when scaling linearly the CH$_4$ loss with the expected X/EUV flux history on TRAPPIST-1b-- through (i) photolysis, then (ii) organic haze formation, and (iii) haze sedimentation on the surface, light hydrogen being lost to space in the process. This mechanism is known as the photochemical atmospheric collapse \citep{Titan:1997science,Turbet:2018aa}.

Knowing that the NH$_3$ high-energy absorption cross-section also peaks around 80~nm and is significantly higher in the 20-150~nm wavelength range \citep{Keller-Rudek:2013}, and that the amplitude of the cross-section is similar to that of CH$_4$ (peak at 5$\times$10$^{-17}$~cm$^2$~molecule$^{-1}$ for CH$_4$ versus 3$\times$10$^{-17}$~cm$^2$~molecule$^{-1}$ for NH$_3$, around 80~nm and at room temperature), the same photochemical atmospheric collapse mechanism should operate for NH$_3$. Unlike CH$_4$, NH$_3$ (with a photodissociation energy threshold $\sim$~301~nm) absorption cross-section is also high in the 160-210~nm wavelength range \citep{Keller-Rudek:2013}. This means that -- comparatively to planets orbiting Sun-like star -- the NH$_3$ photochemical atmospheric collapse could be weaker.

Sustaining continuously a CH$_4$-rich (or NH$_3$-rich) atmosphere over TRAPPIST-1 lifetime would require an extremely large source of methane (or ammonia, respectively). Low concentration of CH$_4$ (up to the $\sim$~0.3$\%$ level) could be sustained in TRAPPIST-1 planetary atmospheres assuming Earth-like CH$_4$ surface production rate from the Earth biosphere \citep{Rugheimer:2015}. CH$_4$ production rates would however have to be extremely high to maintain a CH$_4$-dominated atmosphere. A similar argument could be made with regard to NH$_3$ \citep{Kasting:1982}. We do, however, acknowledge that the formation of high-altitude organic hazes from the NH$_3$ and/or CH$_4$ photolysis \citep{Sagan:1997,Wolf:2010,Arney:2016} may shield these molecules. This could reduce the CH$_4$ and/or NH$_3$ photolysis rate, which would increase their lifetime.

Similarly, large quantities of N$_2$ could be photodissociated, forming HCN \citep{Liang:2007,Tian:2011,Krasnopolsky:2009,Krasnopolsky:2014} and could be lost subsequently in long nitrogen-enriched carbonated chains that could sediment on the surface. This mechanism could in principle remove efficiently N$_2$ from the atmosphere in the long term \citep{Turbet:2018aa}. 
Coupled photochemical Global Climate Models (see e.g. \citealt{Chen:2019}) could be used in the future to further explore these photochemical atmospheric collapse mechanisms to understand when and to what extent it should play a role in shaping TRAPPIST-1 planets-like atmospheres.

\subsection{N$_2$ atmospheres}

If it is not lost in a chemically reduced atmosphere, N$_2$ could also be lost because of ion escape mechanisms.

\citet{Dong:2018} carried out numerical simulations to characterize the stellar wind produced by TRAPPIST-1 and the resulting atmospheric ion escape rates for all of the seven planets, assuming CO$_2$-dominated atmospheres. In their calculations, the stellar wind-driven atmospheric escape rate ranges from 0.1 to 10~bar per billion year, depending on the planet considered (T1b versus T1h). This number could be lowered by up to two orders of magnitude when considering the planet has a strong magnetic field \citep{Dong:2017}. 

\citet{Dong:2019} showed with the case of TRAPPIST-1g that the rate of atmospheric escape due to stellar wind can actually increase by a factor of $\sim$~100 for O$_2$-dominated atmospheres, since O$_2$-dominated exospheres are not expected to cool as efficiently as CO$_2$-dominated ones.
Although to the best of our knowledge no numerical experiments have yet been conducted to study the stellar wind-driven atmospheric escape rate for N$_2$-dominated atmospheres around low mass stars, we argue that quantitatively similar results are expected.
O$_2$ and N$_2$ have different structural, electronic and radiative properties \citep{Fennelly:1992,Heays:2017,Gordon:2017}, resulting in a much higher concentration of O+ than that of N+ in the present-day Earth's upper atmosphere \citep{Bilitza:2017}. However, \citet{Tian:2008,Tian:2008b} showed in fact with a thermosphere model that N+ and O+ ions are 
expected to be present in quite similar abundance in the upper atmosphere of Earth-like (i.e. N$_2$/O$_2$-dominated) planetary atmospheres, 
assuming that they are exposed to large Extreme-UV (EUV) fluxes typical of the young Sun or low-mass stars. 
As a matter of fact, while the upper atmosphere density of N+ in their model is always at least two orders of magnitude 
lower than that of O+ for a present-day Earth incident EUV flux, the density of N+ is very close (and always at least 2.5$\times$ lower) 
to that of O+ for a 10$\times$ higher incident EUV flux (see \citealt{Tian:2008b}, Fig.~7).
Using the prescriptions of \citet{Tian:2008}, \citet{Lichtenegger:2010} evaluated that the stellar wind-driven atmospheric escape rate ranges from 20 to 500~bar of N$_2$ lost per billion year for a planet with an Earth-like atmosphere and exposed to an Extreme-UV from 7 to 20~times that of present-day Earth.

Although the presence of additional coolant gases such as CO$_2$ may lower these rates \citep{Lichtenegger:2010,Johnstone:2018}, we conclude that while an Earth-like atmosphere (i.e. $\sim$~1~bar N$_2$-dominated atmosphere) may be stable against stellar wind-driven escape for the outermost planets of the TRAPPIST-1 system in the event that CO$_2$ is abundant, this is unlikely to be the case for the inner planets in the system. 
We acknowledge however that assessing the amount of nitrogen that may have been lost through this non-thermal escape process 
deserves to be studied in more details with models adapted to the TRAPPIST-1 system (see e.g. \citealt{Dong:2019}). 
\subsection{CO$_2$ atmospheres}
\label{subsection_CO2_atm}

CO$_2$ is the most widespread molecule in the atmospheres of the terrestrial planets of the solar system, and is one of the most likely gases to accumulate in the atmospheres of TRAPPIST-1 planets \citep{Lincowski:2019}.

The strongest argument for the presence (and possibly the accumulation) of CO$_2$ resides in its robustness to atmospheric escape processes. CO$_2$ is a very good radiative coolant of upper atmospheres \citep{Tian:2009,Johnstone:2018} due mostly to strong absorption bands (or emission bands, according to the Kirchhoff's law of radiation) in the thermal infrared wavelengths. This radiative cooling, coupled with a heavy molecular weight, strongly reduces the efficiency of thermal escape. In addition, simulations of \citet{Dong:2017,Dong:2018} show CO$_2$ should be rather robust to ion escape mechanisms too.

\begin{figure*}
\centering
\centerline{\includegraphics[width=\linewidth]{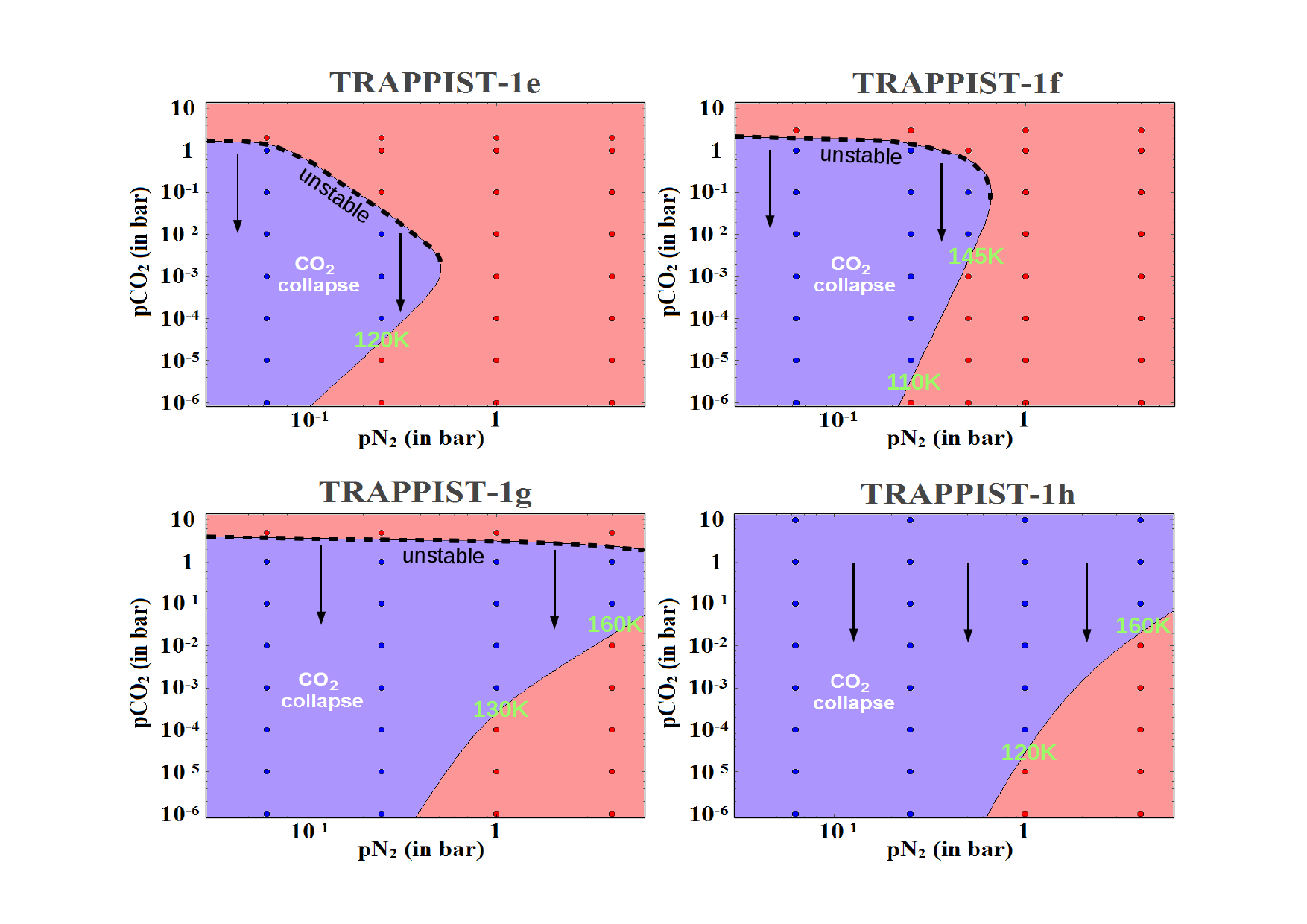}}
\caption{This diagram indicates the range of N$_2$ and CO$_2$ partial pressures for which TRAPPIST-1 planetary atmospheres are robust to CO$_2$ surface condensation collapse (red regions) or not (blue regions). Each dot corresponds to the result of a 3-D Global Climate Model simulation. The black arrows indicate how planets that have an unstable atmosphere (due to CO$_2$ surface condensation) would evolve on the diagram. Temperatures (in green) correspond to the rough estimate (based on GCM simulations) of the surface temperature of the coldest point of the planet, at the stable lower boundary (blue is up; red is down). The figure was taken from \citet{Turbet:2018aa}.}
\label{n2_co2_collapse}
\end{figure*}

For the outer TRAPPIST-1 planets (efgh), CO$_2$ accumulation could be limited by surface condensation, which could be particularly strong if the planets were to be locked in synchronous rotation (see Section~\ref{subsection_tides}). \citet{Turbet:2018aa} explored this possibility with 3-D Global Climate Model simulations taking into account CO$_2$ surface and atmospheric condensation (see Fig.~\ref{n2_co2_collapse}). Specifically, they showed that the CO$_2$ surface collapse mechanism must be all the more important as the planet considered is far from the host star TRAPPIST-1, and the amount of non-condensable gas (that transport heat efficiently ; and broaden the absorption lines of CO$_2$, which increases the greenhouse effect) is low.

Recently, \citet{Hu:2020} used 1-D atmospheric photochemistry	models of TRAPPIST-1 planets assuming CO$_2$-rich atmospheres to show the photodissociation of CO$_2$ may lead to the accumulation of CO and O$_2$, in agreement with previous calculations \citep{Gao:2015}, because the recombination of CO and O would be slow around low mass stars such as TRAPPIST-1. While in Mars and Venus catalytical cycles recombine CO and O$_2$ into CO$_2$ (and thus stabilize CO$_2$), \citet{Hu:2020} showed that this may not be the case for planets orbiting a very low mass star like TRAPPIST-1. This stems from the fact \citep{Hu:2020} that the photodissociation rate of CO$_2$ is higher (due to increased far-UV emission) and catalytical cycles are less efficient (due to decreased near-UV emission) to reform CO$_2$.  We note that while CO and O$_2$ are not radiatively significant greenhouse gases, the build up of multiple bars of these gases can warm the planet through pressure broadening of remaining CO$_2$, and through adiabatic heating (see Table~4 in \citealt{Lincowski:2018}).

Depending on the water vapour abundance in the atmosphere, and thus whether a liquid water ocean is present or not (CO and O$_2$ could also be recombined directly in the ocean), CO and O$_2$ could potentially massively accumulate in a CO$_2$-rich planetary atmosphere around TRAPPIST-1 \citep{Gao:2015,Hu:2020}.

\subsection{Implications for the presence of surface liquid water on TRAPPIST-1 planets}

\begin{figure*}
    \centering
\includegraphics[width=\linewidth]{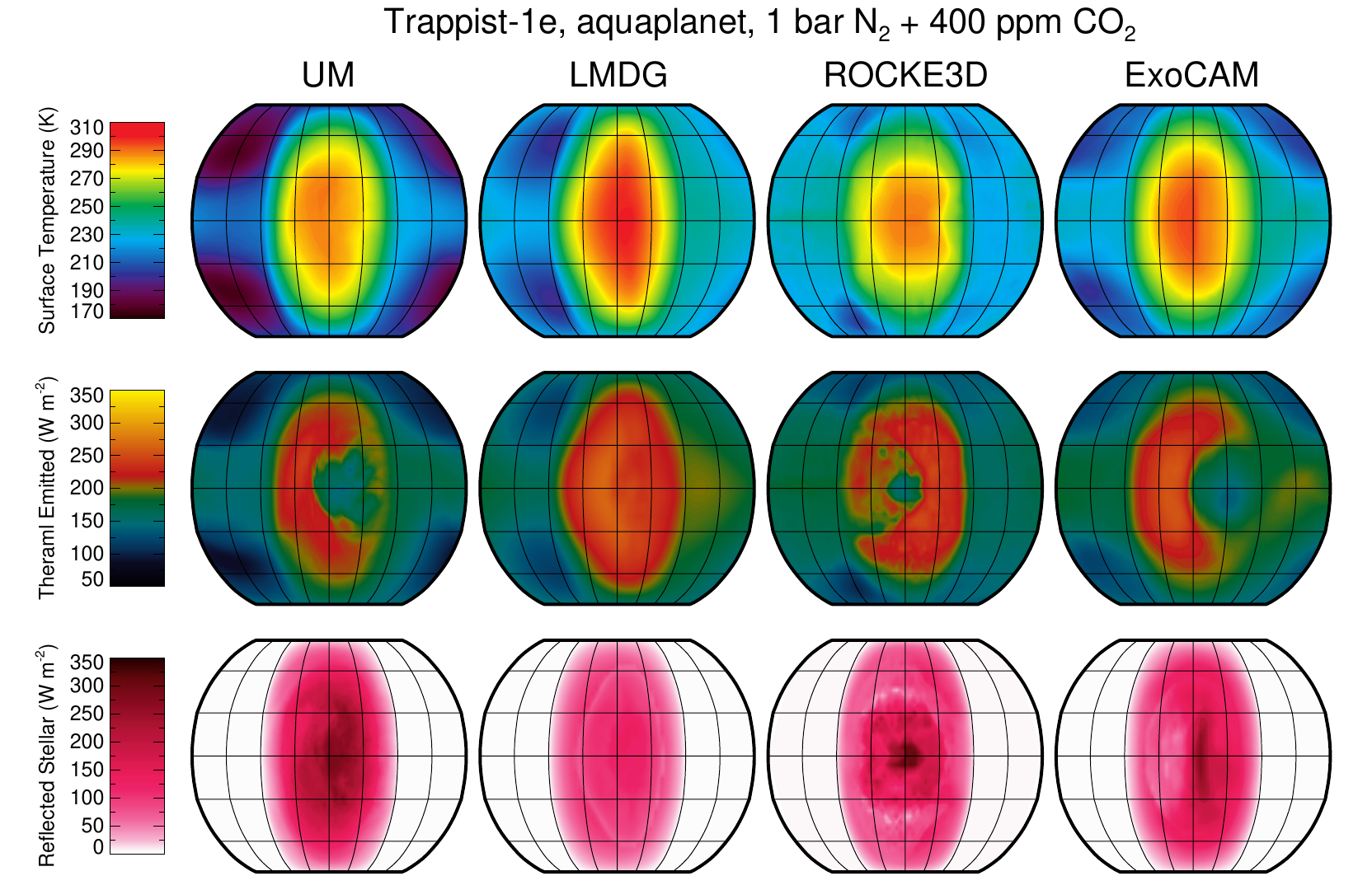}
\caption{Surface contours for surface temperature, thermal emitted radiation at the top of the atmosphere (TOA) and reflected stellar radiation at TOA for "Hab1" scenario (i.e. a scenario of a planet with global ocean and an atmospheric bulk composition similar to present-day Earth) simulated by four of the state-of-the-art exoplanet GCMs: the UK Met Office United Model (UM) \citep{Mayne:2014,Boutle:2017}, the Laboratoire de Météorologie Dynamique Generic model (LMDG) \citep{Wordsworth:2011ajl,Turbet:2018aa}, the Resolving Orbital and Climate Keys of Earth and Extraterrestrial Environments with Dynamics (ROCKE-3D) \citep{Way:2017}, and the National Center for Atmospheric Research Community Atmosphere Model version 4 modified for exoplanets (ExoCAM) \citep{Wolf:2015,Wolf:2017}. The figure was taken from \citet{Fauchez:2020gmd}.}
\label{thai_gcm}
\end{figure*}

Several 3-D Global Climate Models have been used to explore the potential habitability of TRAPPIST-1 planets, i.e. their ability to sustain liquid water on their surface. In short, and looking at Figure~\ref{trappist-1_luminosity_evolution} we see that there are three main possible scenarios
\begin{itemize}
    \item[$\bullet$] Planets b and c (and potentially d) receive more insolation than the runaway greenhouse limit for water. A global surface ocean is therefore unstable, resulting in steam atmospheres and the eventual loss of water to space, pushing the planets into a desiccated state. The presence of surface liquid water would require some very specific circumstances, such as a very thin atmosphere with surface liquid water trapped on the nightside, assuming they are locked in synchronous rotation \citep{Leconte:2013aa,Turbet:2016,Turbet:2018aa}.
    \item[$\bullet$] Planets e, f and g are in a region where only moderate amounts of 
    CO$_2$ would suffice to warm the surface above the melting point of water \citep{Wolf:2018,Turbet:2018aa,Fauchez:2019apj}. TRAPPIST-1e, in particular, should be able to sustain surface liquid water for a wide range of atmospheric compositions and pressures \citep{Turbet:2018aa}. Note presently underway is the TRAPPIST-1 Habitable Atmosphere Intercomparison (THAI) project \citep{Fauchez:2020gmd}, where multiple 3-D GCMs will be compared for habitable iterations of TRAPPIST-1e. This is illustrated in Fig.~\ref{thai_gcm}. While differences in the cloud fields produce variations in the thermal emitted and reflected stellar fluxes, all GCMs tested similarly predict temperate surface conditions and large areas of open ocean on the day-side of the planet, assuming an Earth-like atmospheric composition.
    \item[$\bullet$] Planet h is further out where the classical CO$_2$ runaway collapse is supposed to take place. To reach temperate conditions, in addition to a very thick CO$_2$ atmosphere, the planet would need to retain enough volcanic gases such as CH$_4$ or H$_2$ in order to block CO$_2$ infrared windows and prevent the collapse of the atmosphere \citep{Pierrehumbert:2011h2,Wordsworth:2017,Ramirez:2017b,Lincowski:2018,Turbet:2018aa,Turbet:2019icarus,Turbet:2020ch4-h2}.  
\end{itemize}

A detailed discussion about the potential habitability of TRAPPIST-1 planets is provided in \citet{Wolf:2017,Wolf:2018} as well as in the Section 7 of \citet{Turbet:2018aa}.

\section{Future prospects}
\label{section_future_prospects}

Thanks to its exceptional properties (system very close to Earth, extremely small host star, planets transiting frequently, very compact orbital architecture), the TRAPPIST-1 system and its seven temperate-orbit, terrestrial-size planets is likely our best chance in the future to learn more about potentially habitable planets. There are several ways to learn more about the system and its planets in the future.

First of all, the basic properties of the seven planets (i.e. their masses and radii) need to be determined with more accuracy in order to carry out comparative planetology with high-precision density measurements. Mass measurements of TRAPPIST-1 planets could be improved with the help of a TTV analysis (i) including all existing transit light curves (e.g. all Spitzer light curves, which are the best quality data available so far ; see Agol et al., in preparation) and accounting for subtle processes such as gravitational tides (that may influence TTV in such a compact system; see \citealt{Bolmont:2020}) and for the presence of additional non-detected (yet!) planets in the system. Mass measurements could also be obtained with radial velocity measurements with near-infrared spectrographs mounted on big telescopes such as SPIRou \citep{Klein:2019}, CARMENES \citep{Quirrenbach:2014}, IRD \citep{Kotani:2018} and NIRPS \citep{Wildi:2017}. Near infrared wavelengths are optimal for these observations because this is where TRAPPIST-1 is the brightest. Eventually, the best mass measurements could be derived using a joined fit of the TTV measurements, the RV measurements, and the long-term stability of the system. 

Likewise, radius measurements of TRAPPIST-1 planets could be improved through several ways. First, a more accurate measurement of the stellar radius will help to better constrain the planet's radii because their transit depths measure the planet-to-star radius ratios. TRAPPIST-1 radius has been measured to a $\sim$~3$\%$~precision (at 1~$\sigma$) so far \citep{vangrootel:2018}. The same $\sim$~3$\%$ uncertainty propagates therefore on the absolute radii of the planets. The distance of the star TRAPPIST-1 has been measured -- thanks to Gaia DR2 parallaxes -- to the exquisite accuracy of 0.16$\%$ \citep{Gaia:2018,Lindegren:2018,Kane:2018}, which is low enough that it has a very minor impact on the final radii uncertainty. The $\sim$~3$\%$ stellar uncertainty mostly depends now on the uncertainty on the age of TRAPPIST-1 \citep{Burgasser:2017}, which propagates to stellar radius uncertainty in stellar evolution models such as those of \citet{Filippazzo:2015}. However, we acknowledge that the density should be better constrained than either the mass
or radius, because it can be computed without the stellar uncertainty \citep{Grimm:2018}. Note that the stellar radius estimate may also be biased by (i) magnetic activity effects and/or (ii) tidal interactions of the planets with the star \citep{Burgasser:2017,Gonzales:2019} which are not included in existing stellar evolution models. Secondly, stellar contamination of the photosphere of TRAPPIST-1 by spots need to be further investigated, as it was shown to be a potential source of bias -- up to $\sim$~2.5$\%$ for TRAPPIST-1 in the infrared Spitzer IRAC bands -- for the radius estimates \citep{Rackham:2018}. Last, alternative methods for planetary radius determination such as the ingress/egress duration measurement \citep{Morris:2018b} could be used on larger datasets to derive independent, possibly more precise measurements.

\begin{figure*}
\centering
\centerline{\includegraphics[width=\linewidth]{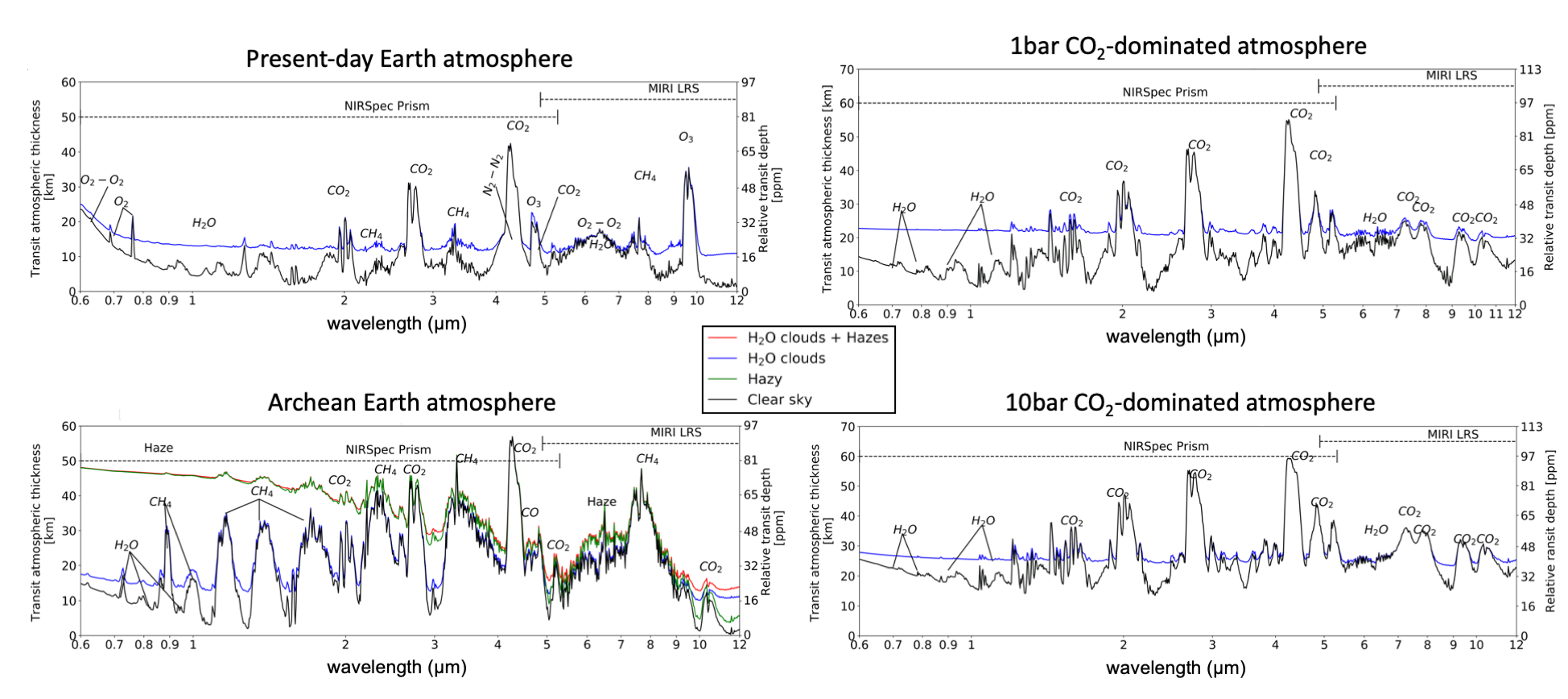}}
\caption{Synthetic transmission spectra simulated for TRAPPIST-1e at the spectral coverage and resolution of JWST NIRSpec and MIRI instruments. Each panel corresponds to a different composition, from top to bottom and left to right: a present-day Earth atmosphere, an Archean Earth (N$_2$/CO$_2$/CH$_4$-dominated) atmosphere, a 1~bar CO$_2$-dominated atmosphere and a 10~bar CO$_2$-dominated atmosphere. While black lines indicate cloud-free transmission spectra, coloured lines take into account the effect of clouds, hazes, and both at the same time. The transmission spectra were computed using coupled 3-D Global Climate Model and 1-D photochemical climate model simulations. The figure was adapted from \citet{Fauchez:2019apj}.}
\label{transmission_spectrum_T1e_jwst}
\end{figure*}

The second main area of progress is atmospheric characterizations techniques: (i) transmission spectroscopy during transits and (ii) thermal infrared secondary eclipses and phase curves. Transmission spectroscopy observations have been initiated on TRAPPIST-1 planets in the near-infrared with HST/WFC3 observations \citep{Dewit:2016}. Meanwhile, significant efforts have been made to prepare transmission spectroscopy observations with forthcoming large-aperture telescopes  \citep{Barstow:2016,Morley:2017,Lincowski:2018,Krissansen-Totton:2018,Wunderlich:2019,Lustig-yaeger:2019,Fauchez:2019apj,Pidhorodetska:2020} with a particular focus on the James Webb Space Telescope \citep{Gillon:2020} which should be in operation within a few years. \citet{Morley:2017} first used a very simplified reverse 1-D radiative-convective model of TRAPPIST-1 planets to determine our ability to characterize Earth-like, Venus-like and Titan-like atmospheres around TRAPPIST-1 planets. They found that $\sim$~20 transits are required for a 5~$\sigma$ detection of molecular spectral features (i.e. to rule out a flat line at 5~$\sigma$ confidence) for most of TRAPPIST-1 planets, but could be as low as 4 transits for some planets if their properties are favourable. \citet{Lincowski:2018} then used a 1-D radiative-convective model coupled with a photochemistry module to show that transit spectroscopy on JWST can be used to distinguish between some of the possible atmospheres (e.g. CO$_2$-dominated and O$_2$-dominated, discussed in Section~\ref{section_numerical_modelling}) expected on TRAPPIST-1 planets. Using the same models, \citet{Lustig-yaeger:2019} estimated that CO$_2$-containing atmospheres could be detected potentially in fewer than 10~transits -- for all seven TRAPPIST-1 planets -- if they lack high-altitude aerosols. \citet{Fauchez:2019apj} calculated similar number of transits using more sophisticated 3-D Global Climate Models. The detection of CO$_2$-containing atmosphere can be done mainly through the strong CO$_2$ 4.3~$\mu$m absorption band \citep{Barstow:2016,Morley:2017,Krissansen-Totton:2018,Lustig-yaeger:2019,Wunderlich:2019,Fauchez:2019apj,Pidhorodetska:2020}. This 4.3~$\mu$m CO$_2$ band is clearly the most promising absorption feature to search for in TRAPPIST-1 planets transmission spectra with JWST. As illustrated in Fig.~\ref{transmission_spectrum_T1e_jwst}, it is weakly affected by clouds and hazes and robust to a wide range of CO$_2$ concentration. Water may be extremely difficult to detect for TRAPPIST-1e and more distant planets (unless the planets are in an unlikely moist greenhouse state) because (i) water vapour should be -- in most cases -- confined in the lower atmosphere (because of a condensation cold trap at the top of the troposphere) where transmission spectra are less sensitive \citep{Lustig-yaeger:2019,Fauchez:2019apj}, but also because (ii) water clouds (forming preferentially at the top of the troposphere) should flatten the transmission spectra \citep{Fauchez:2019apj,Komacek:2020}. O$_2$ could also be detected with transit spectroscopy on JWST through the O$_2$-O$_2$ infrared 6.4~$\mu$m collision-induced absorption \citep{Fauchez:2020} in $\sim$~10~transits for most planets of the system if they have a very dry O$_2$-dominated atmosphere. The presence of O$_2$ could also be indirectly inferred from the detection of O$_3$ (through the 9.6~$\mu$m absorption band), which has been shown to be detectable in tens of transits if present at present-day Earth level \citep{Barstow:2016}. Detecting these two molecular features (O$_2$-O$_2$ at 6.4~$\mu$m; O$_3$ at 9.6~$\mu$m) requires the use of the JWST-MIRI\footnote{MIRI has a wavelength coverage from 4.9 to 28.8~$\mu$m. It can be used with 9 different broad-band filters, and can also be used in a Low Resolution Spectroscopy (LRS) mode of R ($\lambda/{\Delta \lambda}$)~$\sim$~100. Detailed documentation is available at \url{https://jwst-docs.stsci.edu/}.} (Mid-InfraRed Instrument) instrument \citep{Rieke:2015} in Low Resolution Spectroscopy (LRS) mode \citep{Kendrew:2015}. The analysis of \citet{Fauchez:2019apj} suggests that other gases would potentially require hundreds (or thousands) of transits to be detectable, because of clouds and/or hazes flattening the transmission spectra in near-infrared wavelengths. \citet{Lincowski:2019} proposed that JWST-NIRSpec\footnote{NIRSPEC has a wavelength coverage from 0.6 to 5.3 $\mu$m. It can be used in several resolving power modes, including a low resolution prism mode (R ($\lambda/{\Delta \lambda}$)~$\sim$~30-300) and several medium-to-high resolution grism modes (R ($\lambda/{\Delta \lambda}$)~$\sim$~1000-3000) but with a smaller wavelength coverage. Detailed documentation is available at \url{https://jwst-docs.stsci.edu/}} (Near-InfraRed Spectrograph) in particular in prism mode \citep{Bagnasco:2007,Ferruit:2014} transmission spectra could be used to detect CO$_2$ and H$_2$O isotopologues, which would be very informative about isotopic fractionation processes such as atmospheric escape. Isotopologues such as HDO or $^{18}$OCO (at mixing ratios compatible with solar system values) could be detected through their near-infrared absorption bands with as few as 4 to 11~transits at 5~$\sigma$ on JWST-NIRSpec Prism \citep{Lincowski:2019}. However, the presence of clouds and/or hazes would likely preclude detection of these isotopologues with JWST \citep{Fauchez:2019apj}. Moreover, we emphasize that the real impact of stellar contamination by the presence of heterogeneities in the photosphere of TRAPPIST-1 needs to be elucidated, as it can significantly bias the transmission spectra measured with JWST. For that, more information need to be gathered on TRAPPIST-1 during planetary transits (e.g. search for spot crossing events) and out of planetary transit (e.g. characterize the wavelength dependency of the stellar variability). That being said, these promising results motivated altogether Guaranteed Time Observations (GTO) programs on JWST (GTO 1201, PI: David Lafreniere; GTO 1331, PI: Nikole Lewis) to observe 2 transits of TRAPPIST-1d and 4 of TRAPPIST-1e with NIRSpec prism; 4 transits of TRAPPIST-1f and 3 of TRAPPIST-1g with JWST-NIRISS\footnote{NIRISS has a wavelength coverage from 0.6 to 2.8 $\mu$m. In SOSS (Single Object Slitless Spectroscopy) mode, which is the most relevant mode for the characterization of exoplanet atmospheres, it has a medium-resolution (R ($\lambda/{\Delta \lambda}$)~$\sim$~150). Detailed documentation is available at \url{https://jwst-docs.stsci.edu/}.} (Near InfraRed Imager and Slitless Spectrograph) \citep{Doyon:2012}.

In parallel, large-aperture ground-based telescopes coupled to high-resolution spectrographs could be used to infer the properties of the TRAPPIST-1 system and the planetary atmospheres. Such observations could lead to (i) the detection of the Rossiter-McLaughlin effect \citep{Rossiter:1924,McLaughlin:1924,Cloutier:2016} to further constrain the orbital architecture of the system and confirm the 3.3~days rotation period of TRAPPIST-1, which is a crucial assumption in the stellar contamination model of \citealt{Morris:2018a}. \citet{Hirano:2020} have very recently made a first potential detection of the Rossiter-McLaughlin effect with the Subaru-IRD spectrograph and derived a projected rotation velocity of TRAPPIST-1 of 1.49$^{+0.36}_{-0.37}$~km~s$^{-1}$, which corresponds to a maximum stellar rotation period of 3.97$^{+1.32}_{-0.77}$~days, in agreement with the 3.3~days rotation period from the K2 light curves. It is also consistent with the stellar line rotational broadening measurement of TRAPPIST-1 with CARMENES \citep{Reiners:2018} leading to a projected rotation velocity of $\sim$~2~km~s$^{-1}$. The observations of \citet{Hirano:2020} also suggest that the stellar obliquity of TRAPPIST-1 (i.e. the angle between the star’s spin axis and the orbital axis of the planets) is likely small, providing new information on the orbital architecture of the system.; (ii) the detection of molecular absorption lines. At very high spectral resolution,  molecular absorption lines can be individually resolved and their signal can be co-added using the cross-correlation technique \citep{Snellen:2013}. The contamination by resolved telluric lines can be significantly reduced by taking advantage of the Doppler shift arising from the differential speed of the observed system and observers on Earth. Specifically, detection of molecules such as H$_2$O or O$_2$ (e.g. through the 760~nm A-band) could be attempted on the E-ELT \citep{Snellen:2013,Rodler:2014,Serindag:2019}, although the number of required transits may be prohibitively high, especially if clouds and/or hazes are present; (iii) the detection of the high-resolution component (e.g. inverted water vapour feature) of the stellar contamination spectrum (see Fig.~4-5 of \citealt{Ducrot:2018}). The amplitude of inverted water vapour features can be as high as $\sim$~1000~ppm per planet and this signal could be again boosted with the use of the cross-correlation technique. These transit observations could be attempted with near-infrared instruments on existing large telescopes (e.g. CFHT-SPIRou, ESO-3.6m-NIRPS, Calar Alto-CARMENES, Subaru-IRD) and future instruments on extremely large telescopes (e.g. E-ELT/HIRES) ; (iv) the use of Doppler tomography techniques to probe the photosphere of TRAPPIST-1 and put constraints on stellar contamination models. For example, the Zeeman Doppler Imaging (ZDI) tomographic technique could be used to characterize the distribution of magnetically active regions at the stellar surfaces \citep{Hebrard:2016} and thus help to identify the nature of stellar heterogeneities of TRAPPIST-1's photosphere, e.g. the distribution of spots. Such observations could be attempted with near-infrared spectro-polarimeters such as SPIRou or CRIRES+ \citep{Follert:2014}.

In an even more distant future, transit observations in UV could be attempted with UVSPEX, a conceptual design of  Ultraviolet Spectrograph for Exoplanet (UVSPEX) for World Space Observatory Ultraviolet (WSO-UV), to detect the presence of an oxygen-rich exosphere \citep{Kameda:2019}. TRAPPIST-1 would be right at the limit of capability of the instrument/telescope; and with LUVOIR/LUMOS to detect the presence of an hydrogen-rich exosphere in Lyman-$\alpha$ \citep{dosSantos:2019}. As a matter of fact, \citet{dosSantos:2019} calculated that it would be possible to detect the exosphere of an Earth-like planet transiting TRAPPIST-1 at 5~$\sigma$ within 10 transits using LUVOIR-A design with the LUMOS instrument concept.

In parallel, thermal emission observations have been initiated on TRAPPIST-1 planets by observations of the secondary eclipses of TRAPPIST-1b and c using the Spitzer/IRAC 4-5~$\mu$m channel \citep{Ducrot:2020}. While these observations were inconclusive because of the low signal to noise level of the observations \citep{Ducrot:2020}, \citet{Morley:2017} have shown that JWST-MIRI could bring useful constraints in a handful of secondary eclipses for the innermost planets of the system. Besides, emission spectroscopy and photometry are expected to be less affected by the presence of heterogeneities in the photosphere of TRAPPIST-1. \citet{Morley:2017} showed that TRAPPIST-1b -- the most irradiated planet of the system -- is an excellent target for emission spectroscopy with JWST-MIRI in LRS mode, requiring fewer than 10~eclipse observations to detect the band-integrated thermal emission at a 25~$\sigma$ confidence. Inferring the atmospheric composition (CO$_2$-dominated? H$_2$O-dominated? CH$_4$-dominated?) of TRAPPIST-1b may require tens (possibly hundreds) of eclipses for a 5~$\sigma$ detection \citep{Morley:2017,Lustig-yaeger:2019,Koll:2019,Malik:2019}, depending on the atmospheric composition and pressure of the planet. These promising results motivated two Guaranteed Time Observations (GTO) programs on JWST (GTO 1279, PI: Pierre-Olivier Lagage ; GTO 1177, PI: Thomas Greene) to observe 10 secondary eclipses of TRAPPIST-1b with MIRI in imager mode. TRAPPIST-1c and outer planets which are supposedly colder are likely out of reach of eclipse spectroscopy \citep{Lustig-yaeger:2019}. Eclipse photometry could be used to detect an atmosphere on these colder planets in tens of eclipses \citep{Morley:2017,Lustig-yaeger:2019}, but this requires special circumstances (right combination of JWST-MIRI photometric filters and planet atmospheric compositions).

Last but not least, thermal phase curves \citep{Selsis:2011} could be used to further constrain the presence of an atmosphere on TRAPPIST-1 planets as was recently done on the very hot, rocky exoplanet LHS~3844b \citep{Kreidberg:2019}. However, such observations would be time-consuming and potentially difficult to analyze because TRAPPIST-1 is a multiplanetary system and the measured signal would be the result of the superposition of the thermal emission signal of 7 distinct planets.

\section{Conclusions}
\label{section_conclusions}

First of all, we have seen that the host star of the system TRAPPIST-1 is rather harmful to the atmospheric evolution of the planets due to (i) the evolution of its luminosity during the long Pre Main Sequence phase likely causing a runaway greenhouse and (ii) its high non-thermal X and far-UV emission (below $\sim$~150~nm). These two characteristics put together suggest that the TRAPPIST-1 planets likely suffered from intense atmospheric erosion.

However, the ability of the planets today to have an atmosphere depends just as much on their initial volatile reservoir, which strongly depends on how they formed. For now, a scenario where planets were formed far away in the protoplanetary disk (and thus possibly beyond the ice lines of common volatile species such as H$_2$O) and then migrated inwards in resonant chains is preferred, mostly because of the orbital architecture of the system (highly compact and resonant). This is of course not the only possible formation scenario, just as it is not the only possible scenario in which the planets would have an atmosphere today (i.e. secondary outgassing or late cometary delivery). It is important to note that all processes (planet formation, stellar evolution, X/UV and stellar-wind driven erosion, cometary delivery, outgassing) should make it easier for the outer, more massive planets in the system to have an atmosphere than the inner planets.

The most favored formation scenario (formation far in the disk combined with resonant inward migration) suggests that TRAPPIST-1 planets may be enriched in volatiles, possibly water, thus possibly lowering their density in a detectable way. Preliminary results combining radius estimates (based on Spitzer transit light curves, and precise stellar parameters for TRAPPIST-1) with mass estimates (based on TTV analysis, and confirmed with orbital stability analysis) show this may be the case for at least some planets in the system. 

Near-infrared transit observations with the Hubble Space Telescope show that the six inner planets are unlikely to be endowed with a cloud-free hydrogen-dominated envelope. Although the transit observations cannot exclude the case of hydrogen-dominated envelopes with high altitude clouds, measurements of the mass and radius of the planets, combined with atmospheric escape modelling and gas accretion modelling, indicate altogether that this case is unlikely. Despite many observation campaigns with many different telescopes (Spitzer, Hubble, K2, VLT, etc.), existing transit transmission spectra cannot be used to infer the molecular composition of atmospheres (if any) of TRAPPIST-1 planets. This stems from the fact that (i) the signal to noise ratio of single observations is insufficient to detect high mean molecular weight atmospheres, (ii) it is challenging to accurately compare the absolute transit depths between observations made by different instruments at different wavelengths, (iii) last but not least the transit transmission spectra may be significantly affected by the stellar contamination due to the presence of (occulted and/or unocculted) cold and/or bright spots in the photosphere of TRAPPIST-1. Depending on the (still debated) quantity, temperature(s) and spatial distribution of the spots, stellar contamination may (or may not) have a huge impact on the transit spectra of the TRAPPIST-1 planets. By combining information from all existing observations on the system (stellar temporal variability at different wavelengths, search for spot crossing events in the transit light curves, quality of the fit of the stellar contamination spectrum on the combined transit light curves, etc.) it should be possible to better constrain the nature of heterogeneities present in the photosphere of TRAPPIST-1, as well as to build a stellar contamination model that can be used to correct TRAPPIST-1 planets transit transmission spectra. This work is crucial in preparing future transit observations with the James Webb Space Telescope \citep{Gillon:2020}.

In addition to transit observations, density measurements can be used to place constraints on the possible water content of the system's planets. Specifically, possible water mass fractions range from 0 up to 25$\%$, despite expected strong atmospheric water vapour erosion. Upper water mass fraction estimates are however expected to be significantly lowered for TRAPPIST-1 b, c and d. This stems from the fact that on these three planets, water must be in the form of steam, which increases -- for a given amount of water -- the apparent size of their atmosphere.

Last, numerical modeling (3-D global climate models, 1-D photochemical climate models, ion escape models) can be used to further constrain the possible atmospheres of TRAPPIST-1 planets. Some of the most likely atmospheres include:
\begin{enumerate}
    \item H$_2$O-dominated atmospheres. Only for the three inner TRAPPIST-1 planets, because they are more irradiated than the runaway greenhouse radiation limit.
    \item O$_2$-dominated atmospheres. They are the natural remnant of early erosion of H$_2$O-dominated atmospheres.
    \item CO$_2$-dominated atmospheres. This stems from the fact that CO$_2$ is much more resilient to atmospheric escape processes than other common gases. However, photochemistry and surface condensation could significantly reduce the CO$_2$ content of the atmosphere.
\end{enumerate}

CH$_4$ and NH$_3$ are unlikely to be dominant gases because of their sensitivity to atmospheric photochemical collapse. It is unlikely that the planets could have maintained an N$_2$-dominated atmosphere (especially the innermost planets) unless they started with a much higher nitrogen content than that of the present-day Earth. 

There are undoubtedly, at the present stage, so many known unknowns and unknown unknowns that we must remain open to a wide range of possible atmospheric compositions of the TRAPPIST-1 planets. Fortunately, a new wave of observations with the James Webb Space Telescope and near-infrared high-resolution ground-based spectrographs on existing very large and forthcoming extremely large telescopes will bring significant advances in the coming decade.

\section*{Acknowledgements}
This project has received funding from the European Union’s Horizon 2020 research and innovation program under the 
Marie Sklodowska-Curie Grant Agreement No. 832738/ESCAPE, and through the European Research Council (ERC) grant agreements 
n$^\circ$ 679030/WHIPLASH and n$^\circ$ 724427/FOUR ACES. M.T. thanks the Gruber Foundation for its support to this research. 
M.T. thanks Sarah Peacock for sharing her high resolution TRAPPIST-1 spectra. M.T. thanks Nathan Hara and Christophe Lovis for 
useful discussions related to this work. The authors acknowledge support by the Swiss National Science Foundation (SNSF) in the 
frame of the National Centre for Competence in Research "PlanetS". This research has made use of NASA's Astrophysics Data System. 
This project is partly supported by the International Space Science Institute (ISSI) in the framework of an international team entitled 
"Understanding the Diversity of Planetary Atmospheres". 
The authors thank the reviewers for their constructive remarks and comments which helped to improve the manuscript.

\bibliographystyle{apalike} 
\bibliography{biblio} 

\begin{thebibliography}{}

\bibitem[{Agol} et~al., 2005]{Agol:2005}
{Agol}, E., {Steffen}, J., {Sari}, R., and {Clarkson}, W. (2005).
\newblock {On detecting terrestrial planets with timing of giant planet
  transits}.
\newblock {\em Monthly Notices of the Royal Astronomical Society},
  359:567--579.

\bibitem[{Akeson} et~al., 2013]{Akeson:2013}
{Akeson}, R.~L., {Chen}, X., {Ciardi}, D., {Crane}, M., {Good}, J., {Harbut},
  M., {Jackson}, E., {Kane}, S.~R., {Laity}, A.~C., {Leifer}, S., {Lynn}, M.,
  {McElroy}, D.~L., {Papin}, M., {Plavchan}, P., {Ram{\'\i}rez}, S.~V., {Rey},
  R., {von Braun}, K., {Wittman}, M., {Abajian}, M., {Ali}, B., {Beichman}, C.,
  {Beekley}, A., {Berriman}, G.~B., {Berukoff}, S., {Bryden}, G., {Chan}, B.,
  {Groom}, S., {Lau}, C., {Payne}, A.~N., {Regelson}, M., {Saucedo}, M.,
  {Schmitz}, M., {Stauffer}, J., {Wyatt}, P., and {Zhang}, A. (2013).
\newblock {The NASA Exoplanet Archive: Data and Tools for Exoplanet Research}.
\newblock {\em Publications of the Astronomical Society of the Pacific},
  125(930):989.

\bibitem[{Arney} et~al., 2016]{Arney:2016}
{Arney}, G., {Domagal-Goldman}, S.~D., {Meadows}, V.~S., {Wolf}, E.~T.,
  {Schwieterman}, E., {Charnay}, B., {Claire}, M., {H{\'e}brard}, E., and
  {Trainer}, M.~G. (2016).
\newblock {The Pale Orange Dot: The Spectrum and Habitability of Hazy Archean
  Earth}.
\newblock {\em Astrobiology}, 16:873--899.

\bibitem[{Arney} et~al., 2017]{Arney:2017}
{Arney}, G.~N., {Meadows}, V.~S., {Domagal-Goldman}, S.~D., {Deming}, D.,
  {Robinson}, T.~D., {Tovar}, G., {Wolf}, E.~T., and {Schwieterman}, E. (2017).
\newblock {Pale Orange Dots: The Impact of Organic Haze on the Habitability and
  Detectability of Earthlike Exoplanets}.
\newblock {\em The Astrophysical Journal}, 836:49.

\bibitem[{Auclair-Desrotour} and {Heng}, 2020]{Auclair-Desrotour:2020}
{Auclair-Desrotour}, P. and {Heng}, K. (2020).
\newblock {Atmospheric stability and collapse on tidally locked rocky planets}.
\newblock {\em arXiv e-prints}, page arXiv:2004.07134.

\bibitem[{Auclair-Desrotour} et~al., 2017]{Auclair:2017}
{Auclair-Desrotour}, P., {Laskar}, J., {Mathis}, S., and {Correia}, A.~C.~M.
  (2017).
\newblock {The rotation of planets hosting atmospheric tides: from Venus to
  habitable super-Earths}.
\newblock {\em Astronomy $\&$ Astrophysics}, 603:A108.

\bibitem[{Auclair-Desrotour} et~al., 2019]{Auclair:2019}
{Auclair-Desrotour}, P., {Leconte}, J., {Bolmont}, E., and {Mathis}, S. (2019).
\newblock {Final spin states of eccentric ocean planets}.
\newblock {\em Astronomy $\&$ Astrophysics}, 629:A132.

\bibitem[{Bagnasco} et~al., 2007]{Bagnasco:2007}
{Bagnasco}, G., {Kolm}, M., {Ferruit}, P., {Honnen}, K., {Koehler}, J.,
  {Lemke}, R., {Maschmann}, M., {Melf}, M., {Noyer}, G., {Rumler}, P.,
  {Salvignol}, J.-C., {Strada}, P., and {Te Plate}, M. (2007).
\newblock {\em {Overview of the near-infrared spectrograph (NIRSpec) instrument
  on-board the James Webb Space Telescope (JWST)}}, volume 6692 of {\em Society
  of Photo-Optical Instrumentation Engineers (SPIE) Conference Series}, page
  66920M.

\bibitem[{Baraffe} et~al., 1998]{Baraffe:1998}
{Baraffe}, I., {Chabrier}, G., {Allard}, F., and {Hauschildt}, P.~H. (1998).
\newblock {Evolutionary models for solar metallicity low-mass stars:
  mass-magnitude relationships and color-magnitude diagrams}.
\newblock {\em Astronomy $\&$ Astrophysics}, 337:403--412.

\bibitem[{Baraffe} et~al., 2015]{Baraffe:2015}
{Baraffe}, I., {Homeier}, D., {Allard}, F., and {Chabrier}, G. (2015).
\newblock {New evolutionary models for pre-main sequence and main sequence
  low-mass stars down to the hydrogen-burning limit}.
\newblock {\em Astronomy $\&$ Astrophysics}, 577:A42.

\bibitem[{Barr} et~al., 2018]{Barr:2018}
{Barr}, A.~C., {Dobos}, V., and {Kiss}, L.~L. (2018).
\newblock {Interior structures and tidal heating in the TRAPPIST-1 planets}.
\newblock {\em Astronomy $\&$ Astrophysics}, 613:A37.

\bibitem[{Barstow} and {Irwin}, 2016]{Barstow:2016}
{Barstow}, J.~K. and {Irwin}, P.~G.~J. (2016).
\newblock {Habitable worlds with JWST: transit spectroscopy of the TRAPPIST-1
  system?}
\newblock {\em Monthly Notices of the Royal Astronomical Society},
  461:L92--L96.

\bibitem[{Bilitza} et~al., 2017]{Bilitza:2017}
{Bilitza}, D., {Altadill}, D., {Truhlik}, V., {Shubin}, V., {Galkin}, I.,
  {Reinisch}, B., and {Huang}, X. (2017).
\newblock {International Reference Ionosphere 2016: From ionospheric climate to
  real-time weather predictions}.
\newblock {\em Space Weather}, 15(2):418--429.

\bibitem[{Bolmont} et~al., 2020]{Bolmont:2020}
{Bolmont}, E., {Demory}, B.~O., {Blanco-Cuaresma}, S., {Agol}, E., {Grimm},
  S.~L., {Auclair-Desrotour}, P., {Selsis}, F., and {Leleu}, A. (2020).
\newblock {Impact of tides on the transit-timing fits to the TRAPPIST-1
  system}.
\newblock {\em Astronomy $\&$ Astrophysics}, 635:A117.

\bibitem[{Bolmont} et~al., 2017a]{Bolmont:2017}
{Bolmont}, E., {Selsis}, F., {Owen}, J.~E., {Ribas}, I., {Raymond}, S.~N.,
  {Leconte}, J., and {Gillon}, M. (2017a).
\newblock {Water loss from terrestrial planets orbiting ultracool dwarfs:
  implications for the planets of TRAPPIST-1}.
\newblock {\em Monthly Notices of the Royal Astronomical Society},
  464:3728--3741.

\bibitem[{Bolmont} et~al., 2017b]{Bolmont2017}
{Bolmont}, E., {Selsis}, F., {Owen}, J.~E., {Ribas}, I., {Raymond}, S.~N.,
  {Leconte}, J., and {Gillon}, M. (2017b).
\newblock {Water loss from terrestrial planets orbiting ultracool dwarfs:
  implications for the planets of TRAPPIST-1}.
\newblock {\em Monthly Notices of the Royal Astronomical Society},
  464:3728--3741.

\bibitem[{Bourrier} et~al., 2017a]{Bourrier:2017b}
{Bourrier}, V., {de Wit}, J., {Bolmont}, E., {Stamenkovi{\'c}}, V., {Wheatley},
  P.~J., {Burgasser}, A.~J., {Delrez}, L., {Demory}, B.-O., {Ehrenreich}, D.,
  {Gillon}, M., {Jehin}, E., {Leconte}, J., {Lederer}, S.~M., {Lewis}, N.,
  {Triaud}, A.~H.~M.~J., and {Van Grootel}, V. (2017a).
\newblock {Temporal Evolution of the High-energy Irradiation and Water Content
  of TRAPPIST-1 Exoplanets}.
\newblock {\em The Astronomical Journal}, 154:121.

\bibitem[{Bourrier} et~al., 2017b]{Bourrier:2017}
{Bourrier}, V., {Ehrenreich}, D., {Wheatley}, P.~J., {Bolmont}, E., {Gillon},
  M., {de Wit}, J., {Burgasser}, A.~J., {Jehin}, E., {Queloz}, D., and
  {Triaud}, A.~H.~M.~J. (2017b).
\newblock {Reconnaissance of the TRAPPIST-1 exoplanet system in the
  Lyman-{$\alpha$} line}.
\newblock {\em Astronomy $\&$ Astrophysics}, 599:L3.

\bibitem[{Boutle} et~al., 2017]{Boutle:2017}
{Boutle}, I.~A., {Mayne}, N.~J., {Drummond}, B., {Manners}, J., {Goyal}, J.,
  {Hugo Lambert}, F., {Acreman}, D.~M., and {Earnshaw}, P.~D. (2017).
\newblock {Exploring the climate of Proxima B with the Met Office Unified
  Model}.
\newblock {\em Astronomy $\&$ Astrophysics}, 601:A120.

\bibitem[{Brasser} et~al., 2019]{Brasser:2019}
{Brasser}, R., {Barr}, A.~C., and {Dobos}, V. (2019).
\newblock {The tidal parameters of TRAPPIST-1b and c}.
\newblock {\em Monthly Notices of the Royal Astronomical Society},
  487(1):34--47.

\bibitem[{Burdanov} et~al., 2019]{Burdanov:2019}
{Burdanov}, A.~Y., {Lederer}, S.~M., {Gillon}, M., {Delrez}, L., {Ducrot}, E.,
  {de Wit}, J., {Jehin}, E., {Triaud}, A.~H.~M.~J., {Lidman}, C., {Spitler},
  L., {Demory}, B.~O., {Queloz}, D., and {Van Grootel}, V. (2019).
\newblock {Ground-based follow-up observations of TRAPPIST-1 transits in the
  near-infrared}.
\newblock {\em Monthly Notices of the Royal Astronomical Society},
  487(2):1634--1652.

\bibitem[{Burgasser} and {Mamajek}, 2017]{Burgasser:2017}
{Burgasser}, A.~J. and {Mamajek}, E.~E. (2017).
\newblock {On the Age of the TRAPPIST-1 System}.
\newblock {\em The Astrophysical Journal}, 845:110.

\bibitem[{Carter} et~al., 2012]{Carter:2012}
{Carter}, J.~A., {Agol}, E., {Chaplin}, W.~J., {Basu}, S., {Bedding}, T.~R.,
  {Buchhave}, L.~A., {Christensen-Dalsgaard}, J., {Deck}, K.~M., {Elsworth},
  Y., {Fabrycky}, D.~C., {Ford}, E.~B., {Fortney}, J.~J., {Hale}, S.~J.,
  {Handberg}, R., {Hekker}, S., {Holman}, M.~J., {Huber}, D., {Karoff}, C.,
  {Kawaler}, S.~D., {Kjeldsen}, H., {Lissauer}, J.~J., {Lopez}, E.~D., {Lund},
  M.~N., {Lundkvist}, M., {Metcalfe}, T.~S., {Miglio}, A., {Rogers}, L.~A.,
  {Stello}, D., {Borucki}, W.~J., {Bryson}, S., {Christiansen}, J.~L.,
  {Cochran}, W.~D., {Geary}, J.~C., {Gilliland}, R.~L., {Haas}, M.~R., {Hall},
  J., {Howard}, A.~W., {Jenkins}, J.~M., {Klaus}, T., {Koch}, D.~G., {Latham},
  D.~W., {MacQueen}, P.~J., {Sasselov}, D., {Steffen}, J.~H., {Twicken}, J.~D.,
  and {Winn}, J.~N. (2012).
\newblock {Kepler-36: A Pair of Planets with Neighboring Orbits and Dissimilar
  Densities}.
\newblock {\em Science}, 337:556.

\bibitem[{Chabrier} and {Baraffe}, 1997]{Chabrier:1997}
{Chabrier}, G. and {Baraffe}, I. (1997).
\newblock {Structure and evolution of low-mass stars}.
\newblock {\em Astronomy $\&$ Astrophysics}, 327:1039--1053.

\bibitem[{Chadney} et~al., 2015]{Chadney:2015}
{Chadney}, J.~M., {Galand}, M., {Unruh}, Y.~C., {Koskinen}, T.~T., and
  {Sanz-Forcada}, J. (2015).
\newblock {XUV-driven mass loss from extrasolar giant planets orbiting active
  stars}.
\newblock {\em Icarus}, 250:357--367.

\bibitem[{Chapman} and {Lindzen}, 1970]{Chapman:1970}
{Chapman}, S. and {Lindzen}, R. (1970).
\newblock {\em {Atmospheric tides. Thermal and gravitational}}.

\bibitem[{Chen} et~al., 2019]{Chen:2019}
{Chen}, H., {Wolf}, E.~T., {Zhan}, Z., and {Horton}, D.~E. (2019).
\newblock {Habitability and Spectroscopic Observability of Warm M-dwarf
  Exoplanets Evaluated with a 3D Chemistry-Climate Model}.
\newblock {\em The Astrophysical Journal}, 886(1):16.

\bibitem[{Cloutier} and {Triaud}, 2016]{Cloutier:2016}
{Cloutier}, R. and {Triaud}, A. H.~M.~J. (2016).
\newblock {Prospects for detecting the Rossiter-McLaughlin effect of Earth-like
  planets: the test case of TRAPPIST-1b and c}.
\newblock {\em Monthly Notices of the Royal Astronomical Society},
  462(4):4018--4027.

\bibitem[{Coleman} et~al., 2019]{Coleman:2019}
{Coleman}, G.~A.~L., {Leleu}, A., {Alibert}, Y., and {Benz}, W. (2019).
\newblock {Pebbles versus planetesimals: the case of Trappist-1}.
\newblock {\em Astronomy $\&$ Astrophysics}, 631:A7.

\bibitem[{Correia} and {Laskar}, 2001]{Correia:2001}
{Correia}, A.~C.~M. and {Laskar}, J. (2001).
\newblock {The four final rotation states of Venus}.
\newblock {\em Nature}, 411:767--770.

\bibitem[{Cossou} et~al., 2014]{Cossou:2014}
{Cossou}, C., {Raymond}, S.~N., {Hersant}, F., and {Pierens}, A. (2014).
\newblock {Hot super-Earths and giant planet cores from different migration
  histories}.
\newblock {\em Astronomy $\&$ Astrophysics}, 569:A56.

\bibitem[{Davies}, 2013]{Davies:2013}
{Davies}, J.~H. (2013).
\newblock {Global map of solid Earth surface heat flow}.
\newblock {\em Geochemistry, Geophysics, Geosystems}, 14:4608--4622.

\bibitem[{Davies} and {Davies}, 2010]{Davies:2010}
{Davies}, J.~H. and {Davies}, D.~R. (2010).
\newblock {Earth's surface heat flux}.
\newblock {\em Solid Earth}, 1:5--24.

\bibitem[{de Wit} et~al., 2016]{Dewit:2016}
{de Wit}, J., {Wakeford}, H.~R., {Gillon}, M., {Lewis}, N.~K., {Valenti},
  J.~A., {Demory}, B.-O., {Burgasser}, A.~J., {Burdanov}, A., {Delrez}, L.,
  {Jehin}, E., {Lederer}, S.~M., {Queloz}, D., {Triaud}, A.~H.~M.~J., and {Van
  Grootel}, V. (2016).
\newblock {A combined transmission spectrum of the Earth-sized exoplanets
  TRAPPIST-1 b and c}.
\newblock {\em Nature}, 537:69--72.

\bibitem[{de Wit} et~al., 2018]{Dewit:2018}
{de Wit}, J., {Wakeford}, H.~R., {Lewis}, N.~K., {Delrez}, L., {Gillon}, M.,
  {Selsis}, F., {Leconte}, J., {Demory}, B.-O., {Bolmont}, E., {Bourrier}, V.,
  {Burgasser}, A.~J., {Grimm}, S., {Jehin}, E., {Lederer}, S.~M., {Owen},
  J.~E., {Stamenkovi{\'c}}, V., and {Triaud}, A.~H.~M.~J. (2018).
\newblock {Atmospheric reconnaissance of the habitable-zone Earth-sized planets
  orbiting TRAPPIST-1}.
\newblock {\em Nature Astronomy}, 2:214--219.

\bibitem[{Deck} and {Agol}, 2015]{Deck:2015}
{Deck}, K.~M. and {Agol}, E. (2015).
\newblock {Measurement of Planet Masses with Transit Timing Variations Due to
  Synodic {\textquotedblleft}Chopping{\textquotedblright} Effects}.
\newblock {\em The Astrophysical Journal}, 802(2):116.

\bibitem[{Delrez} et~al., 2018]{Delrez:2018}
{Delrez}, L., {Gillon}, M., {Triaud}, A.~H.~M.~J., {Demory}, B.-O., {de Wit},
  J., {Ingalls}, J.~G., {Agol}, E., {Bolmont}, E., {Burdanov}, A., {Burgasser},
  A.~J., {Carey}, S.~J., {Jehin}, E., {Leconte}, J., {Lederer}, S., {Queloz},
  D., {Selsis}, F., and {Van Grootel}, V. (2018).
\newblock {Early 2017 observations of TRAPPIST-1 with Spitzer}.
\newblock {\em Monthly Notices of the Royal Astronomical Society},
  475:3577--3597.

\bibitem[{Dencs} and {Reg{\'a}ly}, 2019]{Dencs:2019}
{Dencs}, Z. and {Reg{\'a}ly}, Z. (2019).
\newblock {Water delivery to the TRAPPIST-1 planets}.
\newblock {\em Monthly Notices of the Royal Astronomical Society},
  487(2):2191--2199.

\bibitem[{Dobos} et~al., 2019]{Dobos:2019}
{Dobos}, V., {Barr}, A.~C., and {Kiss}, L.~L. (2019).
\newblock {Tidal heating and the habitability of the TRAPPIST-1 exoplanets}.
\newblock {\em Astronomy $\&$ Astrophysics}, 624:A2.

\bibitem[{Dong} et~al., 2019]{Dong:2019}
{Dong}, C., {Huang}, Z., and {Lingam}, M. (2019).
\newblock {Role of Planetary Obliquity in Regulating Atmospheric Escape:
  G-dwarf versus M-dwarf Earth-like Exoplanets}.
\newblock {\em The Astrophysical Journal Letters}, 882(2):L16.

\bibitem[{Dong} et~al., 2018]{Dong:2018}
{Dong}, C., {Jin}, M., {Lingam}, M., {Airapetian}, V.~S., {Ma}, Y., and {van
  der Holst}, B. (2018).
\newblock {Atmospheric escape from the TRAPPIST-1 planets and implications for
  habitability}.
\newblock {\em Proceedings of the National Academy of Science}, 115:260--265.

\bibitem[{Dong} et~al., 2017]{Dong:2017}
{Dong}, C., {Lingam}, M., {Ma}, Y., and {Cohen}, O. (2017).
\newblock {Is Proxima Centauri b Habitable? A Study of Atmospheric Loss}.
\newblock {\em The Astrophysical Journal Letters}, 837(2):L26.

\bibitem[{Dorn} et~al., 2018]{Dorn:2018}
{Dorn}, C., {Mosegaard}, K., {Grimm}, S.~L., and {Alibert}, Y. (2018).
\newblock {Interior Characterization in Multiplanetary Systems: TRAPPIST-1}.
\newblock {\em The Astrophysical Journal}, 865:20.

\bibitem[{dos Santos} et~al., 2019]{dosSantos:2019}
{dos Santos}, L.~A., {Bourrier}, V., {Ehrenreich}, D., and {Kameda}, S. (2019).
\newblock {Observability of hydrogen-rich exospheres in Earth-like exoplanets}.
\newblock {\em Astronomy $\&$ Astrophysics}, 622:A46.

\bibitem[{Doyon} et~al., 2012]{Doyon:2012}
{Doyon}, R., {Hutchings}, J.~B., {Beaulieu}, M., {Albert}, L.,
  {Lafreni{\`e}re}, D., {Willott}, C., {Touahri}, D., {Rowlands}, N.,
  {Maszkiewicz}, M., {Fullerton}, A.~W., {Volk}, K., {Martel}, A.~R., {Chayer},
  P., {Sivaramakrishnan}, A., {Abraham}, R., {Ferrarese}, L., {Jayawardhana},
  R., {Johnstone}, D., {Meyer}, M., {Pipher}, J.~L., and {Sawicki}, M. (2012).
\newblock {\em {The JWST Fine Guidance Sensor (FGS) and Near-Infrared Imager
  and Slitless Spectrograph (NIRISS)}}, volume 8442 of {\em Society of
  Photo-Optical Instrumentation Engineers (SPIE) Conference Series}, page
  84422R.

\bibitem[{Ducrot} et~al., 2020]{Ducrot:2020}
{Ducrot}, E., {Gillon}, M., {Delrez}, L., {Agol}, E., {Rimmer}, P., {Turbet},
  M., {G{\"u}nther}, M.~N., {Demory}, B.-O., {Triaud}, A.~H.~M.~J., {Bolmont},
  E., {Burgasser}, A., {Carey}, S.~J., {Ingalls}, J.~G., {Jehin}, E.,
  {Leconte}, J., {Lederer}, S.~M., {Queloz}, D., {Raymond}, S.~N., {Selsis},
  F., {Van Grootel}, V., and {de Wit}, J. (2020).
\newblock {TRAPPIST-1: Global Results of the Spitzer Exploration Science
  Program Red Worlds}.
\newblock {\em arXiv e-prints}, page arXiv:2006.13826.

\bibitem[{Ducrot} et~al., 2018]{Ducrot:2018}
{Ducrot}, E., {Sestovic}, M., {Morris}, B.~M., {Gillon}, M., {Triaud},
  A.~H.~M.~J., {De Wit}, J., {Thimmarayappa}, D., {Agol}, E., {Almleaky}, Y.,
  {Burdanov}, A., {Burgasser}, A.~J., {Delrez}, L., {Demory}, B.-O., {Jehin},
  E., {Leconte}, J., {McCormac}, J., {Murray}, C., {Queloz}, D., {Selsis}, F.,
  {Thompson}, S., and {Van Grootel}, V. (2018).
\newblock {The 0.8-4.5 {$\mu$}m Broadband Transmission Spectra of TRAPPIST-1
  Planets}.
\newblock {\em The Astronomical Journal}, 156:218.

\bibitem[{Erkaev} et~al., 2007]{Erkaev:2007}
{Erkaev}, N.~V., {Kulikov}, Y.~N., {Lammer}, H., {Selsis}, F., {Langmayr}, D.,
  {Jaritz}, G.~F., and {Biernat}, H.~K. (2007).
\newblock {Roche lobe effects on the atmospheric loss from ``Hot Jupiters''}.
\newblock {\em Astronomy $\&$ Astrophysics}, 472:329--334.

\bibitem[Fauchez et~al., 2020]{Fauchez:2020}
Fauchez, T., Villanueva, G., Schwieterman, E.~W., Turbet, M., Arney, G.,
  Pidhorodetska, D., Kopparapu, R.~K., Mandell, A., and Domagal-Goldman, S.~D.
  (2020).
\newblock Sensitive probing of exoplanetary oxygen via mid-infrared collisional
  absorption.
\newblock {\em Nature Astronomy}.

\bibitem[{Fauchez} et~al., 2019]{Fauchez:2019apj}
{Fauchez}, T.~J., {Turbet}, M., {Villanueva}, G.~L., {Wolf}, E.~T., {Arney},
  G., {Kopparapu}, R.~K., {Lincowski}, A., {Mandell}, A., {de Wit}, J.,
  {Pidhorodetska}, D., {Domagal-Goldman}, S.~D., and {Stevenson}, K.~B. (2019).
\newblock {Impact of Clouds and Hazes on the Simulated JWST Transmission
  Spectra of Habitable Zone Planets in the TRAPPIST-1 System}.
\newblock {\em The Astrophysical Journal}, 887(2):194.

\bibitem[{Fauchez} et~al., 2020]{Fauchez:2020gmd}
{Fauchez}, T.~J., {Turbet}, M., {Wolf}, E.~T., {Boutle}, I., {Way}, M.~J., {Del
  Genio}, A.~D., {Mayne}, N.~J., {Tsigaridis}, K., {Kopparapu}, R.~K., {Yang},
  J., {Forget}, F., {Mand ell}, A., and {Domagal Goldman}, S.~D. (2020).
\newblock {TRAPPIST-1 Habitable Atmosphere Intercomparison (THAI): motivations
  and protocol version 1.0}.
\newblock {\em Geoscientific Model Development}, 13(2):707--716.

\bibitem[{Fennelly} and {Torr}, 1992]{Fennelly:1992}
{Fennelly}, J.~A. and {Torr}, D.~G. (1992).
\newblock {Photoionization and Photoabsorption Cross Sections of O, N$_{2}$
  O$_{2}$, and N for Aeronomic Calculations}.
\newblock {\em Atomic Data and Nuclear Data Tables}, 51:321.

\bibitem[{Ferruit} et~al., 2014]{Ferruit:2014}
{Ferruit}, P., {Birkmann}, S., {B{\"o}ker}, T., {Sirianni}, M., {Giardino}, G.,
  {de Marchi}, G., {Alves de Oliveira}, C., and {Dorner}, B. (2014).
\newblock {\em {Observing transiting exoplanets with JWST/NIRSpec}}, volume
  9143 of {\em Society of Photo-Optical Instrumentation Engineers (SPIE)
  Conference Series}, page 91430A.

\bibitem[{Filippazzo} et~al., 2015]{Filippazzo:2015}
{Filippazzo}, J.~C., {Rice}, E.~L., {Faherty}, J., {Cruz}, K.~L., {Van Gordon},
  M.~M., and {Looper}, D.~L. (2015).
\newblock {Fundamental Parameters and Spectral Energy Distributions of Young
  and Field Age Objects with Masses Spanning the Stellar to Planetary Regime}.
\newblock {\em The Astrophysical Journal}, 810(2):158.

\bibitem[{Fleming} et~al., 2020]{Fleming:2020}
{Fleming}, D.~P., {Barnes}, R., {Luger}, R., and {Vand erPlas}, J.~T. (2020).
\newblock {On the XUV Luminosity Evolution of TRAPPIST-1}.
\newblock {\em The Astrophysical Journal}, 891(2):155.

\bibitem[{Follert} et~al., 2014]{Follert:2014}
{Follert}, R., {Dorn}, R.~J., {Oliva}, E., {Lizon}, J.~L., {Hatzes}, A.,
  {Piskunov}, N., {Reiners}, A., {Seemann}, U., {Stempels}, E., {Heiter}, U.,
  {Marquart}, T., {Lockhart}, M., {Anglada-Escude}, G., {L{\"o}winger}, T.,
  {Baade}, D., {Grunhut}, J., {Bristow}, P., {Klein}, B., {Jung}, Y., {Ives},
  D.~J., {Kerber}, F., {Pozna}, E., {Paufique}, J., {Kaeufl}, H.~U., {Origlia},
  L., {Valenti}, E., {Gojak}, D., {Hilker}, M., {Pasquini}, L., {Smette}, A.,
  and {Smoker}, J. (2014).
\newblock {\em {CRIRES+: a cross-dispersed high-resolution infrared
  spectrograph for the ESO VLT}}, volume 9147 of {\em Society of Photo-Optical
  Instrumentation Engineers (SPIE) Conference Series}, page 914719.

\bibitem[{Fraschetti} et~al., 2019]{Fraschetti:2019}
{Fraschetti}, F., {Drake}, J.~J., {Alvarado-G{\'o}mez}, J.~D., {Moschou},
  S.~P., {Garraffo}, C., and {Cohen}, O. (2019).
\newblock {Stellar Energetic Particles in the Magnetically Turbulent Habitable
  Zones of TRAPPIST-1-like Planetary Systems}.
\newblock {\em The Astrophysical Journal}, 874:21.

\bibitem[{Gaia Collaboration} et~al., 2018]{Gaia:2018}
{Gaia Collaboration}, {Brown}, A.~G.~A., {Vallenari}, A., {Prusti}, T., {de
  Bruijne}, J.~H.~J., {Babusiaux}, C., {Bailer-Jones}, C.~A.~L., {Biermann},
  M., {Evans}, D.~W., {Eyer}, L., {Jansen}, F., {Jordi}, C., {Klioner}, S.~A.,
  {Lammers}, U., {Lindegren}, L., {Luri}, X., {Mignard}, F., {Panem}, C.,
  {Pourbaix}, D., {Randich}, S., {Sartoretti}, P., {Siddiqui}, H.~I.,
  {Soubiran}, C., {van Leeuwen}, F., {Walton}, N.~A., {Arenou}, F., {Bastian},
  U., {Cropper}, M., {Drimmel}, R., {Katz}, D., {Lattanzi}, M.~G., {Bakker},
  J., {Cacciari}, C., {Casta{\~n}eda}, J., {Chaoul}, L., {Cheek}, N., {De
  Angeli}, F., {Fabricius}, C., {Guerra}, R., {Holl}, B., {Masana}, E.,
  {Messineo}, R., {Mowlavi}, N., {Nienartowicz}, K., {Panuzzo}, P., {Portell},
  J., {Riello}, M., {Seabroke}, G.~M., {Tanga}, P., {Th{\'e}venin}, F.,
  {Gracia-Abril}, G., {Comoretto}, G., {Garcia-Reinaldos}, M., {Teyssier}, D.,
  {Altmann}, M., {Andrae}, R., {Audard}, M., {Bellas-Velidis}, I., {Benson},
  K., {Berthier}, J., {Blomme}, R., {Burgess}, P., {Busso}, G., {Carry}, B.,
  {Cellino}, A., {Clementini}, G., {Clotet}, M., {Creevey}, O., {Davidson}, M.,
  {De Ridder}, J., {Delchambre}, L., {Dell'Oro}, A., {Ducourant}, C.,
  {Fern{\'a}ndez-Hern{\'a}ndez}, J., {Fouesneau}, M., {Fr{\'e}mat}, Y.,
  {Galluccio}, L., {Garc{\'\i}a-Torres}, M., {Gonz{\'a}lez-N{\'u}{\~n}ez}, J.,
  {Gonz{\'a}lez-Vidal}, J.~J., {Gosset}, E., {Guy}, L.~P., {Halbwachs}, J.~L.,
  {Hambly}, N.~C., {Harrison}, D.~L., {Hern{\'a}ndez}, J., {Hestroffer}, D.,
  {Hodgkin}, S.~T., {Hutton}, A., {Jasniewicz}, G., {Jean-Antoine-Piccolo}, A.,
  {Jordan}, S., {Korn}, A.~J., {Krone-Martins}, A., {Lanzafame}, A.~C.,
  {Lebzelter}, T., {L{\"o}ffler}, W., {Manteiga}, M., {Marrese}, P.~M.,
  {Mart{\'\i}n-Fleitas}, J.~M., {Moitinho}, A., {Mora}, A., {Muinonen}, K.,
  {Osinde}, J., {Pancino}, E., {Pauwels}, T., {Petit}, J.~M., {Recio-Blanco},
  A., {Richards}, P.~J., {Rimoldini}, L., {Robin}, A.~C., {Sarro}, L.~M.,
  {Siopis}, C., {Smith}, M., {Sozzetti}, A., {S{\"u}veges}, M., {Torra}, J.,
  {van Reeven}, W., {Abbas}, U., {Abreu Aramburu}, A., {Accart}, S., {Aerts},
  C., {Altavilla}, G., {{\'A}lvarez}, M.~A., {Alvarez}, R., {Alves}, J.,
  {Anderson}, R.~I., {Andrei}, A.~H., {Anglada Varela}, E., {Antiche}, E.,
  {Antoja}, T., {Arcay}, B., {Astraatmadja}, T.~L., {Bach}, N., {Baker}, S.~G.,
  {Balaguer-N{\'u}{\~n}ez}, L., {Balm}, P., {Barache}, C., {Barata}, C.,
  {Barbato}, D., {Barblan}, F., {Barklem}, P.~S., {Barrado}, D., {Barros}, M.,
  {Barstow}, M.~A., {Bartholom{\'e} Mu{\~n}oz}, S., {Bassilana}, J.~L.,
  {Becciani}, U., {Bellazzini}, M., {Berihuete}, A., {Bertone}, S., {Bianchi},
  L., {Bienaym{\'e}}, O., {Blanco-Cuaresma}, S., {Boch}, T., {Boeche}, C.,
  {Bombrun}, A., {Borrachero}, R., {Bossini}, D., {Bouquillon}, S., {Bourda},
  G., {Bragaglia}, A., {Bramante}, L., {Breddels}, M.~A., {Bressan}, A.,
  {Brouillet}, N., {Br{\"u}semeister}, T., {Brugaletta}, E., {Bucciarelli}, B.,
  {Burlacu}, A., {Busonero}, D., {Butkevich}, A.~G., {Buzzi}, R., {Caffau}, E.,
  {Cancelliere}, R., {Cannizzaro}, G., {Cantat-Gaudin}, T., {Carballo}, R.,
  {Carlucci}, T., {Carrasco}, J.~M., {Casamiquela}, L., {Castellani}, M.,
  {Castro-Ginard}, A., {Charlot}, P., {Chemin}, L., {Chiavassa}, A., {Cocozza},
  G., {Costigan}, G., {Cowell}, S., {Crifo}, F., {Crosta}, M., {Crowley}, C.,
  {Cuypers}, J., {Dafonte}, C., {Damerdji}, Y., {Dapergolas}, A., {David}, P.,
  {David}, M., {de Laverny}, P., {De Luise}, F., {De March}, R., {de Martino},
  D., {de Souza}, R., {de Torres}, A., {Debosscher}, J., {del Pozo}, E.,
  {Delbo}, M., {Delgado}, A., {Delgado}, H.~E., {Di Matteo}, P., {Diakite}, S.,
  {Diener}, C., {Distefano}, E., {Dolding}, C., {Drazinos}, P., {Dur{\'a}n},
  J., {Edvardsson}, B., {Enke}, H., {Eriksson}, K., {Esquej}, P., {Eynard
  Bontemps}, G., {Fabre}, C., {Fabrizio}, M., {Faigler}, S., {Falc{\~a}o},
  A.~J., {Farr{\`a}s Casas}, M., {Federici}, L., {Fedorets}, G., {Fernique},
  P., {Figueras}, F., {Filippi}, F., {Findeisen}, K., {Fonti}, A., {Fraile},
  E., {Fraser}, M., {Fr{\'e}zouls}, B., {Gai}, M., {Galleti}, S., {Garabato},
  D., {Garc{\'\i}a-Sedano}, F., {Garofalo}, A., {Garralda}, N., {Gavel}, A.,
  {Gavras}, P., {Gerssen}, J., {Geyer}, R., {Giacobbe}, P., {Gilmore}, G.,
  {Girona}, S., {Giuffrida}, G., {Glass}, F., {Gomes}, M., {Granvik}, M.,
  {Gueguen}, A., {Guerrier}, A., {Guiraud}, J., {Guti{\'e}rrez-S{\'a}nchez},
  R., {Haigron}, R., {Hatzidimitriou}, D., {Hauser}, M., {Haywood}, M.,
  {Heiter}, U., {Helmi}, A., {Heu}, J., {Hilger}, T., {Hobbs}, D., {Hofmann},
  W., {Holland}, G., {Huckle}, H.~E., {Hypki}, A., {Icardi}, V., {Jan{\ss}en},
  K., {Jevardat de Fombelle}, G., {Jonker}, P.~G., {Juh{\'a}sz}, {\'A}.~L.,
  {Julbe}, F., {Karampelas}, A., {Kewley}, A., {Klar}, J., {Kochoska}, A.,
  {Kohley}, R., {Kolenberg}, K., {Kontizas}, M., {Kontizas}, E., {Koposov},
  S.~E., {Kordopatis}, G., {Kostrzewa-Rutkowska}, Z., {Koubsky}, P., {Lambert},
  S., {Lanza}, A.~F., {Lasne}, Y., {Lavigne}, J.~B., {Le Fustec}, Y., {Le
  Poncin-Lafitte}, C., {Lebreton}, Y., {Leccia}, S., {Leclerc}, N.,
  {Lecoeur-Taibi}, I., {Lenhardt} (2018).
\newblock {Gaia Data Release 2. Summary of the contents and survey properties}.
\newblock {\em Astronomy $\&$ Astrophysics}, 616:A1.

\bibitem[{Gao} et~al., 2015]{Gao:2015}
{Gao}, P., {Hu}, R., {Robinson}, T.~D., {Li}, C., and {Yung}, Y.~L. (2015).
\newblock {Stability of CO2 Atmospheres on Desiccated M Dwarf Exoplanets}.
\newblock {\em The Astrophysical Journal}, 806(2):249.

\bibitem[{Garraffo} et~al., 2017]{Garraffo:2017}
{Garraffo}, C., {Drake}, J.~J., {Cohen}, O., {Alvarado-G{\'o}mez}, J.~D., and
  {Moschou}, S.~P. (2017).
\newblock {The Threatening Magnetic and Plasma Environment of the TRAPPIST-1
  Planets}.
\newblock {\em The Astrophysical Journal Letters}, 843:L33.

\bibitem[{Gillon} et~al., 2013]{Gillon:2013}
{Gillon}, M., {Jehin}, E., {Fumel}, A., {Magain}, P., and {Queloz}, D. (2013).
\newblock {TRAPPIST-UCDTS: A prototype search for habitable planets transiting
  ultra-cool stars}.
\newblock In {\em European Physical Journal Web of Conferences}, volume~47 of
  {\em European Physical Journal Web of Conferences}, page 03001.

\bibitem[{Gillon} et~al., 2016]{Gillon:2016}
{Gillon}, M., {Jehin}, E., {Lederer}, S.~M., {Delrez}, L., {de Wit}, J.,
  {Burdanov}, A., {Van Grootel}, V., {Burgasser}, A.~J., {Triaud}, A.~H.~M.~J.,
  {Opitom}, C., {Demory}, B.-O., {Sahu}, D.~K., {Bardalez Gagliuffi}, D.,
  {Magain}, P., and {Queloz}, D. (2016).
\newblock {Temperate Earth-sized planets transiting a nearby ultracool dwarf
  star}.
\newblock {\em Nature}, 533:221--224.

\bibitem[{Gillon} et~al., 2011]{Gillon:2011}
{Gillon}, M., {Jehin}, E., {Magain}, P., {Chantry}, V., {Hutsem{\'e}kers}, D.,
  {Manfroid}, J., {Queloz}, D., and {Udry}, S. (2011).
\newblock {TRAPPIST: a robotic telescope dedicated to the study of planetary
  systems}.
\newblock In {\em European Physical Journal Web of Conferences}, volume~11 of
  {\em European Physical Journal Web of Conferences}, page 06002.

\bibitem[{Gillon} et~al., 2020]{Gillon:2020}
{Gillon}, M., {Meadows}, V., {Agol}, E., {Burgasser}, A.~J., {Deming}, D.,
  {Doyon}, R., {Fortney}, J., {Kreidberg}, L., {Owen}, J., {Selsis}, F., {de
  Wit}, J., {Lustig-Yaeger}, J., and {Rackham}, B.~V. (2020).
\newblock {The TRAPPIST-1 JWST Community Initiative}.
\newblock {\em arXiv e-prints}, page arXiv:2002.04798.

\bibitem[{Gillon} et~al., 2017]{Gillon:2017}
{Gillon}, M., {Triaud}, A.~H.~M.~J., {Demory}, B.-O., {Jehin}, E., {Agol}, E.,
  {Deck}, K.~M., {Lederer}, S.~M., {de Wit}, J., {Burdanov}, A., {Ingalls},
  J.~G., {Bolmont}, E., {Leconte}, J., {Raymond}, S.~N., {Selsis}, F.,
  {Turbet}, M., {Barkaoui}, K., {Burgasser}, A., {Burleigh}, M.~R., {Carey},
  S.~J., {Chaushev}, A., {Copperwheat}, C.~M., {Delrez}, L., {Fernandes},
  C.~S., {Holdsworth}, D.~L., {Kotze}, E.~J., {Van Grootel}, V., {Almleaky},
  Y., {Benkhaldoun}, Z., {Magain}, P., and {Queloz}, D. (2017).
\newblock {Seven temperate terrestrial planets around the nearby ultracool
  dwarf star TRAPPIST-1}.
\newblock {\em Nature}, 542:456--460.

\bibitem[{Goldblatt} et~al., 2013]{Goldblatt:2013}
{Goldblatt}, C., {Robinson}, T.~D., {Zahnle}, K.~J., and {Crisp}, D. (2013).
\newblock {Low simulated radiation limit for runaway greenhouse climates}.
\newblock {\em Nature Geoscience}, 6:661--667.

\bibitem[{Goldblatt} and {Watson}, 2012]{Goldblatt:2012}
{Goldblatt}, C. and {Watson}, A.~J. (2012).
\newblock {The runaway greenhouse: implications for future climate change,
  geoengineering and planetary atmospheres}.
\newblock {\em Philosophical Transactions of the Royal Society of London Series
  A}, 370:4197--4216.

\bibitem[{Goldreich} and {Peale}, 1966]{Goldreich:1966}
{Goldreich}, P. and {Peale}, S. (1966).
\newblock {Spin-orbit coupling in the solar system}.
\newblock {\em The Astrophysical Journal}, 71:425--+.

\bibitem[{Gonzales} et~al., 2019]{Gonzales:2019}
{Gonzales}, E.~C., {Faherty}, J.~K., {Gagn{\'e}}, J., {Teske}, J., {McWilliam},
  A., and {Cruz}, K. (2019).
\newblock {A Reanalysis of the Fundamental Parameters and Age of TRAPPIST-1}.
\newblock {\em The Astrophysical Journal}, 886(2):131.

\bibitem[{Gordon} et~al., 2017]{Gordon:2017}
{Gordon}, I.~E., {Rothman}, L.~S., {Hill}, C., {Kochanov}, R.~V., {Tan}, Y.,
  {Bernath}, P.~F., {Birk}, M., {Boudon}, V., {Campargue}, A., {Chance}, K.~V.,
  {Drouin}, B.~J., {Flaud}, J.~M., {Gamache}, R.~R., {Hodges}, J.~T.,
  {Jacquemart}, D., {Perevalov}, V.~I., {Perrin}, A., {Shine}, K.~P., {Smith},
  M. A.~H., {Tennyson}, J., {Toon}, G.~C., {Tran}, H., {Tyuterev}, V.~G.,
  {Barbe}, A., {Cs{\'a}sz{\'a}r}, A.~G., {Devi}, V.~M., {Furtenbacher}, T.,
  {Harrison}, J.~J., {Hartmann}, J.~M., {Jolly}, A., {Johnson}, T.~J.,
  {Karman}, T., {Kleiner}, I., {Kyuberis}, A.~A., {Loos}, J., {Lyulin}, O.~M.,
  {Massie}, S.~T., {Mikhailenko}, S.~N., {Moazzen-Ahmadi}, N., {M{\"u}ller},
  H.~S.~P., {Naumenko}, O.~V., {Nikitin}, A.~V., {Polyansky}, O.~L., {Rey}, M.,
  {Rotger}, M., {Sharpe}, S.~W., {Sung}, K., {Starikova}, E., {Tashkun}, S.~A.,
  {Auwera}, J. V.~e., {Wagner}, G., {Wilzewski}, J., {Wcis{\l}o}, P., {Yu}, S.,
  and {Zak}, E.~J. (2017).
\newblock {The HITRAN2016 molecular spectroscopic database}.
\newblock {\em Journal of Quantitative Spectroscopy and Radiative Transfer},
  203:3--69.

\bibitem[{Grimm} et~al., 2018]{Grimm:2018}
{Grimm}, S.~L., {Demory}, B.-O., {Gillon}, M., {Dorn}, C., {Agol}, E.,
  {Burdanov}, A., {Delrez}, L., {Sestovic}, M., {Triaud}, A.~H.~M.~J.,
  {Turbet}, M., {Bolmont}, {\'E}., {Caldas}, A., {de Wit}, J., {Jehin}, E.,
  {Leconte}, J., {Raymond}, S.~N., {Van Grootel}, V., {Burgasser}, A.~J.,
  {Carey}, S., {Fabrycky}, D., {Heng}, K., {Hernandez}, D.~M., {Ingalls},
  J.~G., {Lederer}, S., {Selsis}, F., and {Queloz}, D. (2018).
\newblock {The nature of the TRAPPIST-1 exoplanets}.
\newblock {\em Astronomy and Astrophysics}, 613:A68.

\bibitem[{Hamano} et~al., 2013]{Hamano:2013}
{Hamano}, K., {Abe}, Y., and {Genda}, H. (2013).
\newblock {Emergence of two types of terrestrial planet on solidification of
  magma ocean}.
\newblock {\em Nature}, 497:607--610.

\bibitem[{Hay} and {Matsuyama}, 2019]{Hay:2019}
{Hay}, H. C.~F.~C. and {Matsuyama}, I. (2019).
\newblock {Tides Between the TRAPPIST-1 Planets}.
\newblock {\em The Astrophysical Journal}, 875(1):22.

\bibitem[{Heays} et~al., 2017]{Heays:2017}
{Heays}, A.~N., {Bosman}, A.~D., and {van Dishoeck}, E.~F. (2017).
\newblock {Photodissociation and photoionisation of atoms and molecules of
  astrophysical interest}.
\newblock {\em Astronomy $\&$ Astrophysics}, 602:A105.

\bibitem[{H{\'e}brard} et~al., 2016]{Hebrard:2016}
{H{\'e}brard}, {\'E}.~M., {Donati}, J.~F., {Delfosse}, X., {Morin}, J.,
  {Moutou}, C., and {Boisse}, I. (2016).
\newblock {Modelling the RV jitter of early-M dwarfs using tomographic
  imaging}.
\newblock {\em Monthly Notices of the Royal Astronomical Society},
  461(2):1465--1497.

\bibitem[{Hinkel} et~al., 2014]{Hinkel:2014}
{Hinkel}, N.~R., {Timmes}, F.~X., {Young}, P.~A., {Pagano}, M.~D., and
  {Turnbull}, M.~C. (2014).
\newblock {Stellar Abundances in the Solar Neighborhood: The Hypatia Catalog}.
\newblock {\em The Astronomical Journal}, 148(3):54.

\bibitem[{Hirano} et~al., 2020]{Hirano:2020}
{Hirano}, T., {Gaidos}, E., {Winn}, J.~N., {Dai}, F., {Fukui}, A., {Kuzuhara},
  M., {Kotani}, T., {Tamura}, M., {Hjorth}, M., {Albrecht}, S., {Huber}, D.,
  {Bolmont}, E., {Harakawa}, H., {Hodapp}, K., {Ishizuka}, M., {Jacobson}, S.,
  {Konishi}, M., {Kudo}, T., {Kurokawa}, T., {Nishikawa}, J., {Omiya}, M.,
  {Serizawa}, T., {Ueda}, A., and {Weiss}, L.~M. (2020).
\newblock {Evidence for Spin-Orbit Alignment in the TRAPPIST-1 System}.
\newblock {\em The Astrophysical Journal Letters}, 890(2):L27.

\bibitem[{Holman} et~al., 2010]{Holman:2010}
{Holman}, M.~J., {Fabrycky}, D.~C., {Ragozzine}, D., {Ford}, E.~B., {Steffen},
  J.~H., {Welsh}, W.~F., {Lissauer}, J.~J., {Latham}, D.~W., {Marcy}, G.~W.,
  {Walkowicz}, L.~M., {Batalha}, N.~M., {Jenkins}, J.~M., {Rowe}, J.~F.,
  {Cochran}, W.~D., {Fressin}, F., {Torres}, G., {Buchhave}, L.~A., {Sasselov},
  D.~D., {Borucki}, W.~J., {Koch}, D.~G., {Basri}, G., {Brown}, T.~M.,
  {Caldwell}, D.~A., {Charbonneau}, D., {Dunham}, E.~W., {Gautier}, T.~N.,
  {Geary}, J.~C., {Gilliland}, R.~L., {Haas}, M.~R., {Howell}, S.~B., {Ciardi},
  D.~R., {Endl}, M., {Fischer}, D., {F{\"u}r{\'e}sz}, G., {Hartman}, J.~D.,
  {Isaacson}, H., {Johnson}, J.~A., {MacQueen}, P.~J., {Moorhead}, A.~V.,
  {Morehead}, R.~C., and {Orosz}, J.~A. (2010).
\newblock {Kepler-9: A System of Multiple Planets Transiting a Sun-Like Star,
  Confirmed by Timing Variations}.
\newblock {\em Science}, 330(6000):51.

\bibitem[{Holman} and {Murray}, 2005]{Holman:2005}
{Holman}, M.~J. and {Murray}, N.~W. (2005).
\newblock {The Use of Transit Timing to Detect Terrestrial-Mass Extrasolar
  Planets}.
\newblock {\em Science}, 307:1288--1291.

\bibitem[{Hori} and {Ogihara}, 2020]{Hori:2020}
{Hori}, Y. and {Ogihara}, M. (2020).
\newblock {Do the TRAPPIST-1 Planets Have Hydrogen-rich Atmospheres?}
\newblock {\em The Astrophysical Journal}, 889(2):77.

\bibitem[{Howell} et~al., 2014]{Howell:2014}
{Howell}, S.~B., {Sobeck}, C., {Haas}, M., {Still}, M., {Barclay}, T.,
  {Mullally}, F., {Troeltzsch}, J., {Aigrain}, S., {Bryson}, S.~T., {Caldwell},
  D., {Chaplin}, W.~J., {Cochran}, W.~D., {Huber}, D., {Marcy}, G.~W.,
  {Miglio}, A., {Najita}, J.~R., {Smith}, M., {Twicken}, J.~D., and {Fortney},
  J.~J. (2014).
\newblock {The K2 Mission: Characterization and Early Results}.
\newblock {\em Publications of the Astronomical Society of the Pacific},
  126:398.

\bibitem[{Hu} et~al., 2020]{Hu:2020}
{Hu}, R., {Peterson}, L., and {Wolf}, E.~T. (2020).
\newblock {O$_{2}$- and CO-rich Atmospheres for Potentially Habitable
  Environments on TRAPPIST-1 Planets}.
\newblock {\em The Astrophysical Journal}, 888(2):122.

\bibitem[{Ingersoll}, 1969]{Ingersoll:1969}
{Ingersoll}, A.~P. (1969).
\newblock {The Runaway Greenhouse: A History of Water on Venus.}
\newblock {\em Journal of Atmospheric Sciences}, 26:1191--1198.

\bibitem[{Ingersoll} and {Dobrovolskis}, 1978]{Ingersoll:1978}
{Ingersoll}, A.~P. and {Dobrovolskis}, A.~R. (1978).
\newblock {Venus' rotation and atmospheric tides}.
\newblock {\em Nature}, 275(5675):37--38.

\bibitem[{Izidoro} et~al., 2017]{Izidoro:2017}
{Izidoro}, A., {Ogihara}, M., {Raymond}, S.~N., {Morbidelli}, A., {Pierens},
  A., {Bitsch}, B., {Cossou}, C., and {Hersant}, F. (2017).
\newblock {Breaking the chains: hot super-Earth systems from migration and
  disruption of compact resonant chains}.
\newblock {\em Monthly Notices of the Royal Astronomical Society},
  470(2):1750--1770.

\bibitem[{Johnstone}, 2020]{Johnstone:2019escape}
{Johnstone}, C.~P. (2020).
\newblock {Hydrodynamic Escape of Water Vapor Atmospheres near Very Active
  Stars}.
\newblock {\em The Astrophysical Journal}, 890(1):79.

\bibitem[{Johnstone} et~al., 2018]{Johnstone:2018}
{Johnstone}, C.~P., {G{\"u}del}, M., {Lammer}, H., and {Kislyakova}, K.~G.
  (2018).
\newblock {Upper atmospheres of terrestrial planets: Carbon dioxide cooling and
  the Earth's thermospheric evolution}.
\newblock {\em Astronomy $\&$ Astrophysics}, 617:A107.

\bibitem[{Joshi} et~al., 1997]{Joshi:1997}
{Joshi}, M.~M., {Haberle}, R.~M., and {Reynolds}, R.~T. (1997).
\newblock {Simulations of the Atmospheres of Synchronously Rotating Terrestrial
  Planets Orbiting M Dwarfs: Conditions for Atmospheric Collapse and the
  Implications for Habitability}.
\newblock {\em Icarus}, 129(2):450--465.

\bibitem[{Kameda} et~al., 2019]{Kameda:2019}
{Kameda}, S., {Tavrov}, A., {Murakami}, G., {Enya}, K., {Ikoma}, M., {Narita},
  N., {Fujiwara}, H., {Terada}, N., {Korablev}, O., and {Sachkov}, M. (2019).
\newblock {Observability of oxygen exosphere of an Earth-like exoplanet around
  a low temperature star}.
\newblock In {\em EPSC-DPS Joint Meeting 2019}, volume 2019, pages
  EPSC--DPS2019--1031.

\bibitem[{Kane}, 2018]{Kane:2018}
{Kane}, S.~R. (2018).
\newblock {The Impact of Stellar Distances on Habitable Zone Planets}.
\newblock {\em The Astrophysical Journal Letters}, 861:L21.

\bibitem[{Kasting}, 1982]{Kasting:1982}
{Kasting}, J.~F. (1982).
\newblock {Stability of ammonia in the primitive terrestrial atmosphere}.
\newblock {\em Journal of Geophysical Research}, 87(C4):3091--3098.

\bibitem[{Kasting}, 1988]{Kasting:1988}
{Kasting}, J.~F. (1988).
\newblock {Runaway and moist greenhouse atmospheres and the evolution of earth
  and Venus}.
\newblock {\em Icarus}, 74:472--494.

\bibitem[{Keller-Rudek} et~al., 2013]{Keller-Rudek:2013}
{Keller-Rudek}, H., {Moortgat}, G.~K., {Sander}, R., and {S{\"o}rensen}, R.
  (2013).
\newblock {The MPI-Mainz UV/VIS Spectral Atlas of Gaseous Molecules of
  Atmospheric Interest}.
\newblock {\em Earth System Science Data}, 5(2):365--373.

\bibitem[{Kendrew} et~al., 2015]{Kendrew:2015}
{Kendrew}, S., {Scheithauer}, S., {Bouchet}, P., {Amiaux}, J., {Azzollini}, R.,
  {Bouwman}, J., {Chen}, C.~H., {Dubreuil}, D., {Fischer}, S., {Glasse}, A.,
  {Greene}, T.~P., {Lagage}, P.~O., {Lahuis}, F., {Ronayette}, S., {Wright},
  D., and {Wright}, G.~S. (2015).
\newblock {The Mid-Infrared Instrument for the James Webb Space Telescope, IV:
  The Low-Resolution Spectrometer}.
\newblock {\em Publications of the Astronomical Society of the Pacific},
  127(953):623.

\bibitem[{Kislyakova} et~al., 2017]{Kislyakova:2017}
{Kislyakova}, K.~G., {Noack}, L., {Johnstone}, C.~P., {Zaitsev}, V.~V.,
  {Fossati}, L., {Lammer}, H., {Khodachenko}, M.~L., {Odert}, P., and {Guedel},
  M. (2017).
\newblock {Magma oceans and enhanced volcanism on TRAPPIST-1 planets due to
  induction heating}.
\newblock {\em Nature Astronomy}.

\bibitem[{Klein} and {Donati}, 2019]{Klein:2019}
{Klein}, B. and {Donati}, J.~F. (2019).
\newblock {Simulating radial velocity observations of trappist-1 with SPIRou}.
\newblock {\em Monthly Notices of the Royal Astronomical Society},
  488(4):5114--5126.

\bibitem[{Koll} and {Abbot}, 2016]{Koll:2016}
{Koll}, D. D.~B. and {Abbot}, D.~S. (2016).
\newblock {Temperature Structure and Atmospheric Circulation of Dry Tidally
  Locked Rocky Exoplanets}.
\newblock {\em The Astrophysical Journal}, 825(2):99.

\bibitem[{Koll} et~al., 2019]{Koll:2019}
{Koll}, D. D.~B., {Malik}, M., {Mansfield}, M., {Kempton}, E. M.~R., {Kite},
  E., {Abbot}, D., and {Bean}, J.~L. (2019).
\newblock {Identifying Candidate Atmospheres on Rocky M Dwarf Planets via
  Eclipse Photometry}.
\newblock {\em The Astrophysical Journal}, 886(2):140.

\bibitem[{Komacek} et~al., 2020]{Komacek:2020}
{Komacek}, T.~D., {Fauchez}, T.~J., {Wolf}, E.~T., and {Abbot}, D.~S. (2020).
\newblock {Clouds will Likely Prevent the Detection of Water Vapor in JWST
  Transmission Spectra of Terrestrial Exoplanets}.
\newblock {\em The Astrophysical Journal Letters}, 888(2):L20.

\bibitem[{Kopparapu} et~al., 2013a]{Kopparapu:2013erratum}
{Kopparapu}, R.~K., {Ramirez}, R., {Kasting}, J.~F., {Eymet}, V., {Robinson},
  T.~D., {Mahadevan}, S., {Terrien}, R.~C., {Domagal-Goldman}, S., {Meadows},
  V., and {Deshpande}, R. (2013a).
\newblock {Erratum: ``Habitable Zones around Main-sequence Stars: New
  Estimates'' <A href=``/abs/2013ApJ...765..131K''>(2013, ApJ, 765, 131)</A>}.
\newblock {\em The Astrophysical Journal}, 770:82.

\bibitem[{Kopparapu} et~al., 2013b]{Kopparapu:2013}
{Kopparapu}, R.~K., {Ramirez}, R., {Kasting}, J.~F., {Eymet}, V., {Robinson},
  T.~D., {Mahadevan}, S., {Terrien}, R.~C., {Domagal-Goldman}, S., {Meadows},
  V., and {Deshpande}, R. (2013b).
\newblock {Habitable Zones around Main-sequence Stars: New Estimates}.
\newblock {\em The Astrophysical Journal}, 765:131.

\bibitem[{Kopparapu} et~al., 2016]{Kopparapu:2016}
{Kopparapu}, R.~K., {Wolf}, E.~T., {Haqq-Misra}, J., {Yang}, J., {Kasting},
  J.~F., {Meadows}, V., {Terrien}, R., and {Mahadevan}, S. (2016).
\newblock {The Inner Edge of the Habitable Zone for Synchronously Rotating
  Planets around Low-mass Stars Using General Circulation Models}.
\newblock {\em The Astrophysical Journal}, 819:84.

\bibitem[{Kotani} et~al., 2018]{Kotani:2018}
{Kotani}, T., {Tamura}, M., {Nishikawa}, J., {Ueda}, A., {Kuzuhara}, M.,
  {Omiya}, M., {Hashimoto}, J., {Ishizuka}, M., {Hirano}, T., {Suto}, H.,
  {Kurokawa}, T., {Kokubo}, T., {Mori}, T., {Tanaka}, Y., {Kashiwagi}, K.,
  {Konishi}, M., {Kudo}, T., {Sato}, B., {Jacobson}, S., {Hodapp}, K.~W.,
  {Hall}, D.~B., {Aoki}, W., {Usuda}, T., {Nishiyama}, S., {Nakajima}, T.,
  {Ikeda}, Y., {Yamamuro}, T., {Morino}, J.-I., {Baba}, H., {Hosokawa}, K.,
  {Ishikawa}, H., {Narita}, N., {Kokubo}, E., {Hayano}, Y., {Izumiura}, H.,
  {Kambe}, E., {Kusakabe}, N., {Kwon}, J., {Ikoma}, M., {Hori}, Y., {Genda},
  H., {Fukui}, A., {Fujii}, Y., {Kawahara}, H., {Olivier}, G., {Jovanovic}, N.,
  {Harakawa}, H., {Hayashi}, M., {Hidai}, M., {Machida}, M., {Matsuo}, T.,
  {Nagata}, T., {Ogihara}, M., {Takami}, H., {Takato}, N., {Terada}, H., and
  {Oh}, D. (2018).
\newblock {The infrared Doppler (IRD) instrument for the Subaru telescope:
  instrument description and commissioning results}.
\newblock In {\em Proceedings of the SPIE}, volume 10702 of {\em Society of
  Photo-Optical Instrumentation Engineers (SPIE) Conference Series}, page
  1070211.

\bibitem[{Kral} et~al., 2018]{Kral:2019}
{Kral}, Q., {Wyatt}, M.~C., {Triaud}, A. H.~M.~J., {Marino}, S.,
  {Th{\'e}bault}, P., and {Shorttle}, O. (2018).
\newblock {Cometary impactors on the TRAPPIST-1 planets can destroy all
  planetary atmospheres and rebuild secondary atmospheres on planets f, g, and
  h}.
\newblock {\em Monthly Notices of the Royal Astronomical Society},
  479(2):2649--2672.

\bibitem[{Krasnopolsky}, 2009]{Krasnopolsky:2009}
{Krasnopolsky}, V.~A. (2009).
\newblock {A photochemical model of Titan's atmosphere and ionosphere}.
\newblock {\em Icarus}, 201:226--256.

\bibitem[{Krasnopolsky}, 2014]{Krasnopolsky:2014}
{Krasnopolsky}, V.~A. (2014).
\newblock {Chemical composition of Titan atmosphere and ionosphere:
  Observations and the photochemical model}.
\newblock {\em Icarus}, 236:83--91.

\bibitem[{Kreidberg} et~al., 2019]{Kreidberg:2019}
{Kreidberg}, L., {Koll}, D. D.~B., {Morley}, C., {Hu}, R., {Schaefer}, L.,
  {Deming}, D., {Stevenson}, K.~B., {Dittmann}, J., {Vanderburg}, A.,
  {Berardo}, D., {Guo}, X., {Stassun}, K., {Crossfield}, I., {Charbonneau}, D.,
  {Latham}, D.~W., {Loeb}, A., {Ricker}, G., {Seager}, S., and {Vand erspek},
  R. (2019).
\newblock {Absence of a thick atmosphere on the terrestrial exoplanet LHS
  3844b}.
\newblock {\em Nature}, 573(7772):87--90.

\bibitem[{Krissansen-Totton} et~al., 2018]{Krissansen-Totton:2018}
{Krissansen-Totton}, J., {Garland}, R., {Irwin}, P., and {Catling}, D.~C.
  (2018).
\newblock {Detectability of Biosignatures in Anoxic Atmospheres with the James
  Webb Space Telescope: A TRAPPIST-1e Case Study}.
\newblock {\em The Astronomical Journal}, 156(3):114.

\bibitem[{Lammer} et~al., 2003]{Lammer:2003}
{Lammer}, H., {Selsis}, F., {Ribas}, I., {Guinan}, E.~F., {Bauer}, S.~J., and
  {Weiss}, W.~W. (2003).
\newblock {Atmospheric Loss of Exoplanets Resulting from Stellar X-Ray and
  Extreme-Ultraviolet Heating}.
\newblock {\em ApJl}, 598:L121--L124.

\bibitem[{Leconte} et~al., 2013]{Leconte:2013aa}
{Leconte}, J., {Forget}, F., {Charnay}, B., {Wordsworth}, R., {Selsis}, F.,
  {Millour}, E., and {Spiga}, A. (2013).
\newblock {3D climate modeling of close-in land planets: Circulation patterns,
  climate moist bistability, and habitability}.
\newblock {\em Astronomy $\&$ Astrophysics}, 554:A69.

\bibitem[{Leconte} et~al., 2015]{Leconte:2015}
{Leconte}, J., {Wu}, H., {Menou}, K., and {Murray}, N. (2015).
\newblock {Asynchronous rotation of Earth-mass planets in the habitable zone of
  lower-mass stars}.
\newblock {\em Science}, 347:632--635.

\bibitem[{Liang} et~al., 2007]{Liang:2007}
{Liang}, M.-C., {Heays}, A.~N., {Lewis}, B.~R., {Gibson}, S.~T., and {Yung},
  Y.~L. (2007).
\newblock {Source of Nitrogen Isotope Anomaly in HCN in the Atmosphere of
  Titan}.
\newblock {\em The Astrophysical Journal Letters}, 664:L115--L118.

\bibitem[{Lichtenegger} et~al., 2010]{Lichtenegger:2010}
{Lichtenegger}, H.~I.~M., {Lammer}, H., {Grie{\ss}meier}, J.~M., {Kulikov},
  Y.~N., {von Paris}, P., {Hausleitner}, W., {Krauss}, S., and {Rauer}, H.
  (2010).
\newblock {Aeronomical evidence for higher CO$_{2}$ levels during
  Earth{\textquoteright}s Hadean epoch}.
\newblock {\em Icarus}, 210(1):1--7.

\bibitem[{Lincowski} et~al., 2019]{Lincowski:2019}
{Lincowski}, A.~P., {Lustig-Yaeger}, J., and {Meadows}, V.~S. (2019).
\newblock {Observing Isotopologue Bands in Terrestrial Exoplanet Atmospheres
  with the James Webb Space Telescope: Implications for Identifying Past
  Atmospheric and Ocean Loss}.
\newblock {\em The Astronomical Journal}, 158(1):26.

\bibitem[{Lincowski} et~al., 2018]{Lincowski:2018}
{Lincowski}, A.~P., {Meadows}, V.~S., {Crisp}, D., {Robinson}, T.~D., {Luger},
  R., {Lustig-Yaeger}, J., and {Arney}, G.~N. (2018).
\newblock {Evolved Climates and Observational Discriminants for the TRAPPIST-1
  Planetary System}.
\newblock {\em The Astrophysical Journal}, 867(1):76.

\bibitem[{Lindegren} et~al., 2018]{Lindegren:2018}
{Lindegren}, L., {Hern{\'a}ndez}, J., {Bombrun}, A., {Klioner}, S., {Bastian},
  U., {Ramos-Lerate}, M., {de Torres}, A., {Steidelm{\"u}ller}, H.,
  {Stephenson}, C., {Hobbs}, D., {Lammers}, U., {Biermann}, M., {Geyer}, R.,
  {Hilger}, T., {Michalik}, D., {Stampa}, U., {McMillan}, P.~J.,
  {Casta{\~n}eda}, J., {Clotet}, M., {Comoretto}, G., {Davidson}, M.,
  {Fabricius}, C., {Gracia}, G., {Hambly}, N.~C., {Hutton}, A., {Mora}, A.,
  {Portell}, J., {van Leeuwen}, F., {Abbas}, U., {Abreu}, A., {Altmann}, M.,
  {Andrei}, A., {Anglada}, E., {Balaguer-N{\'u}{\~n}ez}, L., {Barache}, C.,
  {Becciani}, U., {Bertone}, S., {Bianchi}, L., {Bouquillon}, S., {Bourda}, G.,
  {Br{\"u}semeister}, T., {Bucciarelli}, B., {Busonero}, D., {Buzzi}, R.,
  {Cancelliere}, R., {Carlucci}, T., {Charlot}, P., {Cheek}, N., {Crosta}, M.,
  {Crowley}, C., {de Bruijne}, J., {de Felice}, F., {Drimmel}, R., {Esquej},
  P., {Fienga}, A., {Fraile}, E., {Gai}, M., {Garralda}, N.,
  {Gonz{\'a}lez-Vidal}, J.~J., {Guerra}, R., {Hauser}, M., {Hofmann}, W.,
  {Holl}, B., {Jordan}, S., {Lattanzi}, M.~G., {Lenhardt}, H., {Liao}, S.,
  {Licata}, E., {Lister}, T., {L{\"o}ffler}, W., {Marchant}, J.,
  {Martin-Fleitas}, J.~M., {Messineo}, R., {Mignard}, F., {Morbidelli}, R.,
  {Poggio}, E., {Riva}, A., {Rowell}, N., {Salguero}, E., {Sarasso}, M.,
  {Sciacca}, E., {Siddiqui}, H., {Smart}, R.~L., {Spagna}, A., {Steele}, I.,
  {Taris}, F., {Torra}, J., {van Elteren}, A., {van Reeven}, W., and
  {Vecchiato}, A. (2018).
\newblock {Gaia Data Release 2. The astrometric solution}.
\newblock {\em Astronomy $\&$ Astrophysics}, 616:A2.

\bibitem[{Linsky} et~al., 2014]{Linsky:2014}
{Linsky}, J.~L., {Fontenla}, J., and {France}, K. (2014).
\newblock {The Intrinsic Extreme Ultraviolet Fluxes of F5 V TO M5 V Stars}.
\newblock {\em The Astrophysical Journal}, 780:61.

\bibitem[{Lodders} et~al., 2009]{Lodders:2009}
{Lodders}, K., {Palme}, H., and {Gail}, H.~P. (2009).
\newblock {Abundances of the Elements in the Solar System}.
\newblock {\em Landolt B\&ouml;rnstein}, 4B:712.

\bibitem[{Lopez} et~al., 2012]{Lopez:2012}
{Lopez}, E.~D., {Fortney}, J.~J., and {Miller}, N. (2012).
\newblock {How Thermal Evolution and Mass-loss Sculpt Populations of
  Super-Earths and Sub-Neptunes: Application to the Kepler-11 System and
  Beyond}.
\newblock {\em The Astrophysical Journal}, 761(1):59.

\bibitem[{Lorenz} et~al., 1997]{Titan:1997science}
{Lorenz}, R.~D., {McKay}, C.~P., and {Lunine}, J.~I. (1997).
\newblock {Photochemically-induced collapse of Titan's atmosphere}.
\newblock {\em Science}, 275:642--644.

\bibitem[{Luger} and {Barnes}, 2015]{Luger:2015}
{Luger}, R. and {Barnes}, R. (2015).
\newblock {Extreme Water Loss and Abiotic O2Buildup on Planets Throughout the
  Habitable Zones of M Dwarfs}.
\newblock {\em Astrobiology}, 15:119--143.

\bibitem[{Luger} et~al., 2017a]{Luger:2017occultation}
{Luger}, R., {Lustig-Yaeger}, J., and {Agol}, E. (2017a).
\newblock {Planet-Planet Occultations in TRAPPIST-1 and Other Exoplanet
  Systems}.
\newblock {\em The Astrophysical Journal}, 851:94.

\bibitem[{Luger} et~al., 2017b]{Luger:2017}
{Luger}, R., {Sestovic}, M., {Kruse}, E., {Grimm}, S.~L., {Demory}, B.-O.,
  {Agol}, E., {Bolmont}, E., {Fabrycky}, D., {Fernandes}, C.~S., {Van Grootel},
  V., {Burgasser}, A., {Gillon}, M., {Ingalls}, J.~G., {Jehin}, E., {Raymond},
  S.~N., {Selsis}, F., {Triaud}, A.~H.~M.~J., {Barclay}, T., {Barentsen}, G.,
  {Howell}, S.~B., {Delrez}, L., {de Wit}, J., {Foreman-Mackey}, D.,
  {Holdsworth}, D.~L., {Leconte}, J., {Lederer}, S., {Turbet}, M., {Almleaky},
  Y., {Benkhaldoun}, Z., {Magain}, P., {Morris}, B.~M., {Heng}, K., and
  {Queloz}, D. (2017b).
\newblock {A seven-planet resonant chain in TRAPPIST-1}.
\newblock {\em Nature Astronomy}, 1:0129.

\bibitem[{Lustig-Yaeger} et~al., 2019]{Lustig-yaeger:2019}
{Lustig-Yaeger}, J., {Meadows}, V.~S., and {Lincowski}, A.~P. (2019).
\newblock {The Detectability and Characterization of the TRAPPIST-1 Exoplanet
  Atmospheres with JWST}.
\newblock {\em The Astronomical Journal}, 158(1):27.

\bibitem[{MacDonald} and {Dawson}, 2018]{Macdonald:2018}
{MacDonald}, M.~G. and {Dawson}, R.~I. (2018).
\newblock {Three Pathways for Observed Resonant Chains}.
\newblock {\em The Astronomical Journal}, 156(5):228.

\bibitem[{Makarov}, 2012]{Makarov:2012}
{Makarov}, V.~V. (2012).
\newblock {Conditions of Passage and Entrapment of Terrestrial Planets in
  Spin-orbit Resonances}.
\newblock {\em The Astrophysical Journal}, 752(1):73.

\bibitem[{Makarov} et~al., 2018]{Makarov:2018}
{Makarov}, V.~V., {Berghea}, C.~T., and {Efroimsky}, M. (2018).
\newblock {Spin-orbital Tidal Dynamics and Tidal Heating in the TRAPPIST-1
  Multiplanet System}.
\newblock {\em The Astrophysical Journal}, 857(2):142.

\bibitem[{Malik} et~al., 2019]{Malik:2019}
{Malik}, M., {Kempton}, E. M.~R., {Koll}, D. D.~B., {Mansfield}, M., {Bean},
  J.~L., and {Kite}, E. (2019).
\newblock {Analyzing Atmospheric Temperature Profiles and Spectra of M Dwarf
  Rocky Planets}.
\newblock {\em The Astrophysical Journal}, 886(2):142.

\bibitem[{Marino} et~al., 2020]{Marino:2020}
{Marino}, S., {Wyatt}, M.~C., {Kennedy}, G.~M., {Kama}, M., {Matr{\`a}}, L.,
  {Triaud}, A.~H.~M.~J., and {Henning}, T. (2020).
\newblock {Searching for a dusty cometary belt around TRAPPIST-1 with ALMA}.
\newblock {\em Monthly Notices of the Royal Astronomical Society},
  492(4):6067--6073.

\bibitem[{Masset} et~al., 2006]{Masset:2006}
{Masset}, F.~S., {Morbidelli}, A., {Crida}, A., and {Ferreira}, J. (2006).
\newblock {Disk Surface Density Transitions as Protoplanet Traps}.
\newblock {\em The Astrophysical Journal}, 642(1):478--487.

\bibitem[{Mayne} et~al., 2014]{Mayne:2014}
{Mayne}, N.~J., {Baraffe}, I., {Acreman}, D.~M., {Smith}, C., {Wood}, N.,
  {Amundsen}, D.~S., {Thuburn}, J., and {Jackson}, D.~R. (2014).
\newblock {Using the UM dynamical cores to reproduce idealised 3-D flows}.
\newblock {\em Geoscientific Model Development}, 7(6):3059--3087.

\bibitem[{Mayor} and {Queloz}, 1995]{Mayor:1995}
{Mayor}, M. and {Queloz}, D. (1995).
\newblock {A Jupiter-mass companion to a solar-type star}.
\newblock {\em Nature}, 378:355--359.

\bibitem[{McLaughlin}, 1924]{McLaughlin:1924}
{McLaughlin}, D.~B. (1924).
\newblock {Some results of a spectrographic study of the Algol system.}
\newblock {\em The Astrophysical Journal}, 60:22--31.

\bibitem[{Meftah} et~al., 2018]{Meftah:2018}
{Meftah}, M., {Dam{\'e}}, L., {Bols{\'e}e}, D., {Hauchecorne}, A., {Pereira},
  N., {Sluse}, D., {Cessateur}, G., {Irbah}, A., {Bureau}, J., {Weber}, M.,
  {Bramstedt}, K., {Hilbig}, T., {Thi{\'e}blemont}, R., {Marchand}, M.,
  {Lef{\`e}vre}, F., {Sarkissian}, A., and {Bekki}, S. (2018).
\newblock {SOLAR-ISS: A new reference spectrum based on SOLAR/SOLSPEC
  observations}.
\newblock {\em Astronomy $\&$ Astrophysics}, 611:A1.

\bibitem[{Moran} et~al., 2018]{Moran:2018}
{Moran}, S.~E., {H{\"o}rst}, S.~M., {Batalha}, N.~E., {Lewis}, N.~K., and
  {Wakeford}, H.~R. (2018).
\newblock {Limits on Clouds and Hazes for the TRAPPIST-1 Planets}.
\newblock {\em The Astronomical Journal}, 156:252.

\bibitem[{Morley} et~al., 2017]{Morley:2017}
{Morley}, C.~V., {Kreidberg}, L., {Rustamkulov}, Z., {Robinson}, T., and
  {Fortney}, J.~J. (2017).
\newblock {Observing the Atmospheres of Known Temperate Earth-sized Planets
  with JWST}.
\newblock {\em The Astrophysical Journal}, 850:121.

\bibitem[{Morris} et~al., 2018a]{Morris:2018a}
{Morris}, B.~M., {Agol}, E., {Davenport}, J. R.~A., {Hebb}, L., and {Hawley},
  S.~L. (2018a).
\newblock Possible bright starspots on {TRAPPIST}-1.
\newblock {\em The Astrophysical Journal}, 857(1):39.

\bibitem[{Morris} et~al., 2018b]{Morris:2018c}
{Morris}, B.~M., {Agol}, E., {Hebb}, L., and {Hawley}, S.~L. (2018b).
\newblock {Robust Transiting Exoplanet Radii in the Presence of Starspots from
  Ingress and Egress Durations}.
\newblock {\em The Astronomical Journal}, 156(3):91.

\bibitem[{Morris} et~al., 2018c]{Morris:2018b}
{Morris}, B.~M., {Agol}, E., {Hebb}, L., {Hawley}, S.~L., {Gillon}, M.,
  {Ducrot}, E., {Delrez}, L., {Ingalls}, J., and {Demory}, B.-O. (2018c).
\newblock Non-detection of contamination by stellar activity in the spitzer
  transit light curves of {TRAPPIST}-1.
\newblock {\em The Astrophysical Journal}, 863(2):L32.

\bibitem[{Murray-Clay} et~al., 2009]{Murray-Clay:2009}
{Murray-Clay}, R.~A., {Chiang}, E.~I., and {Murray}, N. (2009).
\newblock {Atmospheric Escape From Hot Jupiters}.
\newblock {\em The Astrophysical Journal}, 693(1):23--42.

\bibitem[{Ogihara} and {Ida}, 2009]{Ogihara:2009}
{Ogihara}, M. and {Ida}, S. (2009).
\newblock {N-Body Simulations of Planetary Accretion Around M Dwarf Stars}.
\newblock {\em The Astrophysical Journal}, 699(1):824--838.

\bibitem[{O'Malley-James} and {Kaltenegger}, 2017]{Omalley:2017}
{O'Malley-James}, J.~T. and {Kaltenegger}, L. (2017).
\newblock {UV surface habitability of the TRAPPIST-1 system}.
\newblock {\em Monthly Notices of the Royal Astronomical Society},
  469:L26--L30.

\bibitem[{Ormel} et~al., 2017]{Ormel:2017}
{Ormel}, C.~W., {Liu}, B., and {Schoonenberg}, D. (2017).
\newblock {Formation of TRAPPIST-1 and other compact systems}.
\newblock {\em Astronomy $\&$ Astrophysics}, 604:A1.

\bibitem[{Owen} and {Mohanty}, 2016]{Owen:2016}
{Owen}, J.~E. and {Mohanty}, S. (2016).
\newblock {Habitability of terrestrial-mass planets in the HZ of M Dwarfs - I.
  H/He-dominated atmospheres}.
\newblock {\em Monthly Notices of the Royal Astronomical Society},
  459:4088--4108.

\bibitem[{Papaloizou} et~al., 2018]{Papaloizou:2018}
{Papaloizou}, J.~C.~B., {Szuszkiewicz}, E., and {Terquem}, C. (2018).
\newblock {The TRAPPIST-1 system: orbital evolution, tidal dissipation,
  formation and habitability}.
\newblock {\em Monthly Notices of the Royal Astronomical Society},
  476(4):5032--5056.

\bibitem[{Peacock} et~al., 2019]{Peacock:2019}
{Peacock}, S., {Barman}, T., {Shkolnik}, E.~L., {Hauschildt}, P.~H., and
  {Baron}, E. (2019).
\newblock {Predicting the Extreme Ultraviolet Radiation Environment of
  Exoplanets around Low-mass Stars: The TRAPPIST-1 System}.
\newblock {\em The Astrophysical Journal}, 871:235.

\bibitem[{Pidhorodetska} et~al., 2020]{Pidhorodetska:2020}
{Pidhorodetska}, D., {Fauchez}, T., {Villanueva}, G., and {Domagal-Goldman}, S.
  (2020).
\newblock {Detectability of Molecular Signatures on TRAPPIST-1e through
  Transmission Spectroscopy Simulated for Future Space-Based Observatories}.
\newblock {\em arXiv e-prints}, page arXiv:2001.01338.

\bibitem[{Pierrehumbert} and {Gaidos}, 2011]{Pierrehumbert:2011h2}
{Pierrehumbert}, R. and {Gaidos}, E. (2011).
\newblock {Hydrogen Greenhouse Planets Beyond the Habitable Zone}.
\newblock {\em The Astrophysical Journal Letters}, 734:L13.

\bibitem[{Pierrehumbert}, 2010]{Pier:10book}
{Pierrehumbert}, R.~T. (2010).
\newblock {\em {Principles of Planetary Climate}}.

\bibitem[{Quarles} et~al., 2017]{Quarles:2017}
{Quarles}, B., {Quintana}, E.~V., {Lopez}, E., {Schlieder}, J.~E., and
  {Barclay}, T. (2017).
\newblock {Plausible Compositions of the Seven TRAPPIST-1 Planets Using
  Long-term Dynamical Simulations}.
\newblock {\em The Astrophysical Journal Letters}, 842(1):L5.

\bibitem[{Quirrenbach} et~al., 2014]{Quirrenbach:2014}
{Quirrenbach}, A., {Amado}, P.~J., {Caballero}, J.~A., {Mundt}, R., {Reiners},
  A., {Ribas}, I., {Seifert}, W., {Abril}, M., {Aceituno}, J.,
  {Alonso-Floriano}, F.~J., {Ammler-von Eiff}, M., {Antona Jim{\'e}nez}, R.,
  {Anwand -Heerwart}, H., {Azzaro}, M., {Bauer}, F., {Barrado}, D., {Becerril},
  S., {B{\'e}jar}, V.~J.~S., {Ben{\'\i}tez}, D., {Berdi{\~n}as}, Z.~M.,
  {C{\'a}rdenas}, M.~C., {Casal}, E., {Claret}, A., {Colom{\'e}}, J.,
  {Cort{\'e}s-Contreras}, M., {Czesla}, S., {Doellinger}, M., {Dreizler}, S.,
  {Feiz}, C., {Fern{\'a}ndez}, M., {Galad{\'\i}}, D., {G{\'a}lvez-Ortiz},
  M.~C., {Garc{\'\i}a-Piquer}, A., {Garc{\'\i}a-Vargas}, M.~L., {Garrido}, R.,
  {Gesa}, L., {G{\'o}mez Galera}, V., {Gonz{\'a}lez {\'A}lvarez}, E.,
  {Gonz{\'a}lez Hern{\'a}ndez}, J.~I., {Gr{\"o}zinger}, U., {Gu{\`a}rdia}, J.,
  {Guenther}, E.~W., {de Guindos}, E., {Guti{\'e}rrez-Soto}, J., {Hagen},
  H.~J., {Hatzes}, A.~P., {Hauschildt}, P.~H., {Helmling}, J., {Henning}, T.,
  {Hermann}, D., {Hern{\'a}ndez Casta{\~n}o}, L., {Herrero}, E., {Hidalgo}, D.,
  {Holgado}, G., {Huber}, A., {Huber}, K.~F., {Jeffers}, S., {Joergens}, V.,
  {de Juan}, E., {Kehr}, M., {Klein}, R., {K{\"u}rster}, M., {Lamert}, A.,
  {Lalitha}, S., {Laun}, W., {Lemke}, U., {Lenzen}, R., {L{\'o}pez del Fresno},
  M., {L{\'o}pez Mart{\'\i}}, B., {L{\'o}pez-Santiago}, J., {Mall}, U.,
  {Mandel}, H., {Mart{\'\i}n}, E.~L., {Mart{\'\i}n-Ruiz}, S.,
  {Mart{\'\i}nez-Rodr{\'\i}guez}, H., {Marvin}, C.~J., {Mathar}, R.~J.,
  {Mirabet}, E., {Montes}, D., {Morales Mu{\~n}oz}, R., {Moya}, A., {Naranjo},
  V., {Ofir}, A., {Oreiro}, R., {Pall{\'e}}, E., {Panduro}, J., {Passegger},
  V.~M., {P{\'e}rez-Calpena}, A., {P{\'e}rez Medialdea}, D., {Perger}, M.,
  {Pluto}, M., {Ram{\'o}n}, A., {Rebolo}, R., {Redondo}, P., {Reffert}, S.,
  {Reinhardt}, S., {Rhode}, P., {Rix}, H.~W., {Rodler}, F., {Rodr{\'\i}guez},
  E., {Rodr{\'\i}guez-L{\'o}pez}, C., {Rodr{\'\i}guez-P{\'e}rez}, E.,
  {Rohloff}, R.~R., {Rosich}, A., {S{\'a}nchez-Blanco}, E., {S{\'a}nchez
  Carrasco}, M.~A., {Sanz-Forcada}, J., {Sarmiento}, L.~F., {Sch{\"a}fer}, S.,
  {Schiller}, J., {Schmidt}, C., {Schmitt}, J.~H.~M.~M., {Solano}, E., {Stahl},
  O., {Storz}, C., {St{\"u}rmer}, J., {Su{\'a}rez}, J.~C., {Ulbrich}, R.~G.,
  {Veredas}, G., {Wagner}, K., {Winkler}, J., {Zapatero Osorio}, M.~R.,
  {Zechmeister}, M., {Abell{\'a}n de Paco}, F.~J., {Anglada-Escud{\'e}}, G.,
  {del Burgo}, C., {Klutsch}, A., {Lizon}, J.~L., {L{\'o}pez-Morales}, M.,
  {Morales}, J.~C., {Perryman}, M.~A.~C., {Tulloch}, S.~M., and {Xu}, W.
  (2014).
\newblock {\em {CARMENES instrument overview}}, volume 9147 of {\em Society of
  Photo-Optical Instrumentation Engineers (SPIE) Conference Series}, page
  91471F.

\bibitem[{Rackham} et~al., 2018]{Rackham:2018}
{Rackham}, B.~V., {Apai}, D., and {Giampapa}, M.~S. (2018).
\newblock {The Transit Light Source Effect: False Spectral Features and
  Incorrect Densities for M-dwarf Transiting Planets}.
\newblock {\em The Astrophysical Journal}, 853:122.

\bibitem[{Ramirez} and {Kaltenegger}, 2014]{Ramirez:2014c}
{Ramirez}, R.~M. and {Kaltenegger}, L. (2014).
\newblock {The Habitable Zones of Pre-main-sequence Stars}.
\newblock {\em The Astrophysical Journal Letters}, 797:L25.

\bibitem[{Ramirez} and {Kaltenegger}, 2017]{Ramirez:2017b}
{Ramirez}, R.~M. and {Kaltenegger}, L. (2017).
\newblock {A Volcanic Hydrogen Habitable Zone}.
\newblock {\em The Astrophysical Journal Letters}, 837:L4.

\bibitem[{Ranjan} et~al., 2017]{Ranjan2017}
{Ranjan}, S., {Wordsworth}, R., and {Sasselov}, D.~D. (2017).
\newblock {The Surface UV Environment on Planets Orbiting M Dwarfs:
  Implications for Prebiotic Chemistry and the Need for Experimental
  Follow-up}.
\newblock {\em The Astrophysical Journal}, 843:110.

\bibitem[{Reiners} et~al., 2018]{Reiners:2018}
{Reiners}, A., {Zechmeister}, M., {Caballero}, J.~A., {Ribas}, I., {Morales},
  J.~C., {Jeffers}, S.~V., {Sch{\"o}fer}, P., {Tal-Or}, L., {Quirrenbach}, A.,
  {Amado}, P.~J., {Kaminski}, A., {Seifert}, W., {Abril}, M., {Aceituno}, J.,
  {Alonso-Floriano}, F.~J., {Ammler-von Eiff}, M., {Antona}, R.,
  {Anglada-Escud{\'e}}, G., {Anwand-Heerwart}, H., {Arroyo-Torres}, B.,
  {Azzaro}, M., {Baroch}, D., {Barrado}, D., {Bauer}, F.~F., {Becerril}, S.,
  {B{\'e}jar}, V.~J.~S., {Ben{\'\i}tez}, D., {Berdinas}, Z.~M., {Bergond}, G.,
  {Bl{\"u}mcke}, M., {Brinkm{\"o}ller}, M., {del Burgo}, C., {Cano}, J.,
  {C{\'a}rdenas V{\'a}zquez}, M.~C., {Casal}, E., {Cifuentes}, C., {Claret},
  A., {Colom{\'e}}, J., {Cort{\'e}s-Contreras}, M., {Czesla}, S.,
  {D{\'\i}ez-Alonso}, E., {Dreizler}, S., {Feiz}, C., {Fern{\'a}ndez}, M.,
  {Ferro}, I.~M., {Fuhrmeister}, B., {Galad{\'\i}-Enr{\'\i}quez}, D.,
  {Garcia-Piquer}, A., {Garc{\'\i}a Vargas}, M.~L., {Gesa}, L., {G{\'o}mez
  Galera}, V., {Gonz{\'a}lez Hern{\'a}ndez}, J.~I., {Gonz{\'a}lez-Peinado}, R.,
  {Gr{\"o}zinger}, U., {Grohnert}, S., {Gu{\`a}rdia}, J., {Guenther}, E.~W.,
  {Guijarro}, A., {de Guindos}, E., {Guti{\'e}rrez-Soto}, J., {Hagen}, H.~J.,
  {Hatzes}, A.~P., {Hauschildt}, P.~H., {Hedrosa}, R.~P., {Helmling}, J.,
  {Henning}, T., {Hermelo}, I., {Hern{\'a}ndez Arab{\'\i}}, R., {Hern{\'a}ndez
  Casta{\~n}o}, L., {Hern{\'a}ndez Hernando}, F., {Herrero}, E., {Huber}, A.,
  {Huke}, P., {Johnson}, E.~N., {de Juan}, E., {Kim}, M., {Klein}, R.,
  {Kl{\"u}ter}, J., {Klutsch}, A., {K{\"u}rster}, M., {Lafarga}, M., {Lamert},
  A., {Lamp{\'o}n}, M., {Lara}, L.~M., {Laun}, W., {Lemke}, U., {Lenzen}, R.,
  {Launhardt}, R., {L{\'o}pez del Fresno}, M., {L{\'o}pez-Gonz{\'a}lez}, J.,
  {L{\'o}pez-Puertas}, M., {L{\'o}pez Salas}, J.~F., {L{\'o}pez-Santiago}, J.,
  {Luque}, R., {Mag{\'a}n Madinabeitia}, H., {Mall}, U., {Mancini}, L., {Mand
  el}, H., {Marfil}, E., {Mar{\'\i}n Molina}, J.~A., {Maroto Fern{\'a}ndez},
  D., {Mart{\'\i}n}, E.~L., {Mart{\'\i}n-Ruiz}, S., {Marvin}, C.~J., {Mathar},
  R.~J., {Mirabet}, E., {Montes}, D., {Moreno-Raya}, M.~E., {Moya}, A.,
  {Mundt}, R., {Nagel}, E., {Naranjo}, V., {Nortmann}, L., {Nowak}, G., {Ofir},
  A., {Oreiro}, R., {Pall{\'e}}, E., {Pand uro}, J., {Pascual}, J.,
  {Passegger}, V.~M., {Pavlov}, A., {Pedraz}, S., {P{\'e}rez-Calpena}, A.,
  {P{\'e}rez Medialdea}, D., {Perger}, M., {Perryman}, M.~A.~C., {Pluto}, M.,
  {Rabaza}, O., {Ram{\'o}n}, A., {Rebolo}, R., {Redondo}, P., {Reffert}, S.,
  {Reinhart}, S., {Rhode}, P., {Rix}, H.~W., {Rodler}, F., {Rodr{\'\i}guez},
  E., {Rodr{\'\i}guez-L{\'o}pez}, C., {Rodr{\'\i}guez Trinidad}, A., {Rohloff},
  R.~R., {Rosich}, A., {Sadegi}, S., {S{\'a}nchez-Blanco}, E., {S{\'a}nchez
  Carrasco}, M.~A., {S{\'a}nchez-L{\'o}pez}, A., {Sanz-Forcada}, J., {Sarkis},
  P., {Sarmiento}, L.~F., {Sch{\"a}fer}, S., {Schmitt}, J.~H.~M.~M.,
  {Schiller}, J., {Schweitzer}, A., {Solano}, E., {Stahl}, O., {Strachan},
  J.~B.~P., {St{\"u}rmer}, J., {Su{\'a}rez}, J.~C., {Tabernero}, H.~M., {Tala},
  M., {Trifonov}, T., {Tulloch}, S.~M., {Ulbrich}, R.~G., {Veredas}, G., {Vico
  Linares}, J.~I., {Vilardell}, F., {Wagner}, K., {Winkler}, J., {Wolthoff},
  V., {Xu}, W., {Yan}, F., and {Zapatero Osorio}, M.~R. (2018).
\newblock {The CARMENES search for exoplanets around M dwarfs. High-resolution
  optical and near-infrared spectroscopy of 324 survey stars}.
\newblock {\em Astronomy $\&$ Astrophysics}, 612:A49.

\bibitem[{Ribas} et~al., 2016]{Ribas:2016}
{Ribas}, I., {Bolmont}, E., {Selsis}, F., {Reiners}, A., {Leconte}, J.,
  {Raymond}, S.~N., {Engle}, S.~G., {Guinan}, E.~F., {Morin}, J., {Turbet}, M.,
  {Forget}, F., and {Anglada-Escud{\'e}}, G. (2016).
\newblock {The habitability of Proxima Centauri b. I. Irradiation, rotation and
  volatile inventory from formation to the present}.
\newblock {\em Astronomy $\&$ Astrophysics}, 596:A111.

\bibitem[{Rieke} et~al., 2015]{Rieke:2015}
{Rieke}, G.~H., {Wright}, G.~S., {B{\"o}ker}, T., {Bouwman}, J., {Colina}, L.,
  {Glasse}, A., {Gordon}, K.~D., {Greene}, T.~P., {G{\"u}del}, M., {Henning},
  T., {Justtanont}, K., {Lagage}, P.~O., {Meixner}, M.~E.,
  {N{\o}rgaard-Nielsen}, H.~U., {Ray}, T.~P., {Ressler}, M.~E., {van Dishoeck},
  E.~F., and {Waelkens}, C. (2015).
\newblock {The Mid-Infrared Instrument for the James Webb Space Telescope, I:
  Introduction}.
\newblock {\em Publications of the Astronomical Society of the Pacific},
  127(953):584.

\bibitem[{Rodler} and {L{\'o}pez-Morales}, 2014]{Rodler:2014}
{Rodler}, F. and {L{\'o}pez-Morales}, M. (2014).
\newblock {Feasibility Studies for the Detection of O$_{2}$ in an Earth-like
  Exoplanet}.
\newblock {\em The Astrophysical Journal}, 781(1):54.

\bibitem[{Roettenbacher} and {Kane}, 2017]{Roettenbacher:2017}
{Roettenbacher}, R.~M. and {Kane}, S.~R. (2017).
\newblock {The Stellar Activity of TRAPPIST-1 and Consequences for the
  Planetary Atmospheres}.
\newblock {\em The Astrophysical Journal}, 851:77.

\bibitem[{Rossiter}, 1924]{Rossiter:1924}
{Rossiter}, R.~A. (1924).
\newblock {On the detection of an effect of rotation during eclipse in the
  velocity of the brigher component of beta Lyrae, and on the constancy of
  velocity of this system.}
\newblock {\em The Astrophysical Journal}, 60:15--21.

\bibitem[{Rugheimer} et~al., 2015a]{Rugheimer:2015}
{Rugheimer}, S., {Kaltenegger}, L., {Segura}, A., {Linsky}, J., and {Mohanty},
  S. (2015a).
\newblock {Effect of UV Radiation on the Spectral Fingerprints of Earth-like
  Planets Orbiting M Stars}.
\newblock {\em The Astrophysical Journal}, 809(1):57.

\bibitem[{Rugheimer} et~al., 2015b]{Rugheimer:2015b}
{Rugheimer}, S., {Segura}, A., {Kaltenegger}, L., and {Sasselov}, D. (2015b).
\newblock {UV Surface Environment of Earth-like Planets Orbiting FGKM Stars
  through Geological Evolution}.
\newblock {\em The Astrophysical Journal}, 806(1):137.

\bibitem[{Sagan} and {Chyba}, 1997]{Sagan:1997}
{Sagan}, C. and {Chyba}, C. (1997).
\newblock {The early faint sun paradox: Organic shielding of ultraviolet-labile
  greenhouse gases}.
\newblock {\em Science}, 276:1217--1221.

\bibitem[{Saumon} et~al., 1995]{Saumon:1995}
{Saumon}, D., {Chabrier}, G., and {van Horn}, H.~M. (1995).
\newblock {An Equation of State for Low-Mass Stars and Giant Planets}.
\newblock {\em The Astrophysical Journals}, 99:713.

\bibitem[{Schaefer} et~al., 2016]{Schaefer:2016}
{Schaefer}, L., {Wordsworth}, R.~D., {Berta-Thompson}, Z., and {Sasselov}, D.
  (2016).
\newblock {Predictions of the Atmospheric Composition of GJ 1132b}.
\newblock {\em The Astrophysical Journal}, 829(2):63.

\bibitem[{Schneider} and {Shkolnik}, 2018]{Schneider:2018}
{Schneider}, A.~C. and {Shkolnik}, E.~L. (2018).
\newblock {HAZMAT. III. The UV Evolution of Mid- to Late-M Stars with GALEX}.
\newblock {\em The Astronomical Journal}, 155(3):122.

\bibitem[{Schneider} et~al., 2011]{Schneider:2011}
{Schneider}, J., {Dedieu}, C., {Le Sidaner}, P., {Savalle}, R., and
  {Zolotukhin}, I. (2011).
\newblock {Defining and cataloging exoplanets: the exoplanet.eu database}.
\newblock {\em Astronomy $\&$ Astrophysics}, 532:A79.

\bibitem[{Schoonenberg} et~al., 2019]{Schoonenberg:2019}
{Schoonenberg}, D., {Liu}, B., {Ormel}, C.~W., and {Dorn}, C. (2019).
\newblock {Pebble-driven planet formation for TRAPPIST-1 and other compact
  systems}.
\newblock {\em Astronomy $\&$ Astrophysics}, 627:A149.

\bibitem[{Seager} et~al., 2007]{Seager:2007}
{Seager}, S., {Kuchner}, M., {Hier-Majumder}, C.~A., and {Militzer}, B. (2007).
\newblock {Mass-Radius Relationships for Solid Exoplanets}.
\newblock {\em The Astrophysical Journal}, 669:1279--1297.

\bibitem[{Segura} et~al., 2010]{Segura:2010}
{Segura}, A., {Walkowicz}, L.~M., {Meadows}, V., {Kasting}, J., and {Hawley},
  S. (2010).
\newblock {The Effect of a Strong Stellar Flare on the Atmospheric Chemistry of
  an Earth-like Planet Orbiting an M Dwarf}.
\newblock {\em Astrobiology}, 10(7):751--771.

\bibitem[{Selsis} et~al., 2011]{Selsis:2011}
{Selsis}, F., {Wordsworth}, R.~D., and {Forget}, F. (2011).
\newblock {Thermal phase curves of nontransiting terrestrial exoplanets. I.
  Characterizing atmospheres}.
\newblock {\em Astronomy $\&$ Astrophysics}, 532:A1.

\bibitem[{Serindag} and {Snellen}, 2019]{Serindag:2019}
{Serindag}, D.~B. and {Snellen}, I. A.~G. (2019).
\newblock {Testing the Detectability of Extraterrestrial O$_{2}$ with the
  Extremely Large Telescopes Using Real Data with Real Noise}.
\newblock {\em The Astrophysical Journal Letters}, 871(1):L7.

\bibitem[{Shibata} and {Magara}, 2011]{Shibata:2011}
{Shibata}, K. and {Magara}, T. (2011).
\newblock {Solar Flares: Magnetohydrodynamic Processes}.
\newblock {\em Living Reviews in Solar Physics}, 8(1):6.

\bibitem[{Snellen} et~al., 2013]{Snellen:2013}
{Snellen}, I.~A.~G., {de Kok}, R.~J., {le Poole}, R., {Brogi}, M., and
  {Birkby}, J. (2013).
\newblock {Finding Extraterrestrial Life Using Ground-based High-dispersion
  Spectroscopy}.
\newblock {\em The Astrophysical Journal}, 764(2):182.

\bibitem[{Tamayo} et~al., 2017]{Tamayo:2017}
{Tamayo}, D., {Rein}, H., {Petrovich}, C., and {Murray}, N. (2017).
\newblock {Convergent Migration Renders TRAPPIST-1 Long-lived}.
\newblock {\em The Astrophysical Journal Letters}, 840(2):L19.

\bibitem[{Terquem} and {Papaloizou}, 2007]{Terquem:2007}
{Terquem}, C. and {Papaloizou}, J. C.~B. (2007).
\newblock {Migration and the Formation of Systems of Hot Super-Earths and
  Neptunes}.
\newblock {\em The Astrophysical Journal}, 654(2):1110--1120.

\bibitem[{Tian}, 2009]{Tian:2009}
{Tian}, F. (2009).
\newblock {Thermal Escape from Super Earth Atmospheres in the Habitable Zones
  of M Stars}.
\newblock {\em The Astrophysical Journal}, 703(1):905--909.

\bibitem[{Tian}, 2015]{Tian:2015b}
{Tian}, F. (2015).
\newblock {History of water loss and atmospheric O$_{2}$ buildup on rocky
  exoplanets near M dwarfs}.
\newblock {\em Earth and Planetary Science Letters}, 432:126--132.

\bibitem[{Tian} and {Ida}, 2015]{Tian:2015}
{Tian}, F. and {Ida}, S. (2015).
\newblock {Water contents of Earth-mass planets around M dwarfs}.
\newblock {\em Nature Geoscience}, 8:177--180.

\bibitem[{Tian} et~al., 2008a]{Tian:2008}
{Tian}, F., {Kasting}, J.~F., {Liu}, H.-L., and {Roble}, R.~G. (2008a).
\newblock {Hydrodynamic planetary thermosphere model: 1. Response of the
  Earth's thermosphere to extreme solar EUV conditions and the significance of
  adiabatic cooling}.
\newblock {\em Journal of Geophysical Research (Planets)}, 113(E5):E05008.

\bibitem[{Tian} et~al., 2011]{Tian:2011}
{Tian}, F., {Kasting}, J.~F., and {Zahnle}, K. (2011).
\newblock {Revisiting HCN formation in Earth's early atmosphere}.
\newblock {\em Earth and Planetary Science Letters}, 308:417--423.

\bibitem[{Tian} et~al., 2008b]{Tian:2008b}
{Tian}, F., {Solomon}, S.~C., {Qian}, L., {Lei}, J., and {Roble}, R.~G.
  (2008b).
\newblock {Hydrodynamic planetary thermosphere model: 2. Coupling of an
  electron transport/energy deposition model}.
\newblock {\em Journal of Geophysical Research (Planets)}, 113(E7):E07005.

\bibitem[{Turbet} et~al., 2020a]{Turbet:2020aa}
{Turbet}, M., {Bolmont}, E., {Ehrenreich}, D., {Gratier}, P., {Leconte}, J.,
  {Selsis}, F., {Hara}, N., and {Lovis}, C. (2020a).
\newblock {Revised mass-radius relationships for water-rich rocky planets more
  irradiated than the runaway greenhouse limit }.
\newblock {\em Astronomy $\&$ Astrophysics}.

\bibitem[{Turbet} et~al., 2018]{Turbet:2018aa}
{Turbet}, M., {Bolmont}, E., {Leconte}, J., {Forget}, F., {Selsis}, F.,
  {Tobie}, G., {Caldas}, A., {Naar}, J., and {Gillon}, M. (2018).
\newblock {Modeling climate diversity, tidal dynamics and the fate of volatiles
  on TRAPPIST-1 planets}.
\newblock {\em Astronomy $\&$ Astrophysics}, 612:A86.

\bibitem[{Turbet} et~al., 2020b]{Turbet:2020ch4-h2}
{Turbet}, M., {Boulet}, C., and {Karman}, T. (2020b).
\newblock {Measurements and semi-empirical calculations of CO2+CH4 and CO2+H2
  collision-induced absorption across a wide range of wavelengths and
  temperatures. Application for the prediction of early Mars surface
  temperature}.
\newblock {\em Icarus}, page 113762.

\bibitem[{Turbet} et~al., 2019a]{Turbet:2019aa}
{Turbet}, M., {Ehrenreich}, D., {Lovis}, C., {Bolmont}, E., and {Fauchez}, T.
  (2019a).
\newblock The runaway greenhouse radius inflation effect - an observational
  diagnostic to probe water on earth-sized planets and test the habitable zone
  concept.
\newblock {\em A\&A}, 628:A12.

\bibitem[{Turbet} et~al., 2016]{Turbet:2016}
{Turbet}, M., {Leconte}, J., {Selsis}, F., {Bolmont}, E., {Forget}, F.,
  {Ribas}, I., {Raymond}, S.~N., and {Anglada-Escud{\'e}}, G. (2016).
\newblock {The habitability of Proxima Centauri b. II. Possible climates and
  observability}.
\newblock {\em Astronomy $\&$ Astrophysics}, 596:A112.

\bibitem[{Turbet} et~al., 2019b]{Turbet:2019icarus}
{Turbet}, M., {Tran}, H., {Pirali}, O., {Forget}, F., {Boulet}, C., and
  {Hartmann}, J.-M. (2019b).
\newblock {Far infrared measurements of absorptions by CH$_{4}$ + CO$_{2}$ and
  H$_{2}$ + CO$_{2}$ mixtures and implications for greenhouse warming on early
  Mars}.
\newblock {\em Icarus}, 321:189--199.

\bibitem[{Unterborn} et~al., 2018]{Unterborn:2018}
{Unterborn}, C.~T., {Desch}, S.~J., {Hinkel}, N.~R., and {Lorenzo}, A. (2018).
\newblock {Inward migration of the TRAPPIST-1 planets as inferred from their
  water-rich compositions}.
\newblock {\em Nature Astronomy}, 2:297--302.

\bibitem[{Van Grootel} et~al., 2018]{vangrootel:2018}
{Van Grootel}, V., {Fernandes}, C.~S., {Gillon}, M., {Jehin}, E., {Manfroid},
  J., {Scuflaire}, R., {Burgasser}, A.~J., {Barkaoui}, K., {Benkhaldoun}, Z.,
  {Burdanov}, A., {Delrez}, L., {Demory}, B.-O., {de Wit}, J., {Queloz}, D.,
  and {Triaud}, A.~H.~M.~J. (2018).
\newblock {Stellar Parameters for Trappist-1}.
\newblock {\em The Astrophysical Journal}, 853:30.

\bibitem[{Vida} et~al., 2017]{Vida:2017}
{Vida}, K., {K{\H o}v{\'a}ri}, Z., {P{\'a}l}, A., {Ol{\'a}h}, K., and
  {Kriskovics}, L. (2017).
\newblock {Frequent Flaring in the TRAPPIST-1 System - Unsuited for Life?}
\newblock {\em The Astrophysical Journal}, 841:124.

\bibitem[{Vidal-Madjar} et~al., 2003]{Vidal-Madjar:2003}
{Vidal-Madjar}, A., {Lecavelier des Etangs}, A., {D{\'e}sert}, J.-M.,
  {Ballester}, G.~E., {Ferlet}, R., {H{\'e}brard}, G., and {Mayor}, M. (2003).
\newblock {An extended upper atmosphere around the extrasolar planet
  HD209458b}.
\newblock {\em Nature}, 422:143--146.

\bibitem[{Vinson} and {Hansen}, 2017]{Vinson:2017}
{Vinson}, A.~M. and {Hansen}, B. M.~S. (2017).
\newblock {On the spin states of habitable zone exoplanets around M dwarfs: the
  effect of a near-resonant companion}.
\newblock {\em Monthly Notices of the Royal Astronomical Society},
  472(3):3217--3229.

\bibitem[{Vinson} et~al., 2019]{Vinson:2019}
{Vinson}, A.~M., {Tamayo}, D., and {Hansen}, B. M.~S. (2019).
\newblock {The chaotic nature of TRAPPIST-1 planetary spin states}.
\newblock {\em Monthly Notices of the Royal Astronomical Society},
  488(4):5739--5747.

\bibitem[{Wakeford} et~al., 2019]{Wakeford:2019}
{Wakeford}, H.~R., {Lewis}, N.~K., {Fowler}, J., {Bruno}, G., {Wilson}, T.~J.,
  {Moran}, S.~E., {Valenti}, J., {Batalha}, N.~E., {Filippazzo}, J.,
  {Bourrier}, V., {H{\"o}rst}, S.~M., {Lederer}, S.~M., and {de Wit}, J.
  (2019).
\newblock {Disentangling the Planet from the Star in Late-Type M Dwarfs: A Case
  Study of TRAPPIST-1g}.
\newblock {\em The Astronomical Journal}, 157(1):11.

\bibitem[{Way} et~al., 2017]{Way:2017}
{Way}, M.~J., {Aleinov}, I., {Amundsen}, D.~S., {Chand ler}, M.~A., {Clune},
  T.~L., {Del Genio}, A.~D., {Fujii}, Y., {Kelley}, M., {Kiang}, N.~Y., {Sohl},
  L., and {Tsigaridis}, K. (2017).
\newblock {Resolving Orbital and Climate Keys of Earth and Extraterrestrial
  Environments with Dynamics (ROCKE-3D) 1.0: A General Circulation Model for
  Simulating the Climates of Rocky Planets}.
\newblock {\em The Astrophysical Journals}, 231(1):12.

\bibitem[{Way} et~al., 2020]{Way:2020}
{Way}, M.~J., {Anthony}, and {Del Genio}, D. (2020).
\newblock {Venusian Habitable Climate Scenarios: Modeling Venus through time
  and applications to slowly rotating Venus-Like Exoplanets}.
\newblock {\em arXiv e-prints}, page arXiv:2003.05704.

\bibitem[{Wheatley} et~al., 2017]{Wheatley:2017}
{Wheatley}, P.~J., {Louden}, T., {Bourrier}, V., {Ehrenreich}, D., and
  {Gillon}, M. (2017).
\newblock {Strong XUV irradiation of the Earth-sized exoplanets orbiting the
  ultracool dwarf TRAPPIST-1}.
\newblock {\em Monthly Notices of the Royal Astronomical Society},
  465:L74--L78.

\bibitem[{Wildi} et~al., 2017]{Wildi:2017}
{Wildi}, F., {Blind}, N., {Reshetov}, V., {Hernandez}, O., {Genolet}, L.,
  {Conod}, U., {Sordet}, M., {Segovilla}, A., {Rasilla}, J.~L., {Brousseau},
  D., {Thibault}, S., {Delabre}, B., {Bandy}, T., {Sarajlic}, M., {Cabral}, A.,
  {Bovay}, S., {Vall{\'e}e}, P., {Bouchy}, F., {Doyon}, R., {Artigau}, E.,
  {Pepe}, F., {Hagelberg}, J., {Melo}, C., {Delfosse}, X., {Figueira}, P.,
  {Santos}, N.~C., {Gonz{\'a}lez Hern{\'a}ndez}, J.~I., {de Medeiros}, J.~R.,
  {Rebolo}, R., {Broeg}, C., {Benz}, W., {Boisse}, I., {Malo}, L., {K{\"a}ufl},
  U., and {Saddlemyer}, L. (2017).
\newblock {NIRPS: an adaptive-optics assisted radial velocity spectrograph to
  chase exoplanets around M-stars}.
\newblock In {\em Proceedings of the SPIE}, volume 10400 of {\em Society of
  Photo-Optical Instrumentation Engineers (SPIE) Conference Series}, page
  1040018.

\bibitem[{Wolf}, 2017]{Wolf:2017}
{Wolf}, E.~T. (2017).
\newblock {Assessing the Habitability of the TRAPPIST-1 System Using a 3D
  Climate Model}.
\newblock {\em The Astrophysical Journal Letters}, 839:L1.

\bibitem[{Wolf}, 2018]{Wolf:2018}
{Wolf}, E.~T. (2018).
\newblock {Erratum: Assessing the Habitability of the TRAPPIST-1 System Using a
  3D Climate Model}.
\newblock {\em The Astrophysical Journal Letters}, 855(1):L14.

\bibitem[{Wolf} and {Toon}, 2010]{Wolf:2010}
{Wolf}, E.~T. and {Toon}, O.~B. (2010).
\newblock {Fractal Organic Hazes Provided an Ultraviolet Shield for Early
  Earth}.
\newblock {\em Science}, 328:1266.

\bibitem[{Wolf} and {Toon}, 2015]{Wolf:2015}
{Wolf}, E.~T. and {Toon}, O.~B. (2015).
\newblock {The evolution of habitable climates under the brightening Sun}.
\newblock {\em Journal of Geophysical Research (Atmospheres)}, 120:5775--5794.

\bibitem[{Wordsworth}, 2015]{Wordsworth:2015apj}
{Wordsworth}, R. (2015).
\newblock {Atmospheric Heat Redistribution and Collapse on Tidally Locked Rocky
  Planets}.
\newblock {\em The Astrophysical Journal}, 806(2):180.

\bibitem[{Wordsworth} et~al., 2017]{Wordsworth:2017}
{Wordsworth}, R., {Kalugina}, Y., {Lokshtanov}, S., {Vigasin}, A., {Ehlmann},
  B., {Head}, J., {Sanders}, C., and {Wang}, H. (2017).
\newblock {Transient reducing greenhouse warming on early Mars}.
\newblock {\em Geophysical Research Letters}, 44:665--671.

\bibitem[{Wordsworth} et~al., 2011]{Wordsworth:2011ajl}
{Wordsworth}, R.~D., {Forget}, F., {Selsis}, F., {Millour}, E., {Charnay}, B.,
  and {Madeleine}, J.-B. (2011).
\newblock {Gliese 581d is the First Discovered Terrestrial-mass Exoplanet in
  the Habitable Zone}.
\newblock {\em The Astrophysical Journal Letters}, 733:L48.

\bibitem[{Wordsworth} et~al., 2018]{Wordsworth:2018}
{Wordsworth}, R.~D., {Schaefer}, L.~K., and {Fischer}, R.~A. (2018).
\newblock {Redox Evolution via Gravitational Differentiation on Low-mass
  Planets: Implications for Abiotic Oxygen, Water Loss, and Habitability}.
\newblock {\em The Astronomical Journal}, 155:195.

\bibitem[{Wright}, 2018]{Wright:2018}
{Wright}, J.~T. (2018).
\newblock {Planet-Planet Tides in the TRAPPIST-1 System}.
\newblock {\em Research Notes of the American Astronomical Society}, 2(3):175.

\bibitem[{Wunderlich} et~al., 2019]{Wunderlich:2019}
{Wunderlich}, F., {Godolt}, M., {Grenfell}, J.~L., {St{\"a}dt}, S., {Smith}, A.
  M.~S., {Gebauer}, S., {Schreier}, F., {Hedelt}, P., and {Rauer}, H. (2019).
\newblock {Detectability of atmospheric features of Earth-like planets in the
  habitable zone around M dwarfs}.
\newblock {\em Astronomy $\&$ Astrophysics}, 624:A49.

\bibitem[{Yang} et~al., 2014a]{Yang:2014b}
{Yang}, J., {Bou{\'e}}, G., {Fabrycky}, D.~C., and {Abbot}, D.~S. (2014a).
\newblock {Strong Dependence of the Inner Edge of the Habitable Zone on
  Planetary Rotation Rate}.
\newblock {\em The Astrophysical Journal Letters}, 787:L2.

\bibitem[{Yang} et~al., 2013]{Yang:2013}
{Yang}, J., {Cowan}, N.~B., and {Abbot}, D.~S. (2013).
\newblock {Stabilizing Cloud Feedback Dramatically Expands the Habitable Zone
  of Tidally Locked Planets}.
\newblock {\em The Astrophysical Journal Letters}, 771:L45.

\bibitem[{Yang} et~al., 2014b]{Yang:2014}
{Yang}, J., {Liu}, Y., {Hu}, Y., and {Abbot}, D.~S. (2014b).
\newblock {Water Trapping on Tidally Locked Terrestrial Planets Requires
  Special Conditions}.
\newblock {\em The Astrophysical Journal Letters}, 796:L22.

\bibitem[{Yung} et~al., 1984]{Yung:1984}
{Yung}, Y.~L., {Allen}, M., and {Pinto}, J.~P. (1984).
\newblock {Photochemistry of the atmosphere of Titan - Comparison between model
  and observations}.
\newblock {\em The Astrophysical Journal Supplement Series}, 55:465--506.

\bibitem[{Zeng} et~al., 2016]{Zeng:2016}
{Zeng}, L., {Sasselov}, D.~D., and {Jacobsen}, S.~B. (2016).
\newblock {Mass-Radius Relation for Rocky Planets Based on PREM}.
\newblock {\em The Astrophysical Journal}, 819:127.

\bibitem[{Zhang} et~al., 2018]{Zhang:2018}
{Zhang}, Z., {Zhou}, Y., {Rackham}, B.~V., and {Apai}, D. (2018).
\newblock {The Near-infrared Transmission Spectra of TRAPPIST-1 Planets b, c,
  d, e, f, and g and Stellar Contamination in Multi-epoch Transit Spectra}.
\newblock {\em The Astronomical Journal}, 156(4):178.

\end{thebibliography}

\end{document}